\documentclass[final]{siamltex}
\usepackage{amsfonts}
\usepackage{amssymb}
\usepackage[T1]{fontenc}
\usepackage[latin1]{inputenc}
\usepackage{graphicx}
\usepackage{geometry}
\usepackage{url}
\usepackage{epsfig}
\usepackage{verbatim}
\usepackage{euscript}
\usepackage{afterpage}
\usepackage{graphicx}
\usepackage{amsmath}
\usepackage{amsmath}

\input{tcilatex}
\begin{document}

\title{A new approximate mathematical model for global convergence for a
coefficient inverse problem with backscattering data}
\author{Larisa Beilina\thanks{
Department of Mathematical Sciences, Chalmers University of Technology and
Gothenburg University, SE-42196 Gothenburg, Sweden, (\texttt{%
larisa@chalmers.se}).} \and Michael V. Klibanov \thanks{
Department of Mathematics and Statistics, University of North Carolina at
Charlotte, Charlotte, NC 28223, USA, (\texttt{mklibanv{\@@}uncc.edu}).}}
\maketitle

\begin{abstract}
An approximately globally convergent numerical method for a 3d Coefficient
Inverse Problem for a hyperbolic equation\ with backscattering data is
presented. A new approximate mathematical model is presented. An
approximation is used only on the first iteration and amounts to the
truncation of a certain asymptotic series. A significantly new element of
the convergence analysis is that the so-called \textquotedblleft tail
functions\textquotedblright\ are estimated. Numerical results in 2d and 3d
cases are presented, including the one for a quite heterogeneous medium.
\end{abstract}

%\graphicspath{
%{/chalmers/users/larisa/PAPERS/BackScat_2D/FIGURES/Two_mines/}
%{/chalmers/users/larisa/PAPERS/BackScat_3D/FIGURES/3Dgeometry/}
%{/chalmers/users/larisa/PAPERS/BackScat_3D/FIGURES/LaplaceTr_3Dtransmitted_h0_04s1_21/}
%{/chalmers/users/larisa/PAPERS/BackScat_3D/FIGURES/LaplaceTr_3Dbackscat_h0_04s1_21/}
%{/chalmers/users/larisa/PAPERS/BackScat_3D/FIGURES/Rec_freqstep1/}
%{/chalmers/users/larisa/PAPERS/BackScat_3D/FIGURES/Model2/}
%{/chalmers/users/larisa/PAPERS/BackScat_3D/FIGURES/Ground80/}
% {pics/}}

\pagestyle{myheadings} \thispagestyle{plain} 
\markboth{L.BEILINA AND
M.V.KLIBANOV}{Approximate mathematical model for global convergence}

\section{Introduction}

\label{sec:1}

In this paper we work with a Multidimensional Coefficient Inverse Problem
(MCIP) for a hyperbolic PDE with the data resulting from a single
measurement event. This means that the data are generated by either a single
location of the point source or by a single direction of the incident plane
wave. These MCIPs are non-overdetermined ones. For example, in military
applications the best way is to collect backscattering data resulting from a
single measurement event.\ This is because an installation of each new
source means a life treating risk on the battlefield.

Even though MCIPs have been studied by many researchers since 1960-ies, the
topic of reliable numerical methods for them is still in its infancy. This
is because of \emph{enormous challenges}\textbf{\ }one inevitably faces when
trying to study this topic\textbf{.} Those challenges are caused by two
factors combined: nonlinearity and ill-posedness of MCIPs. In the case of
single measurement the third complicating factor is the minimal amount of
available information. It is well known that conventional least squares
Tikhonov functionals for MCIPs suffer from the phenomenon of multiple local
minima and ravines. Hence, to minimize such a functional, one should apply a
locally convergent numerical method, such as, e.g. Newton-like or
gradient-like method. Convergence of such algorithms can be guaranteed only
if the starting point of iterations is located in a sufficiently small
neighborhood of the exact solution. However, the case when a good
approximation about the solution is known in advance is rare in real
applications.

In a series of recent publications \cite%
{AB,BK,BK1,BK2,BK3,BK4,BKK,KFBPS,KBB,KBK,KBKSNF,IEEE} the authors have used
some properties of underlying PDE operators instead of least squares
functionals. A very important feature of our numerical method is that it
does not require any knowledge of neither the medium inside of the domain of
interest nor of any point in a small neighborhood of the true solution. In
all these publications convergence analysis was confirmed by numerical
examples. Both computationally simulated and experimental data were
considered. In particular, the most challenging case of blind real data
(i.e. when the solution is unknown in advance) was successfully handled in 
\cite{KFBPS,KBKSNF,IEEE} as well as in chapter 5 and section 6.9 of the book 
\cite{BK}.

For the first time, the following two goals were simultaneously achieved for
MCIPs for a hyperbolic PDE with single measurement data:

\textbf{Goal 1}. The development of such a numerical method, which would
have a rigorous guarantee of obtaining at least one point in a small
neighborhood of the exact solution without any advanced knowledge of that
neighborhood.

\textbf{Goal 2}. This numerical method should have a good performance on
computationally simulated data. In addition, if experimental data are
available, then this method should also demonstrate a good performance on
these data.

It is important to achieve both these goals \emph{simultaneously} rather
than just only one of them. Because of the above mentioned difficulties, one
inevitably faces a tough dilemma in an attempt to achieve both Goals 1 and
2: either (1) ignore these goals, or (2) still try to achieve both of them.
Because of this dilemma, it is natural to have the rigorous guarantee of
Goal 1 within the framework of a reasonable approximate mathematical model
(see subsection 3.5). Since convergence is guaranteed in the framework of
that model, then we call our numerical method \emph{approximately globally
convergent}. This model is verified via a six-step procedure described in
section 2.

We are unaware about other numerical methods for MCIPs which would: (a)
simultaneously achieve Goals 1 and 2 and, at the same time, (b) would not
rely on some reasonable approximations, which cannot be rigorously justified.

Compared with \cite{AB,BK1,BK2,BK3,BK4,BKK,KFBPS,KBB,KBK,KBKSNF,IEEE}, there
are three main elements of this paper: (1) We propose a new and more
convenient than before approximate mathematical model, (2) This model leads
to a new convergence analysis, and (3) We test this model on computationally
simulated backscattering data, both in 2d and 3d. We point out that we use
the approximation of our model only on the first iteration of our method,
see the first Remark 3.2 in subsection 3.5.

In the majority of the above cited publications we have considered the case
when the data are given at the entire boundary, i.e. the case of complete
data collection. In the analytical study of this paper we also work with the
case of complete data collection. However, in our numerical studies of
section 6 we assume that only the backscattering Dirichlet data are given.
Next, we use a numerical observation to assign the Dirichlet boundary
condition at the rest of the boundary. Although this condition is an
approximate one, numerical results show a good performance.

The case of the backscattering data was also considered in \cite{KBK} and in
sections 6.1-6.7 of the book \cite{BK}. The 1-d case of blind experimental
backscattering data was considered in \cite{KBKSNF,IEEE} and in section 6.9
of the book \cite{BK}. However, in all references cited in this paragraph
both Dirichlet and Neumann boundary conditions were known at the
backscattering part of the boundary. This led to the Quasi-Reversibility
Method.

In our first publications about this method the so-called \textquotedblleft
tail functions" (subsection 3.2) were not estimated \cite{BK1,BK2}. Unlike
this, in the new approximate model of the current paper tail functions are
estimated on each iteration. Let $C^{k+\alpha }$ be Hölder spaces, where $%
k\geq 0$ is an integer and $\alpha \in \left( 0,1\right) .$ Estimating $%
C^{k+\alpha }$ norms of tail functions requires all results of section 4,
and this is the most\emph{\ }difficult part of our convergence analysis.
Indeed, we estimate functions associated with the fundamental solution of a
certain elliptic PDE, which is valid in the entire space $\mathbb{R}^{3}.$
However, the classical theory provides such estimates only in bounded
domains \cite{LU}. Although lemmata of section 4 and the main theorems 4.2
and 5.1 were published in the book \cite{BK}, we believe that it is worthy
to publish their complete proofs here as well. This is because journal
publications are often better available and for wider audiences of readers
than books.

We search for the spatially distributed dielectric constant. We refer to 
\cite{Smirnov} for a different numerical method for an MCIP of calculating
the dielectric constant. Another non-local numerical method for a 2-d MCIP
for a hyperbolic PDE was developed in \cite{KabS1,KabS2}. It is based on a
2-d analog of the Gel'fand-Levitan-Krein equation. A different version of
our approximately globally convergent numerical method was developed in
parallel with the above publications for the case of a 2d MCIP for an
elliptic PDE with the running source, see \cite{KSu1} and references cited
there. The asymptotic behavior of the tail function in this case is
radically different from ours, which led to a different approximation of
tail functions. This problem has an application in medical optical imaging
of brains, see, e.g. \cite{KSu2} for imaging from an experimental data set
for a phantom medium. A theory of a non-local reconstruction technique for
an MCIP for an elliptic PDE with the data given in the form of the
scattering amplitude was developed in \cite{Nov1,Nov2}. We refer to \cite%
{bib2} for a numerical implementation of this theory. Note that in numerical
studies of non-local reconstruction techniques in \cite{bib2,KabS1,KabS2}
some reasonable approximations were used, which cannot be rigorously
justified. This is similar with our approximate global convergence concept.

In section 2 we present the notion of the approximate global convergence. In
section 3 we describe our algorithm. In particular, we present our
approximate mathematical model in subsection 3.5. In section 4 we prove some
estimates for the function which is the Laplace transform of the solution of
the originating hyperbolic PDE. These estimates are used in section 5 then,
where we prove the approximate global convergence theorem 5.1, which is the
central analytical result of this paper. In section 6 we present results of
our numerical experiments. Summary is given in section 7.

\section{Approximate Global Convergence}

\label{sec:2}

To verify our approximate mathematical models, we use the six step procedure:

\textbf{Step 1.} A reasonable approximate mathematical model is proposed.
The accuracy of this model cannot be rigorously estimated.

\textbf{Step 2.} A numerical method is developed, which works within the
framework of this model.

\textbf{Step 3.} A theorem is proven, which guarantees that, within the
framework of this model, the numerical method of Step 2 indeed delivers a
point in a sufficiently small neighborhood of the exact solution, provided
that the following natural condition is in place: the error, both in the
data and in some \textquotedblleft secondary" additional approximations, is
sufficiently small.

\textbf{Step 4.} The numerical method of Step 2 is tested on computationally
simulated data.

\textbf{Step 5} (optional). The numerical method of Step 2 is tested on
experimental data. To have a truly unbiased case, \emph{blind} data are
preferable. This step is optional because it is usually not easy to actually
get experimental data.

\textbf{Step 6.} Finally, if results of Step 4 and (optionally) Step 5 are
good ones, then Goals 1,2 are simultaneously achieved, and that approximate
mathematical model is proclaimed as a valid one.\ 

It is sufficient to achieve that small neighborhood of the exact solution
after a finite (rather than infinite) number of iterations. Next, because of
approximations in the mathematical model, the resulting solution can be
refined via a locally convergent numerical method. We have chosen the
Adaptive Finite Element Method (adaptivity) for the latter, see \cite%
{BK2,BK3,BK4,BKK} and chapter 4 of \cite{BK}. The algorithms of our previous
publications were successfully verified on two types of \emph{blind}
experimental data, see \cite{BK,KFBPS,KBKSNF,IEEE}. However, since the
authors do not posses a proper experimental data for the algorithm of this
paper, it is verified here only on computationally simulated data.

The common perception of the term \textquotedblleft global convergence" is
that one can choose almost any point as the starting point for iterations,
and still the process would converge to the correct solution. Actually,
however, it is sufficient to start from such a reasonable point, which would
not contain any information about a small neighborhood of the exact
solution. In addition, it is not necessary to converge to the solution. In
fact, it would be sufficient to reach at least one this is going along well
with the theory of Ill-Posed problems, see, e.g. Theorem 4.6 of \cite{BKok}
and pages 156, 157 of \cite{EHN}. Therefore, we come up with Definition 2.1.

Consider a\textbf{\ }nonlinear ill-posed problem $P$. Suppose that this
problem has a unique solution $x^{\ast }\in B$ for the noiseless data $%
y^{\ast },$ where \ $B$ is a Banach space with the norm $\left\Vert \cdot
\right\Vert _{B}.$ We call $x^{\ast }$ \textquotedblleft exact
solution\textquotedblright\ or \textquotedblleft correct
solution\textquotedblright . Suppose that a certain approximate mathematical
model $M$ is proposed to solve the problem $P$ numerically. Assume that,
within the framework of the model $M,$ this problem has unique exact
solution $x_{M}^{\ast }$ and let $x_{M}^{\ast }=x^{\ast }.$

\textbf{Definition 2.1 }(approximate global convergence)\textbf{. } Consider
an iterative numerical method for solving the problem $P$. Suppose that this
method produces a sequence of points $\left\{ x_{n}\right\}
_{n=1}^{N}\subset B,$ where the integer $N\in \left[ 1,\infty \right) .$
Furthermore, assume that this sequence is produced without any \emph{a priori%
} knowledge of a sufficiently small neighborhood of $x^{\ast }.$ Let \ a
sufficiently small number $\varepsilon \in \left( 0,1\right) .$ We call this
numerical method \emph{approximately globally convergent of the level }$%
\varepsilon $\emph{, }or shortly\emph{\ globally convergent}, if, within the
framework of the approximate model $M,$ a theorem is proven, which
guarantees that there exists numbers $N_{1},N_{2}\in \left[ 1,N\right]
,N_{1}<N_{2}$ such that 
\begin{equation*}
\left\Vert x_{n}-x^{\ast }\right\Vert _{B}\leq \varepsilon ,n\in \left[
N_{1},N_{2}\right] .
\end{equation*}%
Suppose that iterations are stopped at a certain number $k\in \left[
N_{1},N_{2}\right] .$ Then the point $x_{k}$ is denoted as $x_{k}:=x_{glob}$
and is called \textquotedblleft the approximate solution resulting from this
method".

The validity of the approximate mathematical model $M$ of Definition 2.1
should be verified via above Steps 4,5. Note that the assumption of the
existence of the exact $x^{\ast }$ for the noiseless data $y^{\ast }$ is one
of the key principles of the theory of Ill-Posed problems \cite{BK,T}.

\section{The Approximately Globally Convergent Method}

\label{sec:3}

This method was described in our above cited publications. However, since we
need to prove a new convergence theorem here, then we need to use some
formulas of this method in the proof. Hence, we outline it here while still
omitting many details for brevity.

\subsection{Statements of forward and inverse problems}

\label{sec:3.1}

Let $\Omega \subset \mathbb{R}^{3}$ be a convex bounded domain with the
boundary $\partial \Omega \in C^{3}.$ Denote $\left\vert f\right\vert
_{k+\alpha }=\left\Vert f\right\Vert _{C^{k+\alpha }\left( \overline{\Omega }%
\right) },$ $\forall f\in C^{k+\alpha }\left( \overline{\Omega }\right) .$
Let $d=const.>2.$ We assume that the coefficient $c\left( x\right) $
satisfies the following conditions 
\begin{equation}
c\left( x\right) \in \lbrack 1,d],~~c\left( x\right) =1\text{ for }x\in 
\mathbb{R}^{3}\diagdown \Omega ,c\in C^{\alpha }\left( \mathbb{R}^{3}\right)
.  \label{3.3}
\end{equation}%
We assume \emph{a priori} knowledge of the constant $d,$ which amounts to
the knowledge of the correctness set in the theory of Ill-Posed problems 
\cite{BKok,BK,EHN,T}. However, we do \emph{not} assume that the number $d-1$
is small, i.e. we do not impose smallness assumptions on the unknown
coefficient $c\left( x\right) $. Consider the Cauchy problem for the
hyperbolic equation 
\begin{eqnarray}
c\left( x\right) u_{tt} &=&\Delta u\text{ in }\mathbb{R}^{3}\times \left(
0,\infty \right) ,  \label{3.1} \\
u\left( x,0\right) &=&0,\text{ }u_{t}\left( x,0\right) =\delta \left(
x-x_{0}\right) .  \label{3.2}
\end{eqnarray}%
Equation (\ref{3.1}) governs, e.g. propagation of acoustic and
electromagnetic waves. In the acoustical case $c(x)=b^{-2}(x),$ where $%
b\left( x\right) $ is the sound speed. In the 2-D case of EM waves
propagation, the dimensionless coefficient is $c(x)=\varepsilon _{r}(x),$
where $\varepsilon _{r}(x)$ is the spatially distributed dielectric constant
of the medium. In \ the latter case the assumption $c\left( x\right) =1$ for 
$x\in \mathbb{R}^{3}\diagdown \Omega $ in (\ref{3.3}) means that we have air
outside the medium of interest $\Omega .$ And the assumption $c\left(
x\right) \geq 1$ reflects the fact that the dielectric constants of almost
all materials exceed the one of the air. Equation (\ref{3.1}) was
successfully used in \cite{BK,BK4,KFBPS} to work with experimental data,
which are obviously in 3-d. The latter was recently explained in \cite{BM},
where the Maxwell's system was solved in time domain. It was shown in Test 4
of \cite{BM} that the component of the electric field $E\left( x,t\right)
=\left( E_{1},E_{2},E_{3}\right) \left( x,t\right) ,$ which was originally
initialized, strongly dominates two other components.

We now formulate the CIP for the case when the data are given at the entire
boundary $\partial \Omega $ of the domain $\Omega .$ We show in section 6
how we reduce the problem with backscattering data to this one.

\textbf{Coefficient Inverse Problem (CIP).} \emph{Assume that the
coefficient }$c\left( x\right) $\emph{\ of equation (\ref{3.1}) satisfies
condition (\ref{3.3}) and is unknown in the domain }$\Omega $\emph{.
Determine the function }$c\left( x\right) $\emph{\ for }$x\in \Omega ,$\emph{%
\ assuming that the following function }$g\left( x,t\right) $\emph{\ is
known for a single source position }$x_{0}\notin \overline{\Omega }$ 
\begin{equation}
u\left( x,t\right) =g\left( x,t\right) ,\forall \left( x,t\right) \in
\partial \Omega \times \left( 0,\infty \right) .  \label{3.4}
\end{equation}

The function $g\left( x,t\right) $ models time dependent measurements of the
wave field at the boundary of the domain of interest. Practical measurements
are calculated at a number of detectors, of course. In this case the
function $g\left( x,t\right) $ can be obtained via one of standard
interpolation procedures. The assumption of the infinite time interval in (%
\ref{3.4}) is not a restrictive one, because we work with the Laplace
transform of the function $u\left( x,t\right) $ and the kernel of this
transform decays rapidly as $t\rightarrow \infty .$ Hence, the integral over
the interval $\left( T,\infty \right) $ is actually discounted in practical
computations. Thus, when generating the data for our CIP, we compute the
forward problem for $t\in \left( 0,T\right) ,$ where $T>0$ is a finite
number. Another argument here is that in our work with experimental data 
\cite{BK,BK4,KFBPS,KBKSNF,IEEE} we have actually used only a small portion
of these data after the data pre-processing procedure.

Global uniqueness theorems for MCIPs with single measurement data are
currently known only under the assumption that at least one of initial
conditions does not equal zero in the entire domain $\overline{\Omega },$
which is not our case. All these theorems were proven by the method, which
was proposed in 1981 by Bukhgeim and Klibanov in \cite{BukhKlib,Bukh,Klib1};
also see, e.g. \cite{Bukh1,Klib2,Klib3} for some follow up publications of
these authors and references in books \cite{BK,KT} for publications of many
other researchers about this method. This method is based on Carleman
estimates. Actually the idea of our approximately globally convergent method
of working with an integral differential equation, which does not contain
the unknown coefficient $c\left( x\right) $, has roots in the method of \cite%
{BukhKlib,Bukh,Bukh1,Klib1,Klib2,Klib3,KT}. There are also some uniqueness
theorems for MCIPs with single measurement data for the case when the
unknown coefficient has the form $const.+a\left( x\right) ,$ where $%
\left\Vert a\right\Vert <<1,$ where $\left\Vert \cdot \right\Vert $ is a
certain norm, see, e.g. \cite{Rom2}.\ Our theory below does not rely on any
smallness assumptions imposed on $c\left( x\right) $. Although we image
inclusions of small geometrical sizes in computations, the
inclusions/background contrasts are not small. Thus, we have no choice but
to assume everywhere below that uniqueness theorem is valid for our CIP.

\subsection{Integral differential equation}

\label{sec:3.2}

Consider the Laplace transform of the function $u$, 
\begin{equation}
w(x,s)=\int\limits_{0}^{\infty }u(x,t)e^{-st}dt,\text{ for }s>\underline{s}%
=const.>0.  \label{3.5}
\end{equation}%
We assume that the number $\underline{s}$ is sufficiently large, so that the
integral (\ref{3.5}) converges absolutely and the same is valid for the
derivatives $D^{k}u,k=0,1,2$. We call the parameter $s$ \emph{pseudo
frequency}. It follows from (\ref{3.1}), (\ref{3.2}) and (\ref{3.5}) that
the function $w$ is the solution of the following problem 
\begin{eqnarray}
\Delta w-s^{2}c\left( x\right) w &=&-\delta \left( x-x_{0}\right) ,\text{ }%
x\in \mathbb{R}^{3},  \label{3.6} \\
\lim_{\left\vert x\right\vert \rightarrow \infty }w\left( x,s\right) &=&0,
\label{3.7}
\end{eqnarray}%
see Theorem 4.1 below about (\ref{3.7}). Since $x_{0}\notin \overline{\Omega 
}$, then Theorem 4.1 also implies that the function $w\in C^{2+\alpha
}\left( \overline{\Omega }\right) $. Suppose that geodesic lines generated
by the function $c\left( x\right) $ are regular and $c\left( x\right) $ is
sufficiently smooth. Let $\tau \left( x,x_{0}\right) $ be the length of the
geodesic line connecting points $x$ and $x_{0}.$ Then Theorem 4.1 of \cite%
{Rom2} implies that the following asymptotic behavior of the function $%
w(x,s) $ at $s\rightarrow \infty $ takes place \cite{BK1,BK} 
\begin{equation}
\left\vert D_{s}^{k}w(x,s)\right\vert _{2+\alpha }=\left\vert
D_{s}^{k}\left\{ \frac{\exp \left[ -s\tau \left( x,x_{0}\right) \right] }{%
f\left( x,x_{0}\right) }\right\} \right\vert _{2+\alpha }\left[ 1+O\left( 
\frac{1}{s}\right) \right] ,s\rightarrow \infty ,k=0,1,  \label{3.8}
\end{equation}%
where\ $f\left( x,x_{0}\right) $ is a certain function and $f\left(
x,x_{0}\right) \neq 0$ for $x\in \overline{\Omega }.$ It is unclear how to 
\emph{effectively} verify the regularity of geodesic lines for generic
functions $c\left( x\right) $. Therefore, we assume below the asymptotic
behavior (\ref{3.8}) without linking it to the regularity of geodesic lines.
We verify the asymptotics (\ref{3.8}) computationally, see \cite{BK1} and
page 173 of \cite{BK}.

It follows from Theorem 4.1 (below) that the function $w(x,s)>0.$ Denote 
\begin{equation}
v\left( x,s\right) :=\frac{\ln w\left( x,s\right) }{s^{2}}.  \label{3.8_1}
\end{equation}%
Assuming that (\ref{3.8}) holds{, we obtain} 
\begin{equation}
\left\vert v\left( x,s\right) \right\vert _{2+\alpha }=O\left(
s^{-k-1}\right) ,~s\rightarrow \infty ,k=0,1.  \label{3.9}
\end{equation}%
Keeping in mind that the source $x_{0}\notin \overline{\Omega }$, we obtain 
\begin{equation}
\Delta v+s^{2}\left( \nabla v\right) ^{2}=c(x),x\in \Omega .  \label{3.10}
\end{equation}%
Differentiate both sides of (\ref{3.10}) with respect to $s$ and let $%
q\left( x,s\right) =\partial _{s}v\left( x,s\right) .$ Hence, 
\begin{eqnarray}
v\left( x,s\right) &=&-\int\limits_{s}^{\overline{s}}q\left( x,\tau \right)
d\tau +V\left( x,\overline{s}\right) ,  \label{3.11} \\
V\left( x,\overline{s}\right) &=&v\left( x,\overline{s}\right) =\frac{\ln
w\left( x,\overline{s}\right) }{\overline{s}^{2}}.  \label{3.12}
\end{eqnarray}%
Here the truncation pseudo frequency $\overline{s}>\underline{s}$ is a large
number. We call $V\left( x,\overline{s}\right) $ the \textquotedblleft tail
function\textquotedblright ,\ and this function is unknown. By (\ref{3.9}) 
\begin{equation}
\left\vert V\left( x,\overline{s}\right) \right\vert _{2+\alpha }=O\left( 
\overline{s}^{-1}\right) ,\text{ }\left\vert \partial _{\overline{s}}V\left(
x,\overline{s}\right) \right\vert _{2+\alpha }=O\left( \overline{s}%
^{-2}\right) ,\overline{s}\rightarrow \infty .  \label{3.13}
\end{equation}%
The number $\overline{s}$ is the main regularization parameter of our
numerical method. In the computational practice $\overline{s}$ is chosen in
numerical experiments\textbf{. }

Thus, we obtain from (\ref{3.10}), (\ref{3.11}) the following nonlinear
integral differential equation 
\begin{equation}
\begin{split}
& \Delta q-2s^{2}\nabla q\int\limits_{s}^{\overline{s}}\nabla q\left( x,\tau
\right) d\tau +2s\left[ \int\limits_{s}^{\overline{s}}\nabla q\left( x,\tau
\right) d\tau \right] ^{2} \\
& +2s^{2}\nabla q\nabla V-4s\nabla V\int\limits_{s}^{\overline{s}}\nabla
q\left( x,\tau \right) d\tau +2s\left( \nabla V\right) ^{2}=0,x\in \Omega .
\end{split}
\label{3.14}
\end{equation}%
By (\ref{3.4}) the following Dirichlet boundary condition is given for the
function $q$ 
\begin{equation}
q\left( x,s\right) =\psi \left( x,s\right) ,\text{ }\forall \left(
x,s\right) \in \partial \Omega \times \left[ \underline{s},\overline{s}%
\right] ,  \label{3.15}
\end{equation}%
where $\psi \left( x,s\right) =s^{-2}\partial _{s}\ln \varphi -2s^{-3}\ln
\varphi $ and $\varphi \left( x,s\right) $ is the Laplace transform (\ref%
{3.5}) of the function $g\left( x,t\right) $ in (\ref{3.4}). Equation (\ref%
{3.14}) has two unknown functions $q$ and $V$. Therefore, to approximate
both these functions, we approximate the function $q$ in \textquotedblleft
inner" iterations and the function $V$ is approximated in \textquotedblleft
outer" iterations, see Remark 3.1 in subsection 3.4.

Suppose for a moment that functions $q$ and $V$ are approximated in $\Omega $
together with their derivatives $D_{x}^{\beta }q,D_{x}^{\beta }V,\left\vert
\beta \right\vert \leq 2.$ Then the corresponding approximation for the
target coefficient can be found via (\ref{3.10}) as 
\begin{equation}
c\left( x\right) =\Delta v+\underline{s}^{2}\left( \nabla v\right) ^{2},x\in
\Omega ,  \label{3.16}
\end{equation}%
where the function $v$ is approximated via (\ref{3.11}). Although any value
of the pseudo frequency $s\in \left[ \underline{s},\overline{s}\right] $ can
be used in (\ref{3.16}), we have found in our numerical experiments that the
best value is $s:=\underline{s}.$

\subsection{Discretization with respect to $s$}

\label{sec:3.3}

We assume that $q\left( x,s\right) $ is a piecewise constant function with
respect $s.$ Hence, we assume that there exists a partition 
\begin{equation}
\underline{s}=s_{N}<s_{N-1}<...<s_{1}<s_{0}=\overline{s},s_{i-1}-s_{i}=h
\label{3.17}
\end{equation}%
of the interval $\left[ \underline{s},\overline{s}\right] $ with a
sufficiently small grid step size $h$ such that 
\begin{equation}
q\left( x,s\right) =q_{n}\left( x\right) \text{ for }s\in
(s_{n},s_{n-1}],q_{0}\equiv 0.  \label{3.18}
\end{equation}%
We approximate the boundary condition (\ref{3.9}) as a piecewise constant
function, $q_{n}\left( x\right) =\overline{\psi }_{n}\left( x\right) ,x\in
\partial \Omega ,$ where $\overline{\psi }_{n}\left( x\right) $ is the
average of the function $\psi \left( x,s\right) $ over the interval $\left(
s_{n},s_{n-1}\right) .$ Next, a certain system of elliptic equations for
functions $q_{n}\left( x\right) $ is derived from (\ref{3.14}) using the $s-$%
dependent so-called \textquotedblleft Carleman Weight Function" $\exp \left[
\lambda \left( s-s_{n-1}\right) \right] ,s\in \left( s_{n},s_{n-1}\right) ,$
where $\lambda >>1$ is a certain parameter of ones choice. Certain numbers $%
A_{1,n},A_{2,n},I_{1,n},I_{0},$which can be analytically calculated, are
involved in that system, and the following estimates hold%
\begin{eqnarray}
\left\vert A_{1,n}\right\vert +\left\vert A_{2,n}\right\vert &\leq &8%
\overline{s}^{2},  \label{3.22} \\
\left\vert \frac{I_{1,n}}{I_{0}}\right\vert &\leq &\frac{4\overline{s}^{2}}{%
\lambda },\text{ if }\lambda h\geq 1,  \label{3.23}
\end{eqnarray}%
Because of (\ref{3.23}) we choose in our computations the parameter $\lambda
>>1$ so large that we can ignore the nonlinear term $\left( \nabla
q_{n}\right) ^{2}$ in that system. Thus, we set below%
\begin{equation}
2\frac{I_{1,n}}{I_{0}}\left( \nabla q_{n}\right) ^{2}:=0.  \label{3.24}
\end{equation}

Our algorithm reconstructs iterative approximations $c_{n,i}\left( x\right)
\in C^{\alpha }\left( \overline{\Omega }\right) $ of the function $c\left(
x\right) $\ only inside the domain $\Omega .$ To work with our algorithm, we
should extend each function $c_{n,i}\left( x\right) $ outside of the domain $%
\Omega $. To do this, choose a smaller subdomain $\Omega ^{\prime }\subset
\Omega ,\partial \Omega ^{\prime }\cap \partial \Omega =\varnothing .$ Let
the function $\chi \left( x\right) $ be such that 
\begin{equation}
\chi \in C^{1}\left( \mathbb{R}^{3}\right) ,\text{ }\chi \left( x\right)
=\left\{ 
\begin{array}{c}
1\text{ in }\Omega ^{\prime }, \\ 
\in \left[ 0,1\right] \text{ in }\Omega \diagdown \Omega ^{\prime }, \\ 
0\text{ outside of }\Omega .%
\end{array}%
\right.  \label{3.24_1}
\end{equation}%
The existence of such functions $\chi \left( x\right) $ is well known from
the Real Analysis course. Let the number $l\geq d.$\ Consider the set of
functions $Q\left( d,l\right) \subset C^{\alpha }\left( \overline{\Omega }%
\right) $ defined as 
\begin{equation*}
Q\left( d,l\right) =\left\{ c\in C^{\alpha }\left( \overline{\Omega }\right)
:c\in \left[ 1,d\right] ,\left\vert c\right\vert _{\alpha }\leq l\right\} .
\end{equation*}%
We assume in our algorithm that all functions $c_{n,i}\in Q\left( d,l\right)
.$ Consider the function $\overline{c}_{n,i}\left( x\right) ,$ 
\begin{equation}
\overline{c}_{n,i}\left( x\right) :=\left( 1-\chi \left( x\right) \right)
+\chi \left( x\right) c_{n,i}\left( x\right) ,\forall x\in \mathbb{R}^{3}.
\label{3.25}
\end{equation}%
Then (\ref{3.24_1}) and (\ref{3.25}) imply that 
\begin{equation*}
\overline{c}_{n,i}\in \left[ 1,d\right] ,\overline{c}_{n,i}\in C^{\alpha
}\left( \mathbb{R}^{3}\right) ,\overline{c}\left( x\right) =1\text{ for }%
x\in \mathbb{R}^{3}\diagdown \Omega .
\end{equation*}

\subsection{The Algorithm}

\label{sec:3.4}

We now describe our algorithm for approximating functions $q_{n}$ and $V$.
Following (\ref{3.11}), (\ref{3.16}) and (\ref{3.18}), denote 
\begin{eqnarray}
v_{n,i}\left( x\right) &=&-hq_{n,i}\left( x\right)
-h\sum\limits_{j=0}^{n-1}q_{j}\left( x\right) +V_{n,i}\left( x\right) ,x\in
\Omega ,  \label{3.26} \\
c_{n,i}\left( x\right) &=&\left[ \Delta v_{n,i}+s_{n}^{2}\left( \nabla
v_{n,i}\right) ^{2}\right] \left( x\right) ,\text{ }x\in \Omega ,i=1,...,m,
\label{3.27}
\end{eqnarray}%
where functions $q_{j},q_{n,i},V_{n,i}$ are defined in this subsection below
and $m$ is the number of iterations with respect to tails for each given $%
n\geq 1$. The number $m$ is chosen in numerical experiments. Here $%
V_{n,i}\left( x\right) $ is a certain approximation for the tail function.
Let $V_{1,1}\left( x\right) $ be the first guess for the tail function,
which is described in subsection 3.5. Hence, to start our iterative process,
we set 
\begin{equation}
q_{0}:=0.\text{ }  \label{3.28}
\end{equation}

\textbf{Step} $n_{i}$, $i\in \left[ 1,m\right] ,n\geq 1,$ see (\ref{3.28})
for $q_{0}$. For each $n$ we iterate with respect to the tails. Suppose that
functions $q_{j},V_{n,i}\left( x,\overline{s}\right) $ $\in C^{2+\alpha
}\left( \overline{\Omega }\right) ,j\in \left[ 0,n-1\right] $ are
constructed. Then we solve the following Dirichlet boundary value problem
for the function $q_{n,i}$

\begin{equation}
\begin{split}
& \Delta q_{n,i}-A_{1n}\left( h\sum\limits_{j=0}^{n-1}\nabla q_{j}\right)
\cdot \nabla q_{n,i}+A_{1n}\nabla q_{n,i}\cdot \nabla V_{n,i}= \\
& -A_{2n}h^{2}\left( \sum\limits_{j=0}^{n-1}\nabla q_{j}\right)
^{2}+2A_{2n}\nabla V_{n,i}\cdot \left( h\sum\limits_{j=0}^{n-1}\nabla
q_{j}\right) -A_{2n}\left( \nabla V_{n,i}\right) ^{2},\text{ }x\in \Omega ,
\\
q_{n,i}& \mid _{\partial \Omega }=\overline{\psi }_{n}\left( x\right) ,
\end{split}
\label{3.29}
\end{equation}
Because of (\ref{3.24}), the nonlinear term with $\left( \nabla
q_{n,i}\right) ^{2}$ is ignored in (\ref{3.29}). Having the function $%
q_{n,i},$ we reconstruct the next approximation $c_{n,i}\in C^{\alpha }(%
\overline{\Omega })$ for the target coefficient using (\ref{3.26}), (\ref%
{3.27}). Next, we construct the function $\overline{c}_{n,i}\in C^{\alpha }(%
\mathbb{R}^{3})$ via (\ref{3.25}). Next, we calculate the solution $%
u_{n,i}\left( x,t\right) $ of the forward problem (\ref{3.1}), (\ref{3.2})
with $c\left( x\right) :=\overline{c}_{n,i}\left( x\right) .$ Next, we
calculate the Laplace transform $w_{n,i}\left( x,\overline{s}\right) $ (\ref%
{3.5}) of the function $u_{n,i}\left( x,t\right) $ at $s:=\overline{s}$ and
update the tail function using (\ref{3.12}), 
\begin{equation}
\frac{\ln w_{n,i}\left( x,\overline{s}\right) }{\overline{s}^{2}}=\left\{ 
\begin{array}{c}
V_{n,i+1}\left( x\right) \text{ if }i\in \left[ 1,m-1\right] , \\ 
V_{n+1,1}\left( x\right) \text{ if }i=m\text{ and }n\in \left[ 1,N-1\right] .%
\end{array}%
\right.  \label{3.30}
\end{equation}%
We set 
\begin{equation}
q_{n}:=q_{n,m}\in C^{2+\alpha }\left( \overline{\Omega }\right)
,c_{n}:=c_{n,m}\in C^{\alpha }\left( \overline{\Omega }\right) .
\label{3.31}
\end{equation}%
If $i=m$ and $n=N$, then we stop. In fact, we can stop the iterative process
not only at $n:=N$ but at $n:=\overline{N}\in \left[ 1,N\right) $ as well.
The stopping rule is chosen in numerical experiments, see section 6.

\textbf{Remark 3.1}. It is clear from (\ref{3.29})-(\ref{3.31}) that
functions $q_{n,i}$ are updated via inner iterations while \textquotedblleft
being inside" the domain $\Omega $ only. But to update tail functions $%
V_{n,i},$ we \textquotedblleft go outside of $\Omega ",$ thus, using outer
iterations.

\subsection{ The new approximate mathematical model and the first guess $%
V_{1,1}\left( x\right) $ for the tail}

\label{sec:3.5}

Following the Tikhonov concept \cite{BK,T}, we assume that there exists
unique exact solution $c^{\ast }\left( x\right) $ of our CIP with noiseless
data $g^{\ast }\left( x,t\right) $ (\ref{3.4}). We assume that 
\begin{equation}
c^{\ast }\in C^{\alpha }\left( \mathbb{R}^{3}\right) ,c^{\ast }\left(
x\right) \in \left[ 1,d-1\right] \text{ in }\mathbb{R}^{3},c^{\ast }\left(
x\right) =1\text{ for }x\in \mathbb{R}^{3}\diagdown \Omega ^{\prime
},\left\vert c^{\ast }\right\vert _{\alpha }\leq l-1.  \label{3.33}
\end{equation}%
For each function $c\in Q\left( d,l\right) $ denote $w_{\overline{c}}\left(
x,s\right) $ the unique solution of the problem (\ref{3.6}), (\ref{3.7})
with $c:=\overline{c}$ satisfying conditions (\ref{4.2})-(\ref{4.5})
(Theorem 4.1), where the function $\overline{c}$ is defined in (\ref{3.25}).
Let $w^{\ast }\left( x,s\right) $ be the solution of the problem (\ref{3.6}%
), (\ref{3.7}) with $c:=c^{\ast }$satisfying conditions (\ref{4.2}), (\ref%
{4.3}). Then (\ref{4.4}) and (\ref{4.5}) are also valid for $w^{\ast }\left(
x,s\right) $ (see Theorem 4.1 in subsection 4.1). Using (\ref{3.12}), we
define tails $V_{\overline{c}}\left( x,s\right) ,V^{\ast }\left( x,s\right) $
as%
\begin{equation}
V_{\overline{c}}\left( x,s\right) =\frac{\ln w_{\overline{c}}\left(
x,s\right) }{s^{2}},V^{\ast }\left( x,s\right) =\frac{\ln w^{\ast }\left(
x,s\right) }{s^{2}},\forall s\geq \overline{s}.  \label{3.34}
\end{equation}%
We call $V^{\ast }\left( x,s\right) $ the \textquotedblleft exact tail".
Assuming that the asymptotic behavior (\ref{3.8}) holds, we obtain%
\begin{equation}
V^{\ast }\left( x,s\right) =\frac{p^{\ast }\left( x\right) }{s}+O\left( 
\frac{1}{s^{2}}\right) ,s\rightarrow \infty ,x\in \overline{\Omega }.
\label{3.300}
\end{equation}%
for a certain function $p^{\ast }\left( x\right) .$ We truncate the second
term of this asymptotic behavior. Thus, our \textbf{new approximate
mathematical model} consists of the following assumption.

\textbf{Assumption. }There exists a function $p^{\ast }\left( x\right) \in
C^{2+\alpha }\left( \overline{\Omega }\right) $ such that the exact tail
function $V^{\ast }\left( x,s\right) $ has the form 
\begin{equation}
V^{\ast }\left( x,s\right) :=\frac{p^{\ast }\left( x\right) }{s}.\text{
Furthermore, }\frac{p^{\ast }\left( x\right) }{s}=\frac{\ln w^{\ast }\left(
x,s\right) }{s^{2}},\text{ }\forall s\geq \overline{s}.\text{ }  \label{3.35}
\end{equation}

Since $q^{\ast }\left( x,s\right) =\partial _{s}V^{\ast }\left( x,s\right) $
for $s\geq \overline{s},$ we derive from (\ref{3.35}) that 
\begin{equation}
q^{\ast }\left( x,\overline{s}\right) =-\frac{p^{\ast }\left( x\right) }{%
\overline{s}^{2}}\text{.}  \label{3.36}
\end{equation}%
Set in (\ref{3.14}) $s=\overline{s}$. Then, using (\ref{3.35}) and (\ref%
{3.36}), we obtain the following approximate Dirichlet boundary value
problem for the function $p^{\ast }\left( x\right) $ 
\begin{eqnarray}
\Delta p^{\ast } &=&0\text{ in }\Omega ,\text{ }p^{\ast }\in C^{2+\alpha
}\left( \overline{\Omega }\right) ,  \label{3.37} \\
p^{\ast }|_{\partial \Omega } &=&-\overline{s}^{2}\psi ^{\ast }\left( x,%
\overline{s}\right) ,  \label{3.38}
\end{eqnarray}%
where $\psi ^{\ast }\left( x,s\right) $ is the exact function $\psi \left(
x,s\right) ,$ which corresponds to the function $g^{\ast }\left( x,t\right)
. $ The approximate equation (\ref{3.37}) is valid only within the framework
of the above Assumption. Although this equation is linear, formula (\ref%
{3.16}) for the reconstruction of the target coefficient $c^{\ast }$ is
nonlinear.

Recall that by (\ref{3.15}) $q\left( x,s\right) =\psi \left( x,s\right) ,$ $%
\forall \left( x,s\right) \in \partial \Omega \times \left[ \underline{s},%
\overline{s}\right] .$ Assume that 
\begin{equation}
\psi \left( x,s\right) \in C^{2+\alpha }\left( \overline{\Omega }\right)
,\forall s\in \left[ \underline{s},\overline{s}\right] .  \label{3.39}
\end{equation}%
Consider the solution $p\left( x\right) $ of the following boundary value
problem 
\begin{eqnarray}
\Delta p &=&0\text{ in }\Omega ,\text{ }p\in C^{2+\alpha }\left( \overline{%
\Omega }\right) ,  \label{3.40} \\
p|_{\partial \Omega } &=&-\overline{s}^{2}\psi \left( x,\overline{s}\right) .
\label{3.41}
\end{eqnarray}%
As the first guess for the tail function we take 
\begin{equation}
V_{1,1}\left( x\right) :=\frac{p\left( x\right) }{\overline{s}}.
\label{3.43}
\end{equation}%
By the Schauder theorem there exists unique solution $p$ of the problem (\ref%
{3.40}), (\ref{3.41}).\ Furthermore, it follows from (\ref{3.37})-(\ref{3.43}%
) and Schauder theorem that with a number $M=M\left( \Omega \right) >0$ the
following estimates hold 
\begin{eqnarray}
\left\vert \nabla V_{1,1}-\nabla V^{\ast }\right\vert _{1+\alpha } &\leq
&M\left\Vert \psi \left( x,\overline{s}\right) -\psi ^{\ast }\left( x,%
\overline{s}\right) \right\Vert _{C^{2+\alpha }\left( \partial \Omega
\right) },  \label{3.42} \\
\left\vert \nabla V_{1,1}\right\vert _{1+\alpha } &\leq &M\overline{s}%
\left\Vert \psi \left( x,\overline{s}\right) \right\Vert _{C^{2+\alpha
}\left( \partial \Omega \right) }.  \label{3.420}
\end{eqnarray}

\textbf{Remarks 3.2.}

\textbf{1}. The main approximation is the second equality (\ref{3.35}).\emph{%
\ }This approximation amounts to the truncation of the second term $O\left(
s^{-2}\right) $ of the asymptotics (\ref{3.300}). It is made only to obtain
the estimate (\ref{3.42}) on the first iteration of our method to obtain
estimate (\ref{5.322}) in the proof of the convergence Theorem 5.1. On all
follow up iterations in that proof we do not use the second equality (\ref%
{3.35}). Rather, we use the true fact that $V^{\ast }\left( x\right) =%
\overline{s}^{-2}\ln w^{\ast }\left( x,\overline{s}\right) .$

\textbf{2}. It follows from (\ref{3.42}) that, substituting (\ref{3.43}) in (%
\ref{3.26}) and (\ref{3.27}) at $n=1$ and setting $q_{1,i}:=0$, we obtain a
good approximation for the exact solution already on the first iteration of
our method, as long as the error in the boundary data $\psi ^{\ast }\left( x,%
\overline{s}\right) $ is small.\ The smallness of the error is a natural
assumption. Theorem 5.1 guarantees that all other solutions obtained in the
iterative process of subsection 3.4 also provide good approximations, as
long as the number of iterations is not too large. This means that we should
develop numerically a stopping criterion to stop iterations, see section 6.
Suppose now that iterations are stopped before this stopping criterion is
met, e.g. just on the first iteration. In this case we can apply the second
stage of our two-stage numerical procedure \cite{BK,BK2,BK3,BK4,BKK}.
Namely, we could apply a locally convergent numerical method to refine the
solution via taking the solution obtained on the globally convergent stage
as the starting point of iterations. Such a method can be applied indeed,
since Theorem 5.1 guarantees that the iterative solution, at which we have
stopped, is close to the exact solution $c^{\ast }\left( x\right) .$
Numerical confirmations of this can be found in tests 2,3 of \cite{BK2} and
in tests 2,3 of section 4.16.2 of \cite{BK}.

We now establish uniqueness within the framework of our approximate
mathematical model. We refer to Lemma 2.9.2 of \cite{BK} for the proof of
Lemma 3.1.

\textbf{Lemma 3.1.} \emph{Let the above Assumption holds. In addition, let (%
\ref{3.16}) holds for the function }$v^{\ast }\left( x,s\right) ,$\emph{\ }%
\begin{equation*}
v^{\ast }\left( x,s\right) =-\int\limits_{s}^{\overline{s}}q^{\ast }\left(
x,\tau \right) d\tau +V^{\ast }\left( x,\overline{s}\right)
\end{equation*}%
\emph{\ with the tail function }$V^{\ast }\left( x,s\right) $\ \emph{%
satisfying conditions (\ref{3.35}), (\ref{3.37}) and (\ref{3.38}), i.e. }%
\begin{equation*}
c^{\ast }\left( x\right) =\left[ \Delta v^{\ast }+s^{2}\left\vert \nabla
v^{\ast }\right\vert ^{2}\right] \left( x,s\right) ,\left( x,s\right) \in
\Omega \times \left[ \underline{s},\overline{s}\right] .
\end{equation*}%
\emph{Then there exists at most one function }$c^{\ast }\left( x\right) .$

\section{Estimates of Tails}

\label{sec:4}

\subsection{Some estimates of the function $w\left( x,s\right) $}

\label{sec:4.1}

First, we should justify (\ref{3.6}), (\ref{3.7}). Theorem 4.1 is a
combination of Theorems 2.7.1 and 2.7.2 of \cite{BK}.\ Therefore we refer to 
\cite{BK} for the proof.

\textbf{Theorem 4.1}. \emph{Assume that the coefficient }$c\left( x\right) $%
\emph{\ of equation (\ref{3.1}) is such that }%
\begin{equation}
c\in C^{k+\alpha }\left( \mathbb{R}^{3}\right) ,c\left( x\right) \in \lbrack
1,d],~~c\left( x\right) =1\text{ for }x\in \mathbb{R}^{3}\diagdown \Omega .
\label{4.1}
\end{equation}%
\emph{Assume that the solution }$u\left( x,t\right) $\emph{\ of the problem (%
\ref{3.1}), (\ref{3.2}) is such that }%
\begin{equation*}
u\in C^{2}\left( t\geq M_{1}\left( x,x_{0},c\right) \right) ,\left\vert
D^{\gamma }u\left( x,t\right) \right\vert \leq M_{2}\left( x_{0},c\right)
e^{\beta t},\left\vert \gamma \right\vert \leq 2,\forall x\in \mathbb{R}%
^{3},\forall t\geq M_{1}\left( x,x_{0},c\right) ,
\end{equation*}%
\emph{where }$M_{1}\left( x,x_{0},c\right) $\emph{\ and} $\beta =\beta
\left( c\right) $\emph{\ are positive numbers depending on listed parameters}%
.\emph{\ Let the function }$w\left( x,s\right) $ be\emph{\ the Laplace
transform (\ref{3.5}) of the function }$u\left( x,t\right) .$ \emph{Then
there exists a number }$\underline{s}=\underline{s}\left( c\right) >\beta
\left( c\right) $\emph{\ such that for all }$s\geq \underline{s}\left(
c\right) $\emph{\ the function }$w\left( x,s\right) $\emph{\ is the unique
solution of the problem (\ref{3.6}), (\ref{3.7}) of the form}%
\begin{eqnarray}
w\left( x,s\right) &=&\frac{\exp \left( -s\left\vert x-x_{0}\right\vert
\right) }{4\pi \left\vert x-x_{0}\right\vert }+\overline{w}\left( x,s\right)
:=w_{1}\left( x,s\right) +\overline{w}\left( x,s\right) ,  \label{4.2} \\
\overline{w} &\in &C^{k+2+\alpha }\left( \mathbb{R}^{3}\right) .  \label{4.3}
\end{eqnarray}%
\emph{Also, the following inequalities hold}%
\begin{equation}
w_{d}\left( x,s\right) <w\left( x,s\right) \leq w_{1}\left( x,s\right)
,\forall x\neq x_{0},  \label{4.4}
\end{equation}%
\emph{where the function }$w_{d}\left( x,s\right) $\emph{\ is the unique
solution of the problem (\ref{3.6}), (\ref{3.7}) for the case }$c\left(
x\right) \equiv d,$%
\begin{equation}
w_{d}\left( x,s\right) =\frac{\exp \left( -s\sqrt{d}\left\vert
x-x_{0}\right\vert \right) }{4\pi \left\vert x-x_{0}\right\vert }.
\label{4.5}
\end{equation}%
\emph{Furthermore, consider the problem (\ref{3.6}), (\ref{3.7})
irrelevantly to the problem (\ref{3.1}), (\ref{3.2}) while still assuming (%
\ref{4.1}). Then for any }$s>0$\emph{\ there exists unique solution }$%
w\left( x,s\right) $\emph{\ of this problem satisfying conditions (\ref{4.2}%
), (\ref{4.3}). Furthermore, conditions (\ref{4.4}), (\ref{4.5}) hold for
this function }$w\left( x,s\right) .$

\textbf{Lemma 4.1}. \emph{Let }$\chi \left( x\right) $\emph{\ be the
function defined in (\ref{3.24_1}), functions }$c_{1},c_{2}\in Q\left(
d,l\right) $ \emph{and \ functions }$\overline{c}_{1}\left( x\right) ,%
\overline{c}_{1}\left( x\right) $\emph{\ be defined as in (\ref{3.25})}$.$%
\emph{\ Then }$\left\vert \overline{c}_{1}-\overline{c}_{2}\right\vert
_{\alpha }\leq \left\vert \chi \right\vert _{\alpha }\left\vert
c_{1}-c_{2}\right\vert _{\alpha }.$

\textbf{Proof}. By (\ref{3.25}) $\overline{c}_{1}\left( x\right) -\overline{c%
}_{2}\left( x\right) =\chi \left( x\right) \left( c_{1}\left( x\right)
-c_{2}\left( x\right) \right) .$ The rest of the proof follows from $%
\left\vert fg\right\vert _{\alpha }\leq \left\vert f\right\vert _{\alpha
}\left\vert g\right\vert _{\alpha },\forall f,g\in C^{\alpha }\left( 
\overline{\Omega }\right) .$ $\square $

Note that $C^{1}\left( \overline{\Omega }\right) \subset C^{\alpha }\left( 
\overline{\Omega }\right) ,$ and also there exists a constant $C=C\left(
\Omega ,\alpha \right) >0$ such that 
\begin{equation}
\left\vert f\right\vert _{\alpha }\leq C\left\Vert f\right\Vert
_{C^{1}\left( \overline{\Omega }\right) },\forall f\in C^{1}\left( \overline{%
\Omega }\right) .  \label{4.6}
\end{equation}

\textbf{Lemma 4.2.} \emph{Let the source }$x_{0}\notin \overline{\Omega }.$ 
\emph{Then there exists a constant }$Y=Y\left( \Omega ,\overline{s},d,l,\chi
,x_{0},\alpha \right) >0$\emph{\ depending on listed parameters such that }$%
\left\vert w_{\overline{c}}\left( x,\overline{s}\right) \right\vert _{\alpha
}\leq Y,$ $\forall c\in Q\left( d,l\right) .$\emph{\ }

\textbf{Proof.} Below in this proof\textbf{\ }$Y=Y\left( \Omega ,\overline{s}%
,d,l,\chi ,x_{0},\alpha \right) >0$ denotes different constants depending on
listed parameters. By (\ref{4.2}) and (\ref{4.3}) $w_{\overline{c}}\left( x,%
\overline{s}\right) \in C^{2+\alpha }\left( \overline{\Omega }\right) .$
Denote $b\left( x\right) =\overline{c}\left( x\right) -1.$ Then 
\begin{equation}
w_{\overline{c}}\left( x,\overline{s}\right) =w_{0}\left( x,\overline{s}%
\right) -\overline{s}^{2}\int\limits_{\Omega }w_{1}\left( x-\xi ,\overline{s}%
\right) b\left( \xi \right) w_{\overline{c}}\left( \xi ,\overline{s}\right)
d\xi .  \label{4.7}
\end{equation}%
By (\ref{4.4}) and (\ref{4.7}) 
\begin{equation}
\left\vert w_{\overline{c}}\left( x,\overline{s}\right) \right\vert \leq
Y+Y\left\Vert b\right\Vert _{C\left( \overline{\Omega }\right)
}\int\limits_{\Omega }w_{1}\left( x-\xi ,\overline{s}\right) d\xi \leq Y,%
\text{ }x\in \Omega .  \label{4.8}
\end{equation}%
In addition, by (\ref{4.7}) 
\begin{equation}
\nabla w_{\overline{c}}\left( x,\overline{s}\right) =\nabla w_{0}\left( x,%
\overline{s}\right) -\overline{s}^{2}\int\limits_{\Omega }\nabla w_{1}\left(
x-\xi ,\overline{s}\right) b\left( \xi \right) w_{\overline{c}}\left( \xi ,%
\overline{s}\right) d\xi ,x\in \Omega .  \label{4.9}
\end{equation}%
Hence, $\left\vert \nabla w_{\overline{c}}\left( x,\overline{s}\right)
\right\vert \leq Y,$ $x\in \Omega .$ Hence, (\ref{4.6}), (\ref{4.8}) and (%
\ref{4.9}) imply that $\left\vert w_{\overline{c}}\left( x,\overline{s}%
\right) \right\vert _{\alpha }\leq Y.$ $\square $

Consider a bounded domain $\Omega _{1}\subset \mathbb{R}^{3}$ such that 
\begin{equation}
\Omega \subset \Omega _{1},\text{ }\partial \Omega \cap \partial \Omega
_{1}=\varnothing ,\text{ }\partial \Omega _{1}\in C^{3},\text{ }x_{0}\notin 
\overline{\Omega }_{1}.  \label{4.10}
\end{equation}

\textbf{Lemma 4.3.} \emph{Let }$\Omega ^{\prime }\subset \Omega \subset
\Omega _{1}$\emph{\ be above bounded domains in }$\mathbb{R}^{3},$\emph{\
condition (\ref{4.10}) be satisfied and }$\chi \left( x\right) $\emph{\ be
the function in (\ref{3.24_1}). Let }$\overline{s}\geq 1$\emph{. Then the
function }$w_{\overline{c}}\left( x,\overline{s}\right) \in C^{3}\left(
\partial \Omega _{1}\right) ,\forall c\in Q\left( d,l\right) .$\emph{\
Furthermore, there exists a constant }$B=B\left( \Omega ,\Omega ^{\prime
},\Omega _{1},\overline{s},d,l,\chi ,x_{0},\alpha \right) >2$\emph{\
depending only on listed parameters such that} 
\begin{equation}
\left\Vert w_{\overline{c}}\left( x,\overline{s}\right) \right\Vert
_{C^{3}\left( \partial \Omega _{1}\right) }\leq B,\text{ }\forall c\in
Q\left( d,l\right) .  \label{4.11}
\end{equation}%
\emph{For any two functions }$c_{1},c_{2}\in Q\left( d,l\right) $\emph{\
denote }$\widetilde{w}\left( x\right) =w_{\overline{c}_{1}}\left( x,%
\overline{s}\right) -w_{\overline{c}_{2}}\left( x,\overline{s}\right) .$%
\emph{\ Then } 
\begin{equation}
\left\Vert \widetilde{w}\right\Vert _{C^{3}\left( \partial \Omega
_{1}\right) }\leq B\left\vert c_{1}-c_{2}\right\vert _{\alpha },\text{ }%
\forall c_{1},c_{2}\in Q\left( d,l\right) .  \label{4.11_1}
\end{equation}

\textbf{Proof.} Everywhere below in this paper $B$ denotes different
positive constant depending on above parameters. The integrand of formula (%
\ref{4.7}) does not have a singularity for $x\in \Omega _{1}\diagdown 
\overline{\Omega }.$ Hence, (\ref{4.7}) implies that $w_{\overline{c}}\left(
x,\overline{s}\right) \in C^{3}\left( \partial \Omega _{1}\right) .$ Next, (%
\ref{4.11}) follows from (\ref{4.7}) and Lemma 4.2.

Denote 
\begin{equation*}
\widetilde{c}\left( x\right) =c_{1}\left( x\right) -c_{2}\left( x\right)
,b_{1}\left( x\right) =c_{1}\left( x\right) -1,b_{2}\left( x\right)
=c_{2}\left( x\right) -1.
\end{equation*}%
Hence, by (\ref{3.25}) $\overline{c}_{1}\left( x\right) -\overline{c}%
_{2}\left( x\right) =\chi \left( x\right) \widetilde{c}\left( x\right) .$
First, substitute in (\ref{4.7}) $\left( b_{1},w_{\overline{c}_{1}}\right) .$
Next, substitute $\left( b_{2},w_{\overline{c}_{2}}\right) .$ Next, subtract
the second equation from the first one and denote $\widetilde{w}\left(
x\right) =w_{\overline{c}_{1}}\left( x,\overline{s}\right) -w_{\overline{c}%
_{2}}\left( x,\overline{s}\right) $. We obtain 
\begin{equation}
\widetilde{w}\left( x\right) =-\overline{s}^{2}\int\limits_{\Omega
}w_{1}\left( x-\xi ,\overline{s}\right) \chi \left( \xi \right) \widetilde{c}%
\left( \xi \right) w_{\overline{c}_{1}}\left( \xi ,\overline{s}\right) d\xi -%
\overline{s}^{2}\int\limits_{\Omega }w_{1}\left( x-\xi ,\overline{s}\right)
b_{2}\left( \xi \right) \widetilde{w}\left( \xi \right) d\xi .  \label{4.12}
\end{equation}%
Let 
\begin{eqnarray*}
I_{1}\left( x\right) &=&-\overline{s}^{2}\int\limits_{\Omega }w_{1}\left(
x-\xi ,\overline{s}\right) \chi \left( \xi \right) \widetilde{c}\left( \xi
\right) w_{\overline{c}_{1}}\left( \xi ,\overline{s}\right) d\xi , \\
I_{2}\left( x\right) &=&-\overline{s}^{2}\int\limits_{\Omega }w_{1}\left(
x-\xi ,\overline{s}\right) b_{2}\left( \xi \right) \widetilde{w}\left( \xi
\right) d\xi .
\end{eqnarray*}%
Using the same arguments as ones in the proof of (\ref{4.11}), we obtain 
\begin{equation}
\left\Vert I_{1}\right\Vert _{C^{3}\left( \partial \Omega _{1}\right) }\leq
B\left\vert \widetilde{c}\right\vert _{\alpha }.  \label{4.13}
\end{equation}%
Next,%
\begin{equation}
\Delta \widetilde{w}-\overline{s}^{2}\overline{c}_{2}\widetilde{w}=\overline{%
s}^{2}\chi \left( x\right) \widetilde{c}\left( x\right) w_{\overline{c}%
_{1}}\left( x,\overline{s}\right) ,x\in \mathbb{R}^{3}.  \label{4.14}
\end{equation}%
Furthermore, it follows from (\ref{4.12}) that the function $\widetilde{w}%
\left( x\right) $ decays exponentially together with its derivatives as $%
\left\vert x\right\vert \rightarrow \infty .$ Hence, multiplying (\ref{4.14}%
) by $\widetilde{w}$, integrating over $\mathbb{R}^{3}$ and using Lemma 4.2
and the fact that $\chi \left( x\right) =0$ for $x\in \mathbb{R}%
^{3}\diagdown \Omega $, we obtain in a standard manner $\left\Vert 
\widetilde{w}\right\Vert _{L_{2}\left( \Omega \right) }\leq B\left\Vert 
\widetilde{c}\right\Vert _{L_{2}\left( \Omega \right) }\leq B\left\vert 
\widetilde{c}\right\vert _{\alpha }.$ Hence, $\left\Vert I_{2}\right\Vert
_{C^{3}\left( \partial \Omega _{1}\right) }\leq B\left\vert \widetilde{c}%
\right\vert _{\alpha }.$ This estimate combined with (\ref{4.13}) implies (%
\ref{4.11_1}). $\square $

\subsection{Estimates of tails}

\label{sec:4.2}

\textbf{Theorem 4.2}. \emph{Let }$\Omega ^{\prime }\subset \Omega \subset
\Omega _{1}\subset \mathbb{R}^{3}$\emph{\ be above bounded domains with
condition (\ref{4.10}), and let }$\chi \left( x\right) $\emph{\ be the
function in (\ref{3.24_1}). Also, let }$\overline{s}\geq 1$\emph{.\ For each
function }$c\in Q\left( d,l\right) $\emph{\ consider the function }$%
\overline{c}$\emph{\ defined in (\ref{3.25}). Denote }%
\begin{equation}
V_{\overline{c}}\left( x\right) =\frac{\ln w_{\overline{c}}\left( x,%
\overline{s}\right) }{\overline{s}^{2}}.  \label{4.15}
\end{equation}%
\emph{Then there exists such a constant }$B=B\left( \Omega ,\Omega ^{\prime
},\Omega _{1},\overline{s},d,l,\chi ,x_{0},\alpha \right) >2$\emph{\
depending only on listed parameters such that }%
\begin{eqnarray}
\left\vert \nabla V_{\overline{c}}\right\vert _{1+\alpha } &\leq &B,\forall
c\in Q\left( d,l\right) ,  \label{4.16} \\
\left\vert \nabla V_{\overline{c}_{1}}-\nabla V_{\overline{c}%
_{2}}\right\vert _{1+\alpha } &\leq &B\left\vert c_{1}-c_{2}\right\vert
_{\alpha },\forall c_{1},c_{2}\in Q\left( d,l\right) .  \label{4.17}
\end{eqnarray}%
\textbf{Proof}. By (\ref{4.15}) 
\begin{equation}
\nabla V_{\overline{c}}\left( x\right) =\frac{\nabla w_{\overline{c}}\left(
x,\overline{s}\right) }{\overline{s}^{2}w_{\overline{c}}\left( x,\overline{s}%
\right) },\text{ }\partial _{x_{i}}^{2}V_{\overline{c}}\left( x\right) =%
\frac{\partial _{x_{i}}^{2}w_{\overline{c}}\left( x,\overline{s}\right) }{%
\overline{s}^{2}w_{\overline{c}}\left( x,\overline{s}\right) }-\frac{\left(
\partial _{x_{i}}w_{\overline{c}}\left( x,\overline{s}\right) \right) ^{2}}{%
\overline{s}^{2}w_{\overline{c}}^{2}\left( x,\overline{s}\right) },i=1,2,3.
\label{4.18}
\end{equation}
Since by (\ref{4.4}) and (\ref{4.5})%
\begin{equation*}
\frac{4\pi \exp \left( \overline{s}\left\vert x-x_{0}\right\vert \right) }{%
\overline{s}^{2}}\left\vert x-x_{0}\right\vert \leq \frac{1}{\overline{s}%
^{2}w_{\overline{c}}\left( x,\overline{s}\right) }<\frac{4\pi \exp \left( 
\overline{s}\sqrt{d}\left\vert x-x_{0}\right\vert \right) }{\overline{s}^{2}}%
\left\vert x-x_{0}\right\vert \leq B,
\end{equation*}%
then (\ref{4.18}) implies that in order to prove (\ref{4.16}) and (\ref{4.17}%
), it is sufficient to prove that 
\begin{eqnarray}
\left\vert w_{\overline{c}}\left( x,\overline{s}\right) \right\vert
_{2+\alpha } &\leq &B,\forall c\in Q\left( d,l\right) ,  \label{4.19} \\
\left\vert w_{\overline{c}_{1}}\left( x,\overline{s}\right) -w_{c_{2}}\left(
x,\overline{s}\right) \right\vert _{2+\alpha } &\leq &B\left\vert
c_{1}-c_{2}\right\vert _{\alpha }.  \label{4.20}
\end{eqnarray}%
Denote $f_{\overline{c}}\left( x\right) =w_{\overline{c}}\left( x,\overline{s%
}\right) \mid _{\partial \Omega _{1}}$. By Lemma 4.3%
\begin{equation}
f_{\overline{c}}\left( x\right) \in C^{3}\left( \partial \Omega _{1}\right)
,\left\Vert f\right\Vert _{C^{3}\left( \partial \Omega _{1}\right) }\leq B.
\label{4.21}
\end{equation}%
Since by (\ref{4.10}) $x_{0}\notin \overline{\Omega }_{1},$ then (\ref{4.2})
and (\ref{4.3}) imply that $w_{\overline{c}}\left( x,\overline{s}\right) \in
C^{2+\alpha }\left( \overline{\Omega }_{1}\right) .$ On the other hand, the
function $w_{\overline{c}}\left( x,\overline{s}\right) $ solves the
following Dirichlet boundary value problem in $\Omega _{1}$ 
\begin{eqnarray*}
\Delta w_{\overline{c}}-\overline{s}^{2}\overline{c}\left( x\right) w_{%
\overline{c}} &=&0,x\in \Omega _{1}, \\
w_{\overline{c}} &\mid &_{\partial \Omega _{1}}=f_{\overline{c}}\left(
x\right) .
\end{eqnarray*}%
Since the $C^{2+\alpha }\left( \overline{\Omega }_{1}\right) $ solution of
this problem is unique, then Schauder theorem and (\ref{4.21}) imply that $%
\left\vert w_{\overline{c}}\left( x,\overline{s}\right) \right\vert
_{2+\alpha }\leq B\left\Vert f\right\Vert _{C^{2+\alpha }\left( \partial
\Omega _{1}\right) }\leq B,$ which proves (\ref{4.19}). Denote again $%
\widetilde{w}\left( x\right) =w_{\overline{c}_{1}}\left( x,\overline{s}%
\right) -w_{\overline{c}_{2}}\left( x,\overline{s}\right) .$ Then%
\begin{eqnarray*}
\Delta \widetilde{w}-\overline{s}^{2}\overline{c}_{1}\left( x\right) 
\widetilde{w} &=&\overline{s}^{2}\left( \overline{c}_{1}\left( x\right) -%
\overline{c}_{2}\left( x\right) \right) w_{\overline{c}_{2}}, \\
\widetilde{w} &\mid &_{\partial \Omega _{1}}=f_{\overline{c}_{1}}\left(
x\right) -f_{c_{2}}\left( x\right) .
\end{eqnarray*}%
Hence, Schauder theorem and Lemmata 4.1 and 4.3 imply (\ref{4.20}). $\square 
$

\section{Approximate Global Convergence of the Algorithm of Subsection 3.4}

\label{sec:5}

\subsection{Exact solution}

\label{sec:5.1}

Recall that we the existence and uniqueness of the exact solution $c^{\ast
}\left( x\right) $ of our MCIP satisfying (\ref{3.33}). The corresponding
functions $w^{\ast }\left( x,s\right) ,$ $V^{\ast }\left( x,\overline{s}%
\right) $ were defined in subsection 3.5. Let $\underline{s}>\underline{s}%
\left( c^{\ast }\right) ,$ where $\underline{s}\left( c^{\ast }\right) $ is
the number of Theorem 4.1. Let $u^{\ast }\left( x,t\right) $ be the solution
of the problem (\ref{3.1}), (\ref{3.2}) with $c:=c^{\ast }.$ Since one can
differentiate infinitely many times with respect to $s$ under the integral
sign in (\ref{3.5}), then 
\begin{equation}
w^{\ast }\left( x,s\right) \in C^{2+\alpha }\left( \overline{\Omega }\right)
\times C^{2}\left[ \underline{s},\overline{s}\right] ,\text{ if }\underline{s%
}>\underline{s}\left( c^{\ast }\right) .  \label{5.1}
\end{equation}

Denote 
\begin{equation}
q^{\ast }\left( x,s\right) =\frac{\partial }{\partial s}\left[ \frac{\ln %
\left[ w^{\ast }\left( x,s\right) \right] }{s^{2}}\right] ,\psi ^{\ast
}\left( x,s\right) =q^{\ast }\left( x,s\right) \mid _{\partial \Omega },s\in %
\left[ \underline{s},\overline{s}\right] .  \label{5.2}
\end{equation}%
Consider functions $q_{n}^{\ast }\left( x\right) ,\overline{\psi }_{n}^{\ast
}\left( x\right) ,$%
\begin{equation}
q_{n}^{\ast }\left( x\right) =\frac{1}{h}\int\limits_{s_{n}}^{s_{n-1}}q^{%
\ast }\left( x,s\right) ds,\text{ }\overline{\psi }_{n}^{\ast }\left(
x\right) =\frac{1}{h}\int\limits_{s_{n}}^{s_{n-1}}\psi ^{\ast }\left(
x,s\right) ds,\text{ }q_{0}^{\ast }\left( x\right) \equiv 0.  \label{5.3}
\end{equation}%
Then (\ref{5.1}) and (\ref{5.2}) imply that 
\begin{equation}
\left\vert q^{\ast }\left( x,s\right) -q_{n}^{\ast }\left( x\right)
\right\vert _{2+\alpha }\leq C^{\ast }h,\left\Vert \psi ^{\ast }\left(
x,s\right) -\overline{\psi }_{n}^{\ast }\left( x\right) \right\Vert
_{C^{2+\alpha }\left( \partial \Omega \right) }\leq C^{\ast }h,n\in \left[
1,N\right] ,s\in \left[ s_{n},s_{n-1}\right] .  \label{5.4}
\end{equation}%
Here the constant $C^{\ast }=C^{\ast }\left( \left\Vert q^{\ast }\right\Vert
_{C^{2+\alpha }\left( \overline{\Omega }\right) \times C^{1}\left[ 
\underline{s},\overline{s}\right] },\overline{s}\right) >0$ depends only on
the $C^{2+\alpha }\left( \overline{\Omega }\right) \times C^{1}\left[ 
\underline{s},\overline{s}\right] $ norm of the function $q^{\ast }\left(
x,s\right) $. Hence, we can assume that 
\begin{equation}
\max_{1\leq n\leq N}\left\vert q_{n}^{\ast }\right\vert _{2+\alpha }\leq
C^{\ast }.  \label{5.5}
\end{equation}%
Without any loss of generality we assume that 
\begin{equation}
C^{\ast }\geq 1.  \label{5.6}
\end{equation}%
By the one of concepts of Tikhonov (see, e. g. section 1.4 of \cite{BK}) we
assume that the constant $C^{\ast }$ is known \emph{a priori}. By (\ref{5.2}%
) 
\begin{equation}
q_{n}^{\ast }\left( x\right) =\overline{\psi }_{n}^{\ast }\left( x\right) ,%
\text{ }x\in \partial \Omega .  \label{5.7}
\end{equation}%
Hence we obtain the following analogue of equation (\ref{3.29})

\begin{equation}
\begin{split}
& \Delta q_{n}^{\ast }-A_{1,n}\left( h\sum\limits_{j=0}^{n-1}\nabla
q_{j}^{\ast }\right) \nabla q_{n}^{\ast }+A_{1,n}\nabla q_{n}^{\ast }\nabla
V^{\ast }=-A_{2,n}h^{2}\left( \sum\limits_{j=0}^{n-1}\nabla q_{i}^{\ast
}\right) ^{2} \\
& +2A_{2,n}\nabla V^{\ast }\left( h\sum\limits_{j=0}^{n-1}\nabla q_{j}^{\ast
}\right) -A_{2,n}\left\vert \nabla V^{\ast }\right\vert ^{2}+F_{n}\left(
x,h,\lambda \right) .
\end{split}
\label{5.8}
\end{equation}%
Here the function $F_{n}\left( x,h,\lambda \right) \in C^{\alpha }\left( 
\overline{\Omega }\right) $. The term $2I_{1,n}\left( \nabla q_{n}^{\ast
}\right) ^{2}/I_{0}$ is included in $F_{n},$ i.e., unlike (\ref{3.24}), we
do not ignore this term now, since we work now with the exact solution.
Hence, by (\ref{3.23}) 
\begin{equation}
\max_{\lambda h\geq 1}\left\vert F_{n}\left( x,h,\lambda ,\overline{s}%
\right) \right\vert _{\alpha }\leq C^{\ast }h,n\in \left[ 1,N\right] .
\label{5.9}
\end{equation}%
Let 
\begin{equation}
v_{n}^{\ast }\left( x\right) =-hq_{n}^{\ast }\left( x\right)
-h\sum\limits_{j=0}^{n-1}q_{j}^{\ast }\left( x\right) +V^{\ast }\left(
x\right) ,\text{ }x\in \Omega ,\text{ }n\in \left[ 1,N\right] .  \label{5.10}
\end{equation}%
Using (\ref{5.4}), we obtain similarly with (\ref{3.27}) 
\begin{equation}
c^{\ast }\left( x\right) =\left[ \Delta v_{n}^{\ast }+s_{n}^{2}\left\vert
\nabla v_{n}^{\ast }\right\vert ^{2}\right] \left( x\right) +\overline{F}%
_{n}\left( x\right) ,  \label{5.11}
\end{equation}%
where the error function $\overline{F}_{n}$ is such that 
\begin{equation}
\left\vert \overline{F}_{n}\right\vert _{\alpha }\leq C^{\ast }h.
\label{5.12}
\end{equation}

We also assume that the function $g(x,t)$ in (\ref{3.4}) is given with an
error. This naturally produces an error in functions $\overline{\psi }_{n}$
in (\ref{3.29}). Let $\sigma >0$ be a small parameter characterizing the
level of the error in the data $\psi \left( x,s\right) .$ Because of (\ref%
{3.39}), we assume that in (\ref{3.29}) functions $\overline{\psi }%
_{n}\left( x\right) \in C^{2+\alpha }\left( \partial \Omega \right) $ and 
\begin{equation}
\left\Vert \overline{\psi }_{n}^{\ast }\left( x\right) -\overline{\psi }%
_{n}\left( x\right) \right\Vert _{C^{2+\alpha }\left( \partial \Omega
\right) }\leq C^{\ast }\left( \sigma +h\right) .  \label{5.13}
\end{equation}%
In addition, we assume that We now reformulate the estimate of the Schauder
theorem for the specific case we need. Consider the Dirichlet boundary value
problem 
\begin{eqnarray}
\Delta u+\sum\limits_{j=1}^{3}b_{j}(x)u_{x_{j}}-b_{0}(x)u &=&f\left(
x\right) ,x\in \Omega ,  \label{5.14} \\
u &\mid &_{\partial \Omega }=g\left( x\right) \in C^{2+\alpha }\left(
\partial \Omega \right) .  \label{5.140}
\end{eqnarray}%
Assume that the following conditions are in place 
\begin{equation}
b_{j},b_{0},f\in C^{\alpha }\left( \overline{\Omega }\right) ,\text{ }%
b_{0}\left( x\right) \geq 0,\text{ }\max_{j\in \left[ 0,n\right] }\left(
\left\vert b_{j}\right\vert _{\alpha }\right) \leq P,\text{ }P=const.>0.
\label{5.15}
\end{equation}%
Then Schauder theorem\ \cite{LU} claims that there exists unique solution $%
u\in C^{2+\alpha }\left( \overline{\Omega }\right) $ of the boundary value
problem (\ref{5.14}), (\ref{5.140}), and the following estimate holds with a
certain constant $K=K\left( \Omega ,P\right) >2$, depending only on the
domain $\Omega $ and the constant $P$ 
\begin{equation}
\left\vert u\right\vert _{2+\alpha }\leq K\left[ \left\Vert g\right\Vert
_{C^{2+\alpha }\left( \partial \Omega \right) }+\left\vert f\right\vert
_{\alpha }\right] .  \label{5.16}
\end{equation}

\subsection{Approximate global convergence theorem}

\label{sec:5.2}

\textbf{Theorem 5.1. }\emph{Let }$\Omega ^{\prime }\subset \Omega \subset
\Omega _{1}\subset \mathbb{R}^{3}$\emph{\ be above bounded domains with
condition (\ref{4.10}), let }$\chi \left( x\right) $\emph{\ be the function
in (\ref{3.24_1}), }$\overline{s}\geq 1$ \emph{and (\ref{3.39}) be valid.
Let the function }$c^{\ast }\left( x\right) $\emph{\ satisfying conditions (%
\ref{3.33}) be the exact solution of the CIP (\ref{3.1})-(\ref{3.4}), where
constants }$d,l>1$\ \emph{are given.} \emph{Also, let condition (\ref{5.13})
holds, where }$\sigma $\emph{\ is level of the error in the data, in (\ref%
{3.29}) functions }$\overline{\psi }_{n}\in C^{2+\alpha }\left( \partial
\Omega \right) ,$\emph{\ and the constant }$C^{\ast }=C^{\ast }\left(
\left\Vert q^{\ast }\right\Vert _{C^{2+\alpha }\left( \overline{\Omega }%
\right) \times C^{1}\left[ \underline{s},\overline{s}\right] },\overline{s}%
\right) \geq 1$\emph{\ is defined in (\ref{5.4})-(\ref{5.6}). Consider the
algorithm of subsection 3.4 supplied by Assumption of subsection 3.5. Let
the first tail function }$V_{1,1}\left( x\right) $\emph{\ be calculated via (%
\ref{3.40}), (\ref{3.41}) and (\ref{3.43}). In addition, assume that all
functions }$c_{n,i}\left( x\right) $\emph{\ in (\ref{3.27}) are such that }%
\begin{equation}
c_{n,i}\left( x\right) \geq 1,x\in \Omega .  \label{5.17}
\end{equation}%
\emph{Assume that the parameter }$\lambda $ \emph{of the Carleman Weight
Function is so large that }$\lambda h\geq 1$ \emph{(see (\ref{3.23})).\ Let (%
\ref{5.13}) be valid and also}%
\begin{equation}
\left\Vert \psi \left( x,\overline{s}\right) -\psi ^{\ast }\left( x,%
\overline{s}\right) \right\Vert _{C^{2+\alpha }\left( \partial \Omega
\right) }\leq C^{\ast }\sigma .  \label{5.170}
\end{equation}%
\emph{Consider the error parameter }$\eta ,$ 
\begin{equation}
\eta =h+\sigma .  \label{5.18}
\end{equation}%
\emph{Let }$B=B\left( \Omega ,\Omega _{1},\overline{s},d,l,\chi
,x_{0},\alpha \right) >2$\emph{\ be the constant of Theorem 4.2. Consider
the number }$B_{1}=B_{1}\left( \Omega ,\Omega _{1},\overline{s},d,l,C^{\ast
},\chi ,x_{0},\alpha \right) ,$%
\begin{equation}
B_{1}=\max \left( 4B+3C^{\ast },24\overline{s}^{2}\right) >24.
\label{5.19_1}
\end{equation}%
\emph{Let }%
\begin{equation}
K=K\left( \Omega ,\overline{s}^{2}B_{1}\right) \geq B_{1}  \label{5.19_2}
\end{equation}%
\emph{\ be the constant in (\ref{5.16}). Let\ the parameter }$\eta $\emph{\
be so small that} 
\begin{equation}
\eta \in \left( 0,\eta _{0}\right) ,\text{ }\eta _{0}=\frac{1}{KNB_{1}^{3Nm}}%
.  \label{5.20}
\end{equation}%
\emph{\ Then } 
\begin{eqnarray}
c_{n,i} &\in &C^{\alpha }\left( \overline{\Omega }\right) ,\overline{c}%
_{n,i}\in C^{\alpha }\left( \mathbb{R}^{3}\right) ,\text{ }\left( n,i\right)
\in \left[ 1,N\right] \times \left[ 1,m\right] ,  \label{5.21} \\
c_{n,i},\overline{c}_{n,i}\left( x\right) &\in &Q\left( d,l\right) ,\left(
n,i\right) \in \left[ 1,N\right] \times \left[ 1,m\right] .  \label{5.22}
\end{eqnarray}%
\emph{\ In addition, the following estimates hold for }$\left( n,i\right)
\in \left[ 1,N\right] \times \left[ 1,m\right] $%
\begin{eqnarray}
\left\vert \nabla V_{n,i}\right\vert _{1+\alpha } &\leq &B_{1},  \label{5.23}
\\
\left\vert \nabla V_{n,i}-\nabla V^{\ast }\right\vert _{1+\alpha } &\leq
&B_{1}^{3\left[ i-1+\left( n-1\right) m\right] +1}\cdot \eta ,  \label{5.24}
\\
\left\vert q_{n,i}-q_{n}^{\ast }\right\vert _{2+\alpha } &\leq &KB_{1}^{3 
\left[ i+\left( n-1\right) m\right] }\cdot \eta ,  \label{5.25} \\
\left\vert q_{n,i}\right\vert _{2+\alpha },\left\vert q_{n}\right\vert
_{2+\alpha } &\leq &2C^{\ast },\text{ }n\in \left[ 1,N\right] ,  \label{5.26}
\\
\left\vert c_{n,i}-c^{\ast }\right\vert _{\alpha } &\leq &B_{1}^{3\left[
i+\left( n-1\right) m\right] }\cdot \eta .  \label{5.27}
\end{eqnarray}%
\emph{\ } \emph{Define the number }$\omega \in \left( 0,1\right) $ \emph{as }%
\begin{equation}
\omega =\frac{\ln \left( KN\right) }{2\left[ 3Nm\ln B_{1}+\ln \left(
KN\right) \right] },\omega \in \left( 0,1\right) .  \label{5.28}
\end{equation}%
\emph{\ Then (\ref{5.27}) becomes} 
\begin{equation}
\left\vert c_{n,i}-c^{\ast }\right\vert _{\alpha }\leq \eta ^{\omega
}:=\varepsilon \in \left( 0,1\right) .  \label{5.29}
\end{equation}%
\emph{Therefore, by (\ref{5.29}) and Definition 2.1 the algorithm of
subsection 3.4 possesses the approximate globally convergent property of the
level }$\varepsilon $\emph{.}

\textbf{Remarks 5.1:}

\textbf{1}. Since $K=K\left( \Omega ,\overline{s}^{2}B_{1}\right) ,$ one can
incorporate the term $KN$ in (\ref{5.20}) in the term $B_{1}^{3Nm}.$
However, we are not doing this for the convenience of the proof.

\textbf{2}. Condition (\ref{5.20}) provides a linkage between the level of
the error $\eta $ in the data and the total \textquotedblleft allowable"
number of iterations $Nm.$ The fact that the maximal number of iterations $%
Nm $ is limited is going along well with the theory of Ill-Posed Problems.
Indeed, it is well known that the maximal number of iterations and the error
in the data are often connected with each other, see, e.g. pages 156 and 157
of \cite{EHN} and section 1.6 of \cite{BK}. Hence, So that the maximal
number of iterations $Nm$ is a regularization parameter in this case. The
fact that the constant $B_{1}$ depends not only on the domain $\Omega $ but
also on the domain $\Omega _{1}$ does not affect the approximate global
convergence property.

\textbf{3}. It is hard to establish \emph{a priori} the upper limit for the
maximal number of functions $q_{n}.$We have consistently observed in our
numerical tests that certain numbers indicating convergence stabilize a few
iterations before a certain number $\overline{N}\in \left[ 1,N\right] ,$
i.e. at a certain $n<\overline{N}.$ Next, they grow steeply for $n\geq 
\overline{N}.$ This means that the process should be stopped at a certain $n=%
\widetilde{N}<\overline{N}$. Usually we take $\widetilde{N}:=\overline{N}-1,$
see, e.g. pages 178-182 and 311-314 in the book \cite{BK}. This numerical
observation is going along well with (\ref{5.20}), (\ref{5.29}).

\textbf{Proof of Theorem 5.1.} Estimate (\ref{5.29}) follows from estimates (%
\ref{5.20}), (\ref{5.27}) and (\ref{5.28}). Hence, we focus below on the
proof of relations (\ref{5.21})-(\ref{5.27}). Denote 
\begin{eqnarray*}
\widetilde{V}_{n,i} &=&V_{n,i}-V^{\ast },\text{ }\widetilde{q}%
_{n,i}=q_{n,i}-q_{n}^{\ast },\text{ }\widetilde{q}_{n}=q_{n}-q_{n}^{\ast },
\\
\widetilde{v}_{n,i} &=&v_{n,i}-v_{n,i}^{\ast },\text{ }\widetilde{c}%
_{n,i}=c_{n,i}-c^{\ast },\text{ }\widetilde{\psi }_{n}=\overline{\psi }_{n}-%
\overline{\psi }_{n}^{\ast }.
\end{eqnarray*}%
By (\ref{3.24_1}), (\ref{3.25}) and\emph{\ }(\ref{3.33}) $c^{\ast }\left(
x\right) \equiv \overline{c}^{\ast }\left( x\right) .$ Hence, we can apply
Theorem 4.2 to estimate the norm $\left\vert \nabla \widetilde{V}%
_{n,i}\right\vert _{1+\alpha }$. Suppose that estimate (\ref{5.27}) holds.
Then $c_{n,i}\in Q\left( d,l\right) .$ Indeed, using (\ref{3.33}), (\ref%
{5.20}) and (\ref{5.27}), we obtain 
\begin{equation*}
\left\vert c_{n,i}\right\vert _{\alpha }=\left\vert c_{n,i}-c^{\ast
}+c^{\ast }\right\vert _{\alpha }\leq \left\vert c^{\ast }\right\vert
_{\alpha }+\left\vert c_{n,i}-c^{\ast }\right\vert _{\alpha }\leq
l-1+B_{1}^{3\left[ i+\left( n-1\right) m\right] }\cdot \eta <l.
\end{equation*}%
Similarly $c_{n,i}\leq d.$ The latter two estimates and (\ref{5.17}) imply
that $c_{n,i}\in Q\left( d,l\right) .$ Next, since the function $c_{n,i}\in
Q\left( d,l\right) ,$ then, using (\ref{4.16}) and (\ref{5.19_1}), we obtain
(\ref{5.23}). Also, since the function $c_{n,i}\in \left[ 1,d\right] ,$ then
the function $\overline{c}_{n,i}\in \left[ 1,d\right] $. Hence, if (\ref%
{5.27}) is true, then (\ref{5.21}) and (\ref{5.22}) hold.

First, we prove (\ref{5.25})-(\ref{5.27}) for the case $\left( n,i\right)
=\left( 1,1\right) .$ Subtracting equation (\ref{5.8}) from equation (\ref%
{3.29}) at $\left( n,i\right) =\left( 1,1\right) $ and also subtracting (\ref%
{5.7}) from the boundary condition in (\ref{3.29}), we obtain%
\begin{eqnarray}
\Delta \widetilde{q}_{1,1}+A_{1,1}\nabla V_{1,1}\nabla \widetilde{q}_{1,1}
&=&-A_{1,1}\nabla \widetilde{V}_{1,1}\nabla q_{1}^{\ast }-A_{2,1}\nabla 
\widetilde{V}_{1,1}\left( \nabla V_{1,1}+\nabla V^{\ast }\right) -F_{1},
\label{5.30} \\
\widetilde{q}_{1,1}\left( x\right) &=&\widetilde{\psi }_{1}\left( x\right)
,x\in \partial \Omega .  \label{5.31}
\end{eqnarray}%
We now estimate the right hand side of (\ref{5.30}). It follows from (\ref%
{3.33}), (\ref{3.34}) and (\ref{4.16}) that 
\begin{equation}
\left\vert \nabla V^{\ast }\right\vert _{1+\alpha }\leq B.  \label{5.32_1}
\end{equation}%
By (\ref{3.42}) and (\ref{5.170}) 
\begin{equation}
\left\vert \nabla \widetilde{V}_{1,1}\right\vert _{1+\alpha }\leq B\eta .
\label{5.322}
\end{equation}%
Estimates (\ref{5.23}), (\ref{5.24}) for $\left( n,i\right) =\left(
1,1\right) $ with $B_{1}$ being replaced with $B$ follow from (\ref{3.35}), (%
\ref{3.39})-(\ref{3.43}). Using (\ref{5.5}), (\ref{5.23}), (\ref{5.24}) for $%
\left( n,i\right) =\left( 1,1\right) $ with $B_{1}$ being replaced with $B$,
as well as (\ref{3.22}), (\ref{3.420}), (\ref{5.9}), (\ref{5.19_1}), (\ref%
{5.32_1}) and (\ref{5.322}), we obtain 
\begin{eqnarray*}
&&\left\vert A_{1,1}\nabla \widetilde{V}_{1,1}\nabla q_{1}^{\ast
}+A_{2,1}\nabla \widetilde{V}_{1,1}\left( \nabla V_{1,1}+\nabla V^{\ast
}\right) -F_{1}\right\vert _{\alpha } \\
&\leq &8\overline{s}^{2}BC^{\ast }\eta +16\overline{s}^{2}B^{2}\eta +C^{\ast
}\eta =8\overline{s}^{2}B\left( 2B+C^{\ast }+\frac{C^{\ast }}{8\overline{s}%
^{2}}\right) \eta \leq 4\overline{s}^{2}BB_{1}\eta \leq \overline{s}%
^{2}B_{1}^{2}\eta .
\end{eqnarray*}%
Thus, 
\begin{equation}
\left\vert A_{1,1}\nabla \widetilde{V}_{1,1}\nabla q_{1}^{\ast
}+A_{2,1}\nabla \widetilde{V}_{1,1}\left( \nabla V_{1,1}+\nabla V^{\ast
}\right) -F_{1}\right\vert _{\alpha }\leq \overline{s}^{2}B_{1}^{2}\eta .
\label{5.33}
\end{equation}%
By (\ref{5.19_1}) 
\begin{equation}
C^{\ast }<\frac{B_{1}}{3}.  \label{5.34}
\end{equation}%
Next, consider coefficients in the left hand side of equation (\ref{5.30}).
Using (\ref{3.22}) as well as (\ref{5.23}) for $\left( n,i\right) =\left(
1,1\right) ,$ we obtain $\left\vert A_{1,1}\nabla V_{1,1}\right\vert
_{\alpha }\leq 8\overline{s}^{2}B_{1}.$Hence, conditions (\ref{5.15}) are
satisfied with $P=8\overline{s}^{2}B_{1}$. Hence, by (\ref{5.16}) the
solution of the Dirichlet boundary value problem (\ref{5.30}), (\ref{5.31})
can be estimated as 
\begin{equation*}
\left\vert \widetilde{q}_{1,1}\right\vert _{2+\alpha }\leq \overline{s}%
^{2}KB_{1}^{2}\eta +K\left\Vert \widetilde{\psi }_{1}\right\Vert
_{C^{2+\alpha }\left( \partial \Omega \right) },K=K\left( \Omega ,\overline{s%
}^{2}B_{1}\right) =const.>2.
\end{equation*}%
Using (\ref{5.13}), (\ref{5.18}) and (\ref{5.34}), we obtain from this
inequality 
\begin{equation*}
\left\vert \widetilde{q}_{1,1}\right\vert _{2+\alpha }\leq KB_{1}^{2}\left( 
\overline{s}^{2}+\frac{C^{\ast }}{B_{1}^{2}}\right) \eta \leq
KB_{1}^{2}\left( \overline{s}^{2}+\frac{1}{72}\right) \eta \leq 2\overline{s}%
^{2}KB_{1}^{2}\eta .
\end{equation*}%
Hence, applying (\ref{5.19_1}), we obtain 
\begin{equation}
\left\vert \widetilde{q}_{1,1}\right\vert _{2+\alpha }\leq KB_{1}^{3}\eta .
\label{5.37}
\end{equation}%
Estimate (\ref{5.25}) for $\left( n,i\right) =\left( 1,1\right) $ follows
from (\ref{5.37}). Next, using (\ref{5.6}), (\ref{5.20}) and (\ref{5.37}),
we obtain (\ref{5.26}) for $\left( n,i\right) =\left( 1,1\right) ,$ 
\begin{equation}
\left\vert q_{1,1}\right\vert _{2+\alpha }\leq \left\vert \widetilde{q}%
_{1,1}\right\vert _{2+\alpha }+\left\vert q_{1}^{\ast }\right\vert
_{2+\alpha }\leq KB_{1}^{3}\eta +C^{\ast }\leq 2C^{\ast }.  \label{5.38}
\end{equation}%
Since by (\ref{3.40}) and (\ref{3.43}) $V_{1,1}\in C^{2+\alpha }\left( 
\overline{\Omega }\right) ,$ then (\ref{3.26}) and (\ref{5.38}) imply that $%
v_{1,1}\in C^{2+\alpha }\left( \overline{\Omega }\right) .$ Hence, by (\ref%
{3.27}) $c_{1,1}\in C^{\alpha }\left( \overline{\Omega }\right) .$ This, (%
\ref{3.24_1}) and (\ref{3.25}) imply that $\overline{c}_{1,1}\in C^{\alpha
}\left( \mathbb{R}^{3}\right) .$ Hence, (\ref{5.21}) is true for $\left(
n,i\right) =\left( 1,1\right) .$

Now we estimate the norm $\left\vert \widetilde{c}_{1,1}\right\vert _{\alpha
}.$ Subtracting (\ref{5.11}) from (\ref{3.27}) for $\left( n,i\right)
=\left( 1,1\right) ,$ we obtain 
\begin{equation}
\widetilde{c}_{1,1}=\Delta \widetilde{v}_{1,1}+s_{n}^{2}\nabla \widetilde{v}%
_{1,1}\left( \nabla v_{1,1}+\nabla v_{1}^{\ast }\right) -\overline{F}_{1}.
\label{5.39}
\end{equation}%
Hence, using (\ref{5.12}), (\ref{5.18}), (\ref{5.34}) and (\ref{5.39}), we
obtain 
\begin{equation}
\left\vert \widetilde{c}_{1,1}\right\vert _{\alpha }\leq \left\vert \nabla 
\widetilde{v}_{1,1}\right\vert _{1+\alpha }\left[ 1+\overline{s}^{2}\left(
\left\vert \nabla v_{1,1}\right\vert _{\alpha }+\left\vert \nabla
v_{1}^{\ast }\right\vert _{\alpha }\right) \right] +\frac{B_{1}}{3}\eta .
\label{5.40}
\end{equation}%
Subtracting (\ref{5.10}) from (\ref{3.26}), we obtain $\widetilde{v}_{1,1}=-h%
\widetilde{q}_{1,1}+\widetilde{V}_{1,1}.$ Hence, it follows from (\ref{5.20}%
), (\ref{5.24}) at $\left( n,i\right) =\left( 1,1\right) $ and from (\ref%
{5.37}) that 
\begin{equation}
\left\vert \nabla \widetilde{v}_{1,1}\right\vert _{1+\alpha }\leq
KB_{1}^{3}\eta ^{2}+B_{1}\eta \leq 2B_{1}\eta .  \label{5.41}
\end{equation}%
By (\ref{5.5}), (\ref{5.10}), (\ref{5.19_1}), (\ref{5.20}) and (\ref{5.32_1}%
)-(\ref{5.34}) 
\begin{equation}
\left\vert \nabla v_{n}^{\ast }\right\vert _{1+\alpha }\leq C^{\ast }N\eta
+B_{1}\leq 2B_{1}.  \label{5.42}
\end{equation}%
Hence, (\ref{5.41}), (\ref{5.42}) and (\ref{5.20}) imply that 
\begin{equation*}
\left\vert \nabla v_{1,1}\right\vert _{1+\alpha }=\left\vert \nabla 
\widetilde{v}_{1,1}+\nabla v_{1}^{\ast }\right\vert _{1+\alpha }\leq
\left\vert \nabla \widetilde{v}_{1,1}\right\vert _{1+\alpha }+2B_{1}\leq
2B_{1}\eta +2B_{1}\leq 3B_{1}.
\end{equation*}%
Hence, 
\begin{equation}
1+\overline{s}^{2}\left( \left\vert \nabla v_{1,1}\right\vert _{\alpha
}+\left\vert \nabla v_{1}^{\ast }\right\vert _{\alpha }\right) \leq 1+5B_{1}%
\overline{s}^{2}<6\overline{s}^{2}B_{1}.  \label{5.43}
\end{equation}%
Hence, comparing (\ref{5.43}) with (\ref{5.40}) and (\ref{5.41}), we obtain 
\begin{equation}
\left\vert \widetilde{c}_{1,1}\right\vert _{\alpha }\leq 12\overline{s}%
^{2}B_{1}^{2}\eta +C^{\ast }\eta \leq B_{1}^{3}\eta .  \label{5.431}
\end{equation}%
This establishes (\ref{5.27}) for $\left( n,i\right) =\left( 1,1\right) .$
As it was proved above, (\ref{5.27}) for $\left( n,i\right) =\left(
1,1\right) $ implies (\ref{5.21}) and (\ref{5.22}) for $\left( n,i\right)
=\left( 1,1\right) .$ In summary, we have established (\ref{5.21})-(\ref%
{5.27}) for $\left( n,i\right) =\left( 1,1\right) $.

Since we have established relations (\ref{5.21})-(\ref{5.27}) for $\left(
n,i\right) =\left( 1,1\right) $, we can assume now that we have proved (\ref%
{5.21})-(\ref{5.27}) for $\left( n^{\prime },i^{\prime }\right) \in \left[
0,n\right] \times \left[ 0,i-1\right] $, where $n\geq 1,i\geq 2.$ We now
want to prove (\ref{5.21})-(\ref{5.27}) for $\left( n^{\prime },i^{\prime
}\right) =\left( n,i\right) .$ The mathematical induction principle and
formulas (\ref{3.30}) and (\ref{3.31}) imply that this would be sufficient
for the proof of Theorem 5.1.

Subtracting equation (\ref{5.8}) from equation (\ref{3.29}) and taking into
account boundary conditions (see (\ref{5.7})), we obtain 
\begin{equation}
\begin{split}
\Delta \widetilde{q}_{n,i}& -A_{1,n}\left( h\sum\limits_{j=0}^{n-1}\nabla
q_{j}\right) \nabla \widetilde{q}_{n,i}+A_{1,n}\nabla V_{n,i}\cdot \nabla 
\widetilde{q}_{n,i} \\
& =\left( A_{1,n}\nabla q_{n}^{\ast }-A_{2,n}h\sum\limits_{j=0}^{n-1}\left(
\nabla q_{j}+\nabla q_{j}^{\ast }\right) +2A_{2,n}\nabla V_{n,i}\right)
\left( h\sum\limits_{j=0}^{n-1}\nabla \widetilde{q}_{j}\right) \\
& +\left[ 2A_{2,n}h\sum\limits_{j=0}^{n-1}\nabla q_{j}^{\ast }-A_{1,n}\nabla
q_{n}^{\ast }-A_{2,n}\left( \nabla V_{n,i}+\nabla V^{\ast }\right) \right]
\nabla \widetilde{V}_{n,i}-F_{n}, \\
\widetilde{q}_{n,i}& \mid _{\partial \Omega }=\widetilde{\psi }_{n}(x).
\end{split}
\label{5.44}
\end{equation}%
First, we estimate the difference of tails $\widetilde{V}_{n,i}.$ Since
relations (\ref{5.21})-(\ref{5.27}) are valid for $\left( n^{\prime
},i^{\prime }\right) \in \left[ 0,n\right] \times \left[ 0,i-1\right] ,$
then by Theorem 4.2 
\begin{eqnarray}
\left\vert \nabla V_{n,i}\right\vert _{1+\alpha } &\leq &B_{1},
\label{5.45_0} \\
\left\vert \nabla \widetilde{V}_{n,i}\right\vert _{1+\alpha } &\leq
&B\left\vert \widetilde{c}_{n,i-1}\right\vert _{\alpha }\leq B_{1}B_{1}^{3%
\left[ i-1+\left( n-1\right) m\right] }\cdot \eta =B_{1}^{3\left[ i-1+\left(
n-1\right) m\right] +1}\cdot \eta .  \label{5.45_1}
\end{eqnarray}%
\begin{equation*}
\left\vert \nabla \widetilde{V}_{n,i}\right\vert _{1+\alpha }\leq
B\left\vert \widetilde{c}_{n,i-1}\right\vert _{\alpha }\leq B_{1}B_{1}^{3%
\left[ i-1+\left( n-1\right) m\right] }\cdot \eta =B_{1}^{3\left[ i-1+\left(
n-1\right) m\right] +1}\cdot \eta .
\end{equation*}%
These estimates establish (\ref{5.23}) and (\ref{5.24}) for $\left(
n^{\prime },i^{\prime }\right) =\left( n,i\right) .$

We now estimate the right hand side of equation (\ref{5.44}). First, using (%
\ref{3.22}), (\ref{5.5}), (\ref{5.32_1}) and (\ref{5.45_0}), we obtain 
\begin{equation*}
\left\vert A_{1,n}\nabla q_{n}^{\ast }-A_{2,n}h\sum\limits_{j=0}^{n-1}\left(
\nabla q_{j}+\nabla q_{j}^{\ast }\right) +2A_{2,n}\nabla V_{n,i}\right\vert
_{\alpha }\leq 8\overline{s}^{2}\left( C^{\ast }+3C^{\ast }Nh+2B_{1}\right)
\leq 8\overline{s}^{2}\left( C^{\ast }+1+2B_{1}\right) .
\end{equation*}%
This inequality and (\ref{5.34}) lead to

\begin{equation}
\left\vert A_{1,n}\nabla q_{n}^{\ast }-A_{2,n}h\sum\limits_{j=0}^{n-1}\left(
\nabla q_{j}+\nabla q_{j}^{\ast }\right) +2A_{2,n}\nabla V_{n,i}\right\vert
_{\alpha }\leq 24\overline{s}^{2}B_{1}.  \label{5.46}
\end{equation}%
Estimates (\ref{5.25}) hold for functions $\widetilde{q}_{j}=q_{j}-q_{j}^{%
\ast },j\in \left[ 0,n-1\right] .$ Hence, using (\ref{5.20}), we obtain 
\begin{equation*}
\left\vert h\sum\limits_{j=0}^{n-1}\nabla \widetilde{q}_{j}\right\vert
_{\alpha }\leq KB_{1}^{3Nm}N\eta ^{2}\leq \eta .
\end{equation*}%
Combining this with (\ref{5.46}), we obtain the following estimate for the
term in the second raw of (\ref{5.44}) 
\begin{equation}
\left\vert A_{1,n}\nabla q_{n}^{\ast }-A_{2,n}h\sum\limits_{j=0}^{n-1}\left(
\nabla q_{j}+\nabla q_{j}^{\ast }\right) +2A_{2,n}\nabla V_{n,i}\right\vert
_{\alpha }\left\vert h\sum\limits_{j=0}^{n-1}\nabla \widetilde{q}%
_{j}\right\vert _{\alpha }\leq 24\overline{s}^{2}B_{1}\eta .  \label{5.47}
\end{equation}%
Next, using (\ref{3.22}), (\ref{5.5}), (\ref{5.19_1}), (\ref{5.19_2}), (\ref%
{5.32_1}) and (\ref{5.45_0}), we obtain 
\begin{eqnarray*}
&&\left\vert 2A_{2,n}h\sum\limits_{j=0}^{n-1}\nabla q_{j}^{\ast
}-A_{1,n}\nabla q_{n}^{\ast }-A_{2,n}\left( \nabla V_{n,i}+\nabla V^{\ast
}\right) \right\vert _{\alpha } \\
&\leq &16\overline{s}^{2}C^{\ast }N\eta +8\overline{s}^{2}C^{\ast }+16%
\overline{s}^{2}B_{1}\leq \overline{s}^{2}+3\overline{s}^{2}B_{1}+16%
\overline{s}^{2}B_{1}\leq 20\overline{s}^{2}B_{1}.
\end{eqnarray*}%
Hence, using (\ref{5.9}), (\ref{5.34}) and (\ref{5.45_1}), we obtain 
\begin{eqnarray*}
&&\left\vert 2A_{2,n}h\sum\limits_{j=0}^{n-1}\nabla q_{j}^{\ast
}-A_{1,n}\nabla q_{n}^{\ast }-A_{2,n}\left( \nabla V_{n,i}+\nabla V^{\ast
}\right) \right\vert _{\alpha }\left\vert \nabla \widetilde{V}%
_{n,i}\right\vert _{\alpha }+\left\vert F_{n}\right\vert _{\alpha } \\
&\leq &20\overline{s}^{2}B_{1}B_{1}^{3\left[ i-1+\left( n-1\right) m\right]
+1}\cdot \eta +C^{\ast }\eta \leq 20\overline{s}^{2}B_{1}B_{1}^{3\left[
i-1+\left( n-1\right) m\right] +1}\eta +\frac{B_{1}}{3}\eta .
\end{eqnarray*}%
Combining this with (\ref{5.47}) and using (\ref{5.19_1}), we obtain the
following estimate for the right hand side ($rhs$) of (\ref{5.44}) 
\begin{eqnarray*}
\left\vert rhs\right\vert _{\alpha } &\leq &21\overline{s}^{2}B_{1}B_{1}^{3%
\left[ i-1+\left( n-1\right) m\right] +1}\cdot \eta +24\overline{s}%
^{2}B_{1}\eta \leq 21\overline{s}^{2}B_{1}B_{1}^{3\left[ i-1+\left(
n-1\right) m\right] +1}\left( 1+\frac{24}{21B_{1}}\right) \eta \\
&\leq &21\overline{s}^{2}B_{1}B_{1}^{3\left[ i-1+\left( n-1\right) m\right]
+1}\left( 1+\frac{1}{21}\right) \eta =22\overline{s}^{2}B_{1}B_{1}^{3\left[
i-1+\left( n-1\right) m\right] +1}\cdot \eta .
\end{eqnarray*}%
Thus, 
\begin{equation}
\left\vert rhs\right\vert _{\alpha }\leq 22\overline{s}^{2}B_{1}B_{1}^{3%
\left[ i-1+\left( n-1\right) m\right] +1}\cdot \eta .  \label{5.48}
\end{equation}

We now estimate coefficients which are multiplied by $\nabla \widetilde{q}%
_{n,i}$ in the left hand side of (\ref{5.44}). We use (\ref{3.22}), (\ref%
{5.19_1}), (\ref{5.19_2}), (\ref{5.20}) and (\ref{5.34}). By the assumption
of the mathematical induction method we have that inequalities (\ref{5.26})
are valid for functions $q_{j}$ with $j\in \left[ 0,n-1\right] .$ First, 
\begin{equation}
\left\vert A_{1,n}\left( h\sum\limits_{j=0}^{n-1}\nabla q_{j}\left( x\right)
\right) \right\vert _{\alpha }\leq 16\overline{s}^{2}C^{\ast }N\eta \leq 
\frac{6\overline{s}^{2}}{B_{1}^{3Nm}}<\frac{1}{10}.  \label{5.49}
\end{equation}%
Next, using (\ref{3.22}) and (\ref{5.45_0}), we obtain 
\begin{equation}
\left\vert A_{1,n}\nabla V_{n,i}\right\vert _{1+\alpha }\leq 8\overline{s}%
^{2}B_{1}.  \label{5.50}
\end{equation}%
Hence, it follows from (\ref{5.49}) and (\ref{5.50}) that the Dirichlet
boundary value problem (\ref{5.44}) satisfies conditions (\ref{5.14})-(\ref%
{5.16}) with $P=9\overline{s}^{2}B_{1},K=K\left( \Omega ,\overline{s}%
^{2}B_{1}\right) >2.$ Hence, using (\ref{5.13}), (\ref{5.16}), (\ref{5.18}),
(\ref{5.34}) and (\ref{5.48}), we obtain 
\begin{equation*}
\left\vert \widetilde{q}_{n,i}\right\vert _{2+\alpha }\leq K\left[ 22%
\overline{s}^{2}B_{1}B_{1}^{3\left[ i-1+\left( n-1\right) m\right]
+1}+C^{\ast }\right] \eta \leq K\cdot 23\overline{s}^{2}B_{1}B_{1}^{3\left[
i-1+\left( n-1\right) m\right] +1}\cdot \eta .
\end{equation*}%
Since by (\ref{5.19_1}) $B_{1}\geq 24\overline{s}^{2},$ then the last
estimate leads to 
\begin{equation*}
\left\vert \widetilde{q}_{n,i}\right\vert _{2+\alpha }\leq KB_{1}^{3\left[
i+\left( n-1\right) m\right] }\cdot \eta ,
\end{equation*}%
which is (\ref{5.25}). Next, we prove (\ref{5.26}). We use (\ref{5.5}), (\ref%
{5.6}), (\ref{5.20}) and (\ref{5.25}), 
\begin{equation*}
\left\vert q_{n,i}\right\vert _{2+\alpha }\leq \left\vert \widetilde{q}%
_{n,i}\right\vert _{2+\alpha }+\left\vert q_{n}^{\ast }\right\vert
_{2+\alpha }\leq KB_{1}^{3\left[ i+\left( n-1\right) m\right] }\cdot \eta
+C^{\ast }\leq 2C^{\ast }.
\end{equation*}

Estimate now the norm $\left\vert \widetilde{c}_{n,i}\right\vert _{\alpha }.$
We obtain similarly with (\ref{5.40}) 
\begin{equation}
\left\vert \widetilde{c}_{n,i}\right\vert _{\alpha }\leq \left\vert \nabla 
\widetilde{v}_{n,i}\right\vert _{1+\alpha }\left[ 1+\overline{s}^{2}\left(
\left\vert \nabla v_{n,i}\right\vert _{\alpha }+\left\vert \nabla
v_{n}^{\ast }\right\vert _{\alpha }\right) \right] +\frac{B_{1}}{3}\eta .
\label{5.51}
\end{equation}%
We have 
\begin{equation*}
\widetilde{v}_{n,i}\left( x\right) =-h\widetilde{q}_{n,i}\left( x\right)
-h\sum\limits_{j=0}^{n-1}\widetilde{q}_{j}\left( x\right) +\widetilde{V}%
_{n,i}\left( x\right) ,\text{ }x\in \Omega .
\end{equation*}%
Hence, by (\ref{5.18}), (\ref{5.19_1}), (\ref{5.20}), (\ref{5.24}) and (\ref%
{5.25}) 
\begin{equation}
\left\vert \nabla \widetilde{v}_{n,i}\right\vert _{1+\alpha }\leq KNB_{1}^{3%
\left[ i+\left( n-1\right) m\right] }\eta ^{2}+B_{1}^{3\left[ i-1+\left(
n-1\right) m\right] +1}\eta \leq \frac{25}{24}B_{1}^{3\left[ i-1+\left(
n-1\right) m\right] +1}\cdot \eta .  \label{5.52}
\end{equation}%
Next, using (\ref{3.26}) and (\ref{5.42}), we obtain similarly with (\ref%
{5.43}) 
\begin{equation*}
1+\overline{s}^{2}\left( \left\vert \nabla v_{n,i}\right\vert _{\alpha
}+\left\vert \nabla v_{n}^{\ast }\right\vert _{\alpha }\right) \leq 6%
\overline{s}^{2}B_{1}.
\end{equation*}%
Combining this with (\ref{5.51}) and (\ref{5.52}) and using (\ref{5.19_1}),
we obtain 
\begin{eqnarray*}
\left\vert \widetilde{c}_{n,i}\right\vert _{\alpha } &\leq &7\overline{s}%
^{2}B_{1}B_{1}^{3\left[ i-1+\left( n-1\right) m\right] +1}\cdot \eta +\frac{%
B_{1}}{3}\eta \leq 8\overline{s}^{2}B_{1}B_{1}^{3\left[ i-1+\left(
n-1\right) m\right] +1}\cdot \eta \\
&<&B_{1}^{2}B_{1}^{3\left[ i-1+\left( n-1\right) m\right] +1}\cdot \eta
=B_{1}^{3\left[ i+\left( n-1\right) m\right] }\cdot \eta .
\end{eqnarray*}%
Thus, $\left\vert \widetilde{c}_{n,i}\right\vert _{\alpha }\leq B^{3\left[
i+\left( n-1\right) m\right] }\eta ,$ which proves (\ref{5.27}). Thus,
relations (\ref{5.21})-(\ref{5.27}) are valid $\left( n^{\prime },i^{\prime
}\right) =\left( n,i\right) .\square $

\section{Numerical Studies}

\label{sec:6}

\begin{figure}[tbp]
\begin{center}
\begin{tabular}{ccc}
{\includegraphics[scale=0.19,clip=]{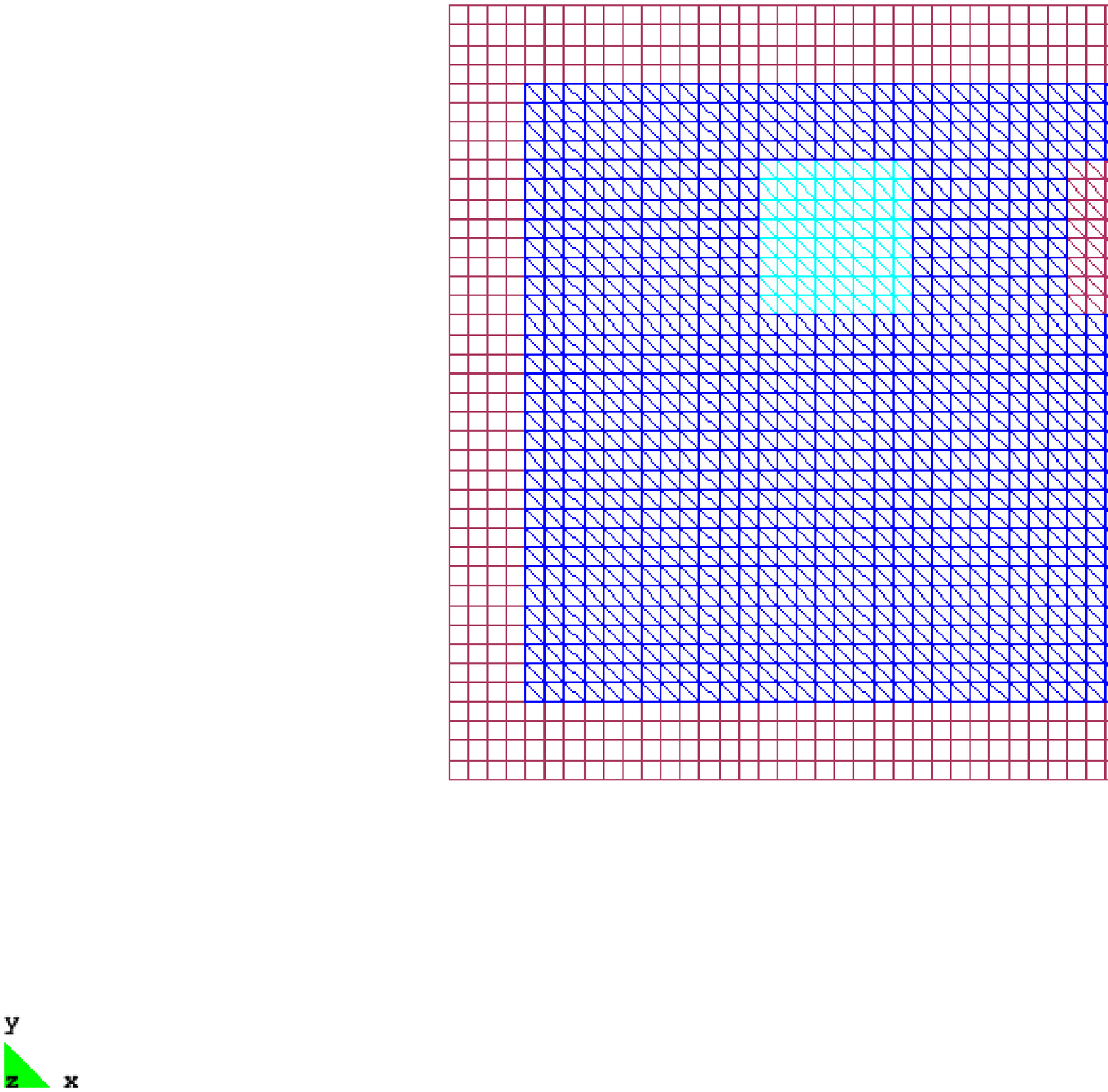}} & {%
\includegraphics[scale=0.19,clip=]{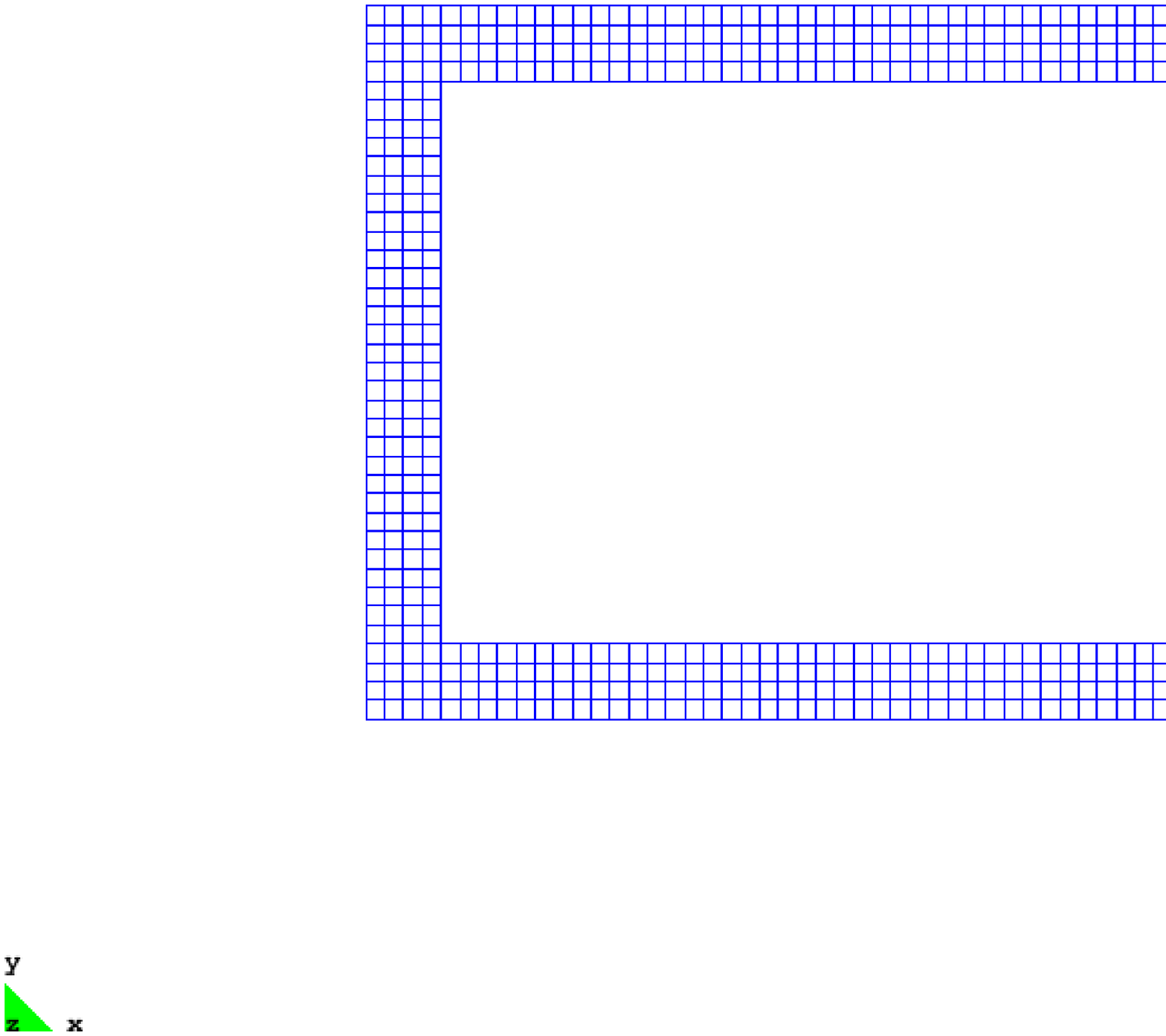}} & {%
\includegraphics[scale=0.19,clip=]{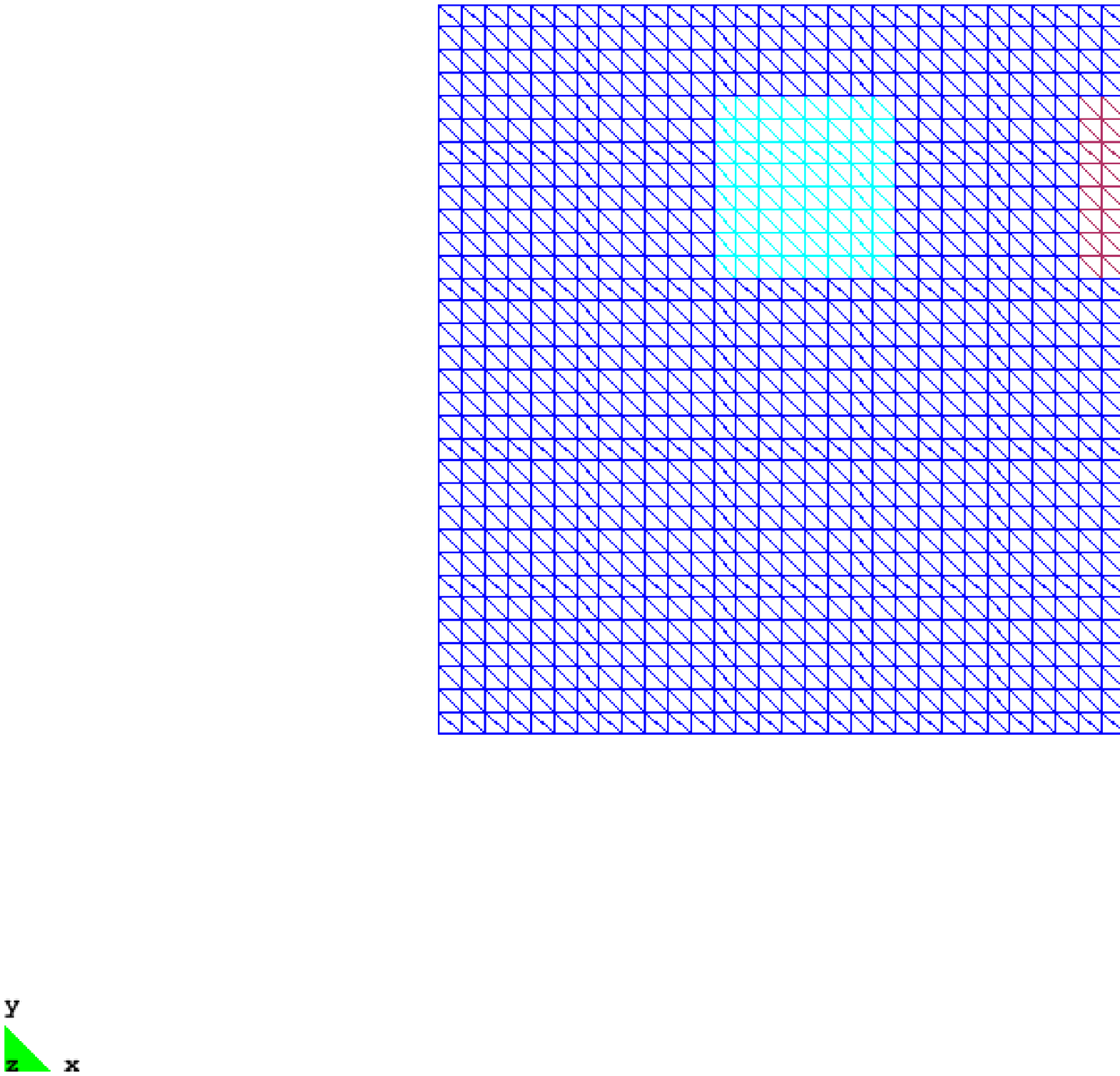}} \\ 
a) $G = G_{FEM} \cup G_{FDM}$ & b) $G_{FDM}$ & c) $G_{FEM}= \Omega$%
\end{tabular}%
\end{center}
\caption{ \emph{\ a) Geometry of the hybrid mesh. This is a combination of
the quadrilateral mesh in the subdomain $G_{FDM}$ b), where we apply FDM,
and the finite element mesh in the inner domain $G_{FEM}= \Omega$ c), where
we use FEM. The solution of the inverse problem is computed in $G_{FEM}=
\Omega$. }}
\label{fig:F1}
\end{figure}

In this section we conduct some numerical experiments in both 2d and 3d
cases. In the 2d case we use specific ranges of parameters for a simplified
mathematical model of imaging of antipersonnel land mines, see \cite{KBK}
and sections 6.8.2 and 6.8.3 of \cite{BK} for this model. In the 3d case we
model imaging of explosives hidden on belts worn by humans. We point out
that in both cases our mathematical models are certainly simplified ones and
further studies are necessary to see how they reflect the reality.

It is well known that there are always some discrepancies between the
theories and numerical implementations of complicated numerical methods. We
now list two discrepancies for our case. First, the above theory was
developed for the case of the point source, because of a convenience of the
analysis. In computations, however, we work with the case of an incident
plane wave with the single direction of incidence. This is because it is
better to operate with a plane wave computationally. Also, in the case when
the point source is far from the domain of interest, it can be approximately
treated as a plane wave. The above theory can be extended to the case of a
plane wave after a purely technical additional effort. Second, to decrease
the complexity of our computations, we replace (\ref{3.24_1}) and (\ref{3.25}%
) with the following simplified formula 
\begin{equation}
\overline{c}_{n,i}\left( x\right) =\left\{ 
\begin{array}{c}
c_{n,i}\left( x\right) \text{ if }c_{n,i}\left( x\right) \geq 1\text{ and }%
x\in \overline{\Omega }, \\ 
1\text{ if either }c_{n,i}\left( x\right) <1\text{ or }x\notin \overline{%
\Omega }.%
\end{array}%
\right.  \label{6.100}
\end{equation}%
When reconstructing functions $c_{n,i}\left( x\right),$ we use a weak
formulation of (\ref{3.27}) via finite elements, see pages 184, 185 of \cite%
{BK} for this formulation. Comparison of Figures 3.12 and 3.13 of \cite{BK}
(pages 182, 183) shows that this formulation provides significantly more
accurate results than the strong formulation (\ref{3.27}).

We now describe our stopping criterion used in computations of sections \ref%
{sec:6.1}, \ref{sec:6.2}. We stop computing functions ${c}_{n,i}$ on every
pseudo-frequency interval $[s_n, s_{n-1})$ when 
\begin{equation}
\text{either }\quad N{_{n}}\geq N{_{n-1}}\text{ or } \quad N{_{n}}\leq \eta ,
\label{crit1}
\end{equation}
where 
\begin{equation}  \label{crit2}
N_n= \frac{||{c}_{n,i} - {c}_{n,i-1}||_{L_2(\Omega)}}{||{c}%
_{n,i}||_{L_2(\Omega)}}.
\end{equation}
Here, $i$ is the number of iterations with respect to the tail on every
pseudo-frequency interval $[s_n, s_{n-1})$. Recall that we  define by $m$
the number when iterations with respect to the tail are  stopped.

To generate data for the CIP, we solve the forward problem for equation (\ref%
{3.1}). Since it is impossible to numerically solve this problem in the
entire space $\mathbb{R}^{n}\left( n=2,3\right) ,$ we solve it in a
rectangle in 2-d and in a rectangular prism in 3-d, just as in \cite{BK}. We
denote this each of these domains $G.$ Thus, $G$ is our computational domain
in which we compute the forward problem, and it replaces $\mathbb{R}%
^{n}\left( n=2,3\right) .$ We impose the first order absorbing boundary
condition \cite{EM} on one part of the boundary $\partial G$ and zero
Neumann boundary condition on another part of $\partial G.$ In all cases the
domain of interest $\Omega \subset G,\partial \Omega \cap \partial
G=\varnothing ,$ see for details below.

\subsection{Our mathematical model of imaging of plastic antipersonnel land
mines: 2d study}

\label{sec:6.1}

The first main simplification of our model is that we consider the 2d case
instead of 3d, although a 3d numerical test is also presented below. Second,
we ignore the air/ground interface, assuming that the governing PDE is valid
on the entire 2d plane. Results of studies of experimental data in \cite%
{KBKSNF,IEEE} as well as of section 6.9 of \cite{BK} indicate that the
influence of the air/ground interface can be significantly decreased via a
data pre-processing procedure.

Let the ground be 
\begin{equation*}
\left\{ (x_{1},x_{2}):x_{2}<a=const.\right\} \subset \mathbb{R}^{2}.
\end{equation*}%
Consider a polarized electric field which is generated by a plane wave,
initialized at the line $\left\{ x_{2}=a^{0}>a,x_{1}\in \mathbb{R}\right\} $
at the moment of time $t=0$. The following hyperbolic equation can be
derived from the Maxwell equations in the 2d case 
\begin{equation}
c(x)u_{tt}=\Delta u,\;\left( x,t\right) \in \mathbb{R}^{2}\times \left(
0,\infty \right) .  \label{6.1}
\end{equation}%
where the function $u(x,t)$ is a component of the electric field and $%
c(x):=\varepsilon _{r}\left( x\right) $ is the spatially distributed
dielectric constant. We assume that the function $c(x)$ satisfies conditions
(\ref{3.3}) in 2d. We model imaging of dielectric constants in plastic land
mines. In doing so, we do not assume a knowledge of the background medium.
So, images of land mines are constructed only on the basis of values of the
dielectric constant $c(x)$ inside of them.

Let $\Omega $ be the domain of interest in the ground, where we search for
land mines. We set 
\begin{equation*}
\Omega =\left\{ \left( x,y\right) \in \left( -0.35,0.35\right) \text{ m}%
\times (-0.05,0.35)\text{ m}\right\} ,
\end{equation*}%
where \textquotedblleft m" stands for meter. Introducing dimensionless
spatial variables $\left( x^{\prime },y^{\prime }\right) =\left( x,y\right)
/\left( 0.1\text{m}\right) $ without changing notations for brevity, we
obtain the dimensionless domain 
\begin{equation}
\Omega =\left( -3.5,3.5\right) \times \left( -0.5,3.5\right) .  \label{6.2}
\end{equation}%
Hence, the ground is at $\left\{ x_{2}=3.5\right\} $ and the depth of the
domain of interest is $4$, which means 40 cm in real dimensions. Our
backreflected signal is measured at the backscattering side, 
\begin{equation}
\text{ the backscattering side is }\Gamma =\left\{ \left( x_{1},x_{2}\right)
:x_{1}\in \left( -0.35,0.35\right) ,x_{2}=3.5\right\} .  \label{6.2_1}
\end{equation}%
It is well known that the maximal depth of an antipersonnel land mine does
not exceed about 10 centimeters. Hence, we model these mines as two small
rectangles with the 0.1 m and 0.2 m length of sides, and 0.1 m width of
sides, respectively. Centers of those rectangles are located at $x_{2}=2.5$
which is of 10 depth cm in variables with dimensions, see Figure \ref{fig:F1}%
.

Tables of dielectric constants \cite{Tables} show that in the dry sand the
dielectric constant $\varepsilon _{r}=5$ and $\varepsilon _{r}=22$ in the
trinitrotoluene (TNT). Hence, the mine/background contrast is $22/5\approx 4$%
. Thus, we consider new parameters $\varepsilon _{r}^{\prime },t^{\prime }$
without changing notations, $\varepsilon _{r}^{\prime }=\varepsilon
_{r}/5,t^{\prime }=t/\sqrt{5}.$ Hence, we obtain the following relative
values of the dielectric constant in our tests 
\begin{equation}
c\left( x\right) =\varepsilon _{r}\text{(dry sand)}=1,\text{ }c\left(
x\right) =\varepsilon _{r}\text{(TNT)}=4.\text{ }  \label{6.3}
\end{equation}

\subsection{Our mathematical model of imaging of explosives hidden in belts
worn by humans: 3d study}

\label{sec:6.2}

\begin{figure}[tbp]
\begin{center}
\begin{tabular}{cc}
{\includegraphics[scale=0.4,clip=]{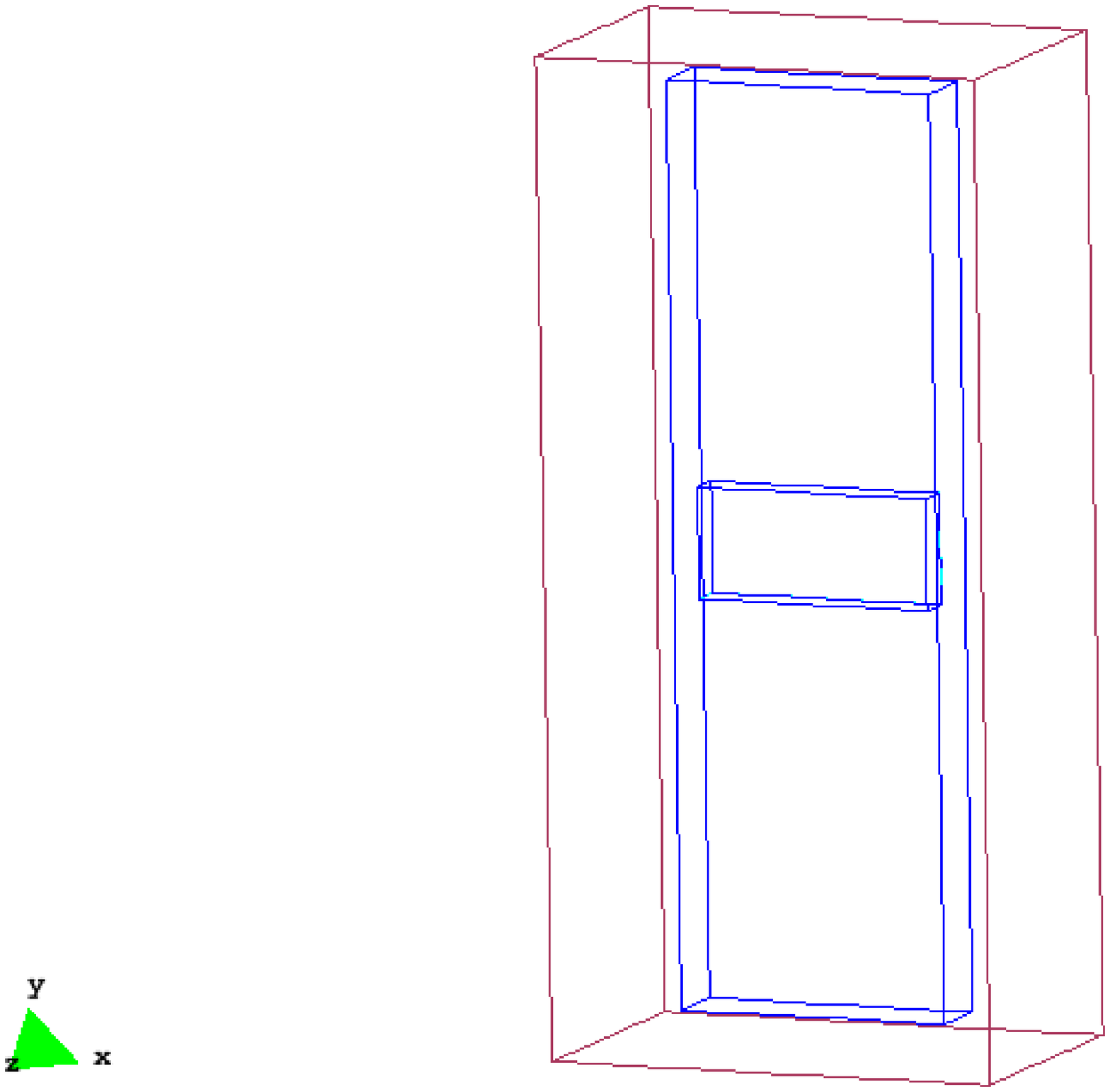}} & {\ %
\includegraphics[scale=0.4,clip=]{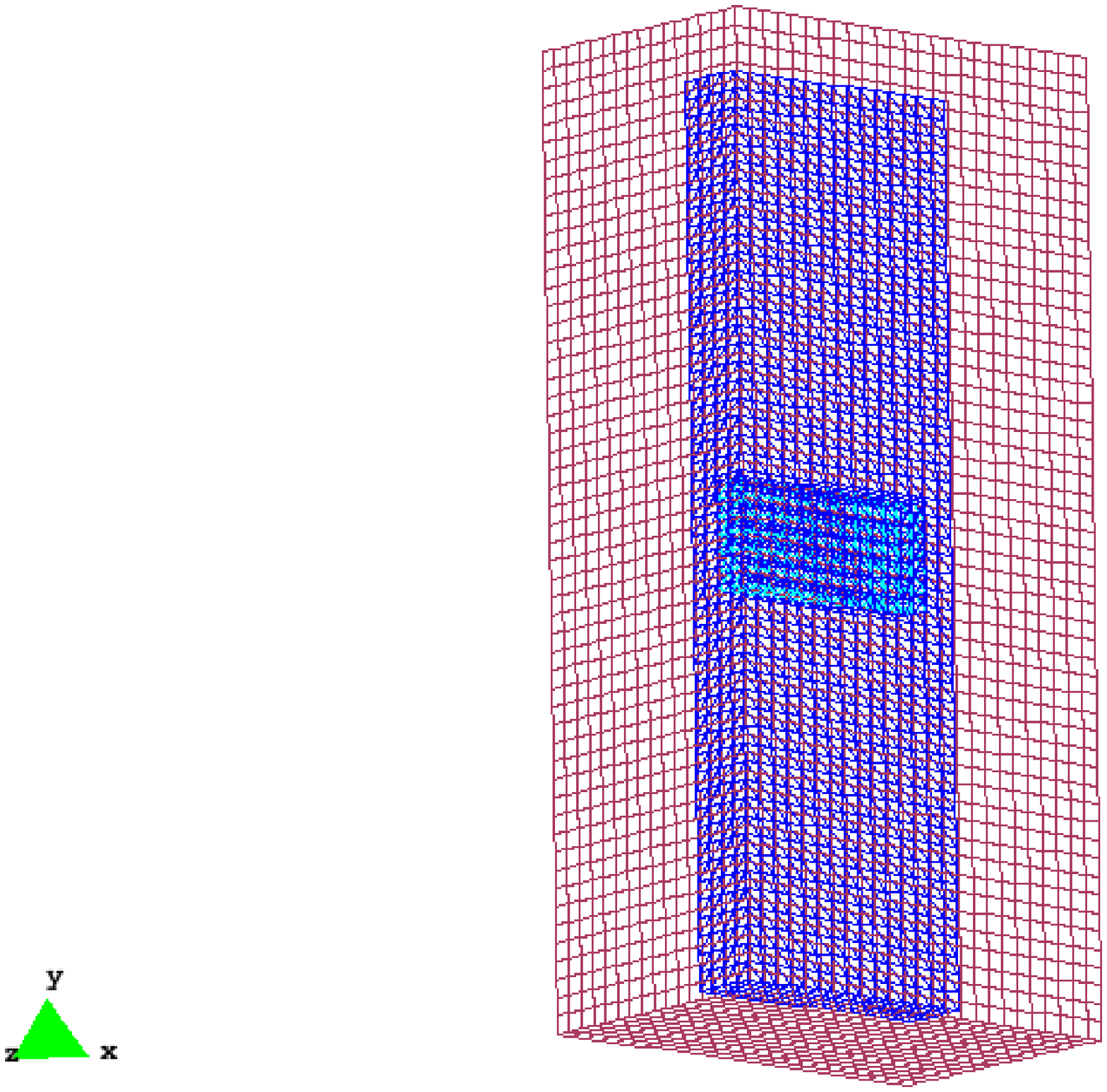}} \\ 
a) & b) \\ 
{\includegraphics[scale=0.4,clip=]{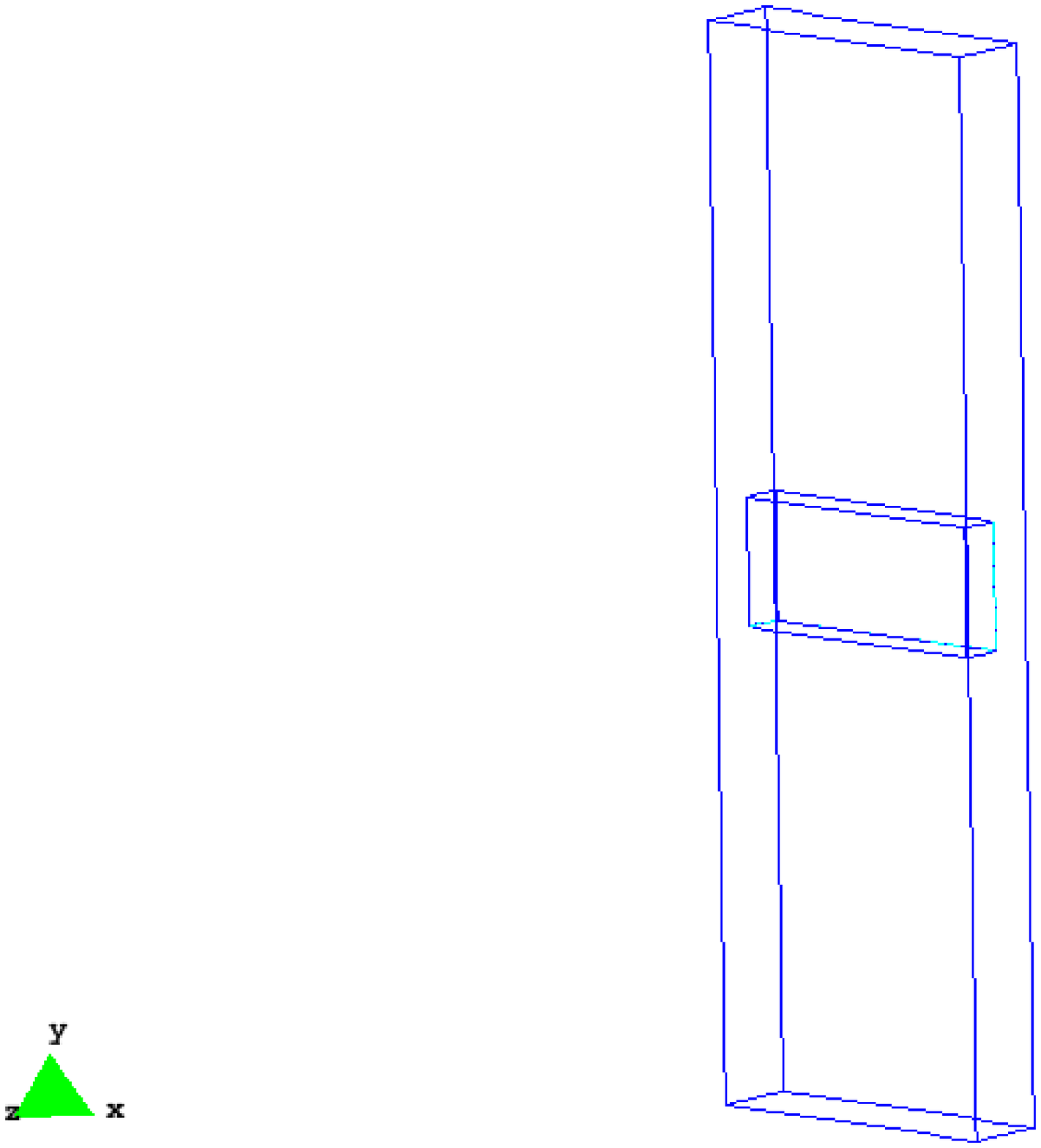}} & {\ %
\includegraphics[scale=0.4,clip=]{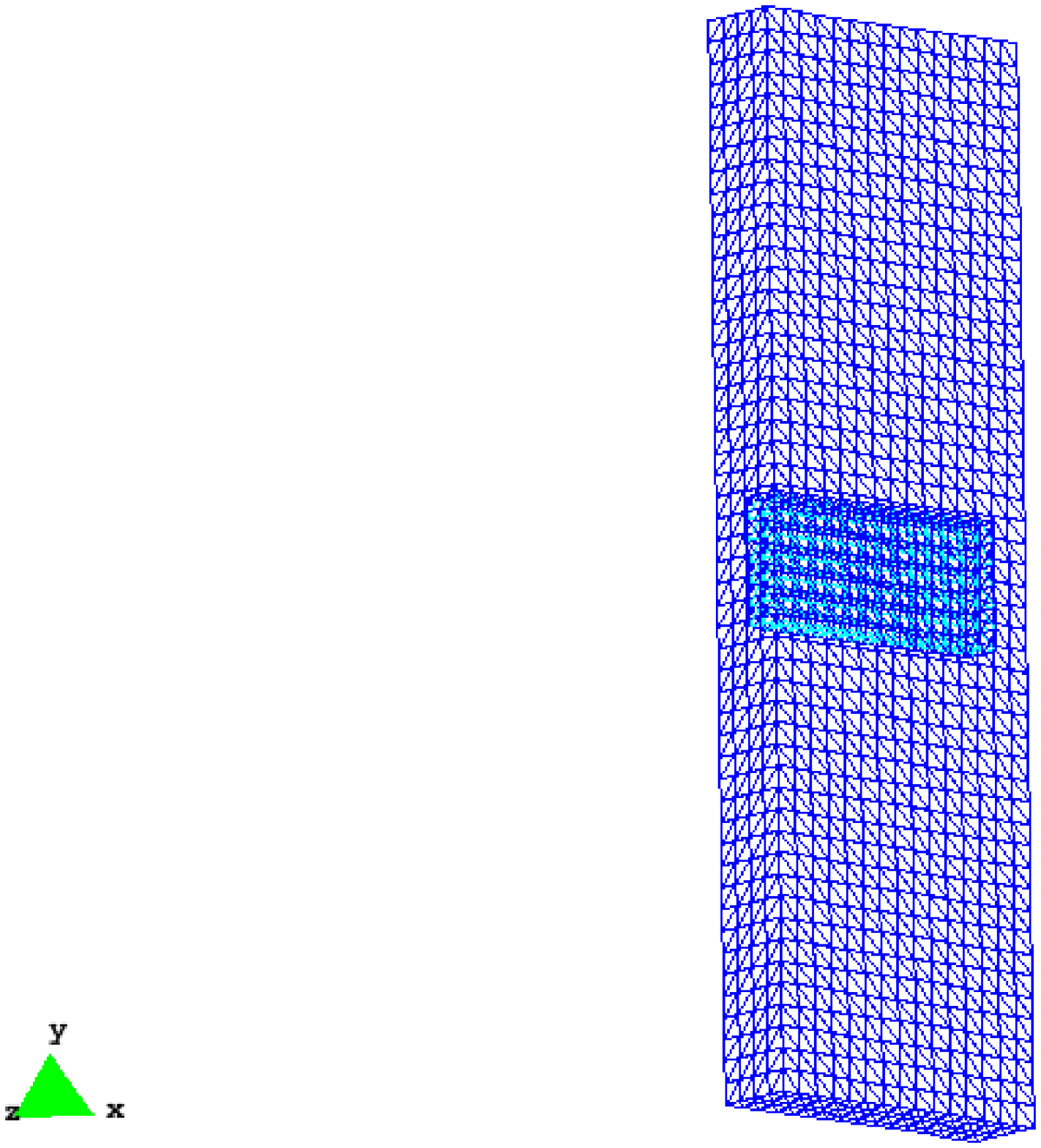}} \\ 
c) & d)%
\end{tabular}%
\end{center}
\caption{{\protect\small \emph{a) Hybrid FEM/FDM geometry $G$; b) Mesh
outlined at the boundary in the Hybrid FEM/FDM geometry $G$; c) Inner FEM
geometry $G_{FEM} = \Omega \subset G$; d) Mesh outlined at the boundary of
the inner FEM geometry $G_{FEM} = \Omega$.}}}
\label{fig:F3D_1}
\end{figure}

In all places below where the 3d case is discussed, we use the same notation
for the vector $x=\left( x,y,z\right) $ and for its first coordinate. This
does not lead to an ambiguity. In the 3d case we model the body of a human
as a rectangular prism of 2 meters tall, 0.6 meters wide and 0.16 meters
"deep". The vertical coordinate is $y$ and $z$ is responsible for the depth.
Hence, in this case computational domain $G$ is 
\begin{equation}
G=\left\{ \left( x,y,z\right) :x\in \left( -0.5,0.5\right) ,y\in \left(
-1.08,1.08\right) ,z\in \left( -0.32,0.32\right) \right\} .  \label{6.23}
\end{equation}%
We model the belt with explosives as the rectangular prism, which is a
subdomain of the first one. Sizes of that \textquotedblleft belt" are 0.3
meters in the vertical direction, 0.52 meters in horizontal direction and
0.08 meters of \textquotedblleft depth". Hence, dividing by 1 meter, we
obtain that these two prisms are respectively dimensionless domains $\Omega $
and $\Omega _{belt},$%
\begin{eqnarray}
\Omega &=&G_{FEM}=\left\{ \left( x,y,z\right) :x\in \left( -0.3,0.3\right)
,y\in \left( -1,1\right) ,z\in \left( -0.08,0.08\right) \right\} ,\Omega
\subset G  \label{6.20} \\
\Omega _{belt} &=&\left\{ \left( x,y,z\right) :x\in \left( -0.26,0.26\right)
,y\in \left( -0.15,0.15\right) ,z\in \left( -0.04,0.04\right) \right\}
\subset \Omega .  \label{6.21}
\end{eqnarray}%
On Figure \ref{fig:F3D_1} the domain $G$ is the largest prism, $\Omega $ is
the smaller prism and $\Omega _{belt}$ is the smallest prism. Our
backscattering signal is measured at the front side $\Gamma $ of the prism $%
\Omega .$ The incident plane wave propagates along the positive direction of
the $z-$axis. Therefore, the front side of the prism $\Omega $ is the
backscattering side. We define different boundaries of $G$ and $\Omega $ as 
\begin{eqnarray}
\text{Left side of }\Omega \text{ is }\Gamma _{l} &=&\left\{ x=-0.3\right\}
\cap \overline{\Omega },  \label{6.24} \\
\text{Right side of }\Omega \text{ is }\Gamma _{r} &=&\left\{ x=0.3\right\}
\cap \overline{\Omega },  \label{6.25} \\
\text{Back side of }\Omega \text{ is }\Gamma _{b} &=&\left\{ z=0.08\right\}
\cap \overline{\Omega },  \label{6.24_1} \\
\text{Front (backscattering) side of }\Omega \text{ is }\Gamma &=&\left\{
z=-0.08\right\} \cap \overline{\Omega },  \label{6.25_1} \\
\text{Top side of }\Omega \text{ is }\Gamma _{t} &=&\left\{ y=1\right\} \cap 
\overline{\Omega },  \label{6.24_2} \\
\text{Bottom side of }\Omega \text{ is }\Gamma _{bot} &=&\left\{
y=-1\right\} \cap \overline{\Omega },  \label{6.25_2} \\
\text{Front side of }G\text{ is }\partial _{1}G &=&\left\{ z=-0.32\right\}
\cap \overline{G},  \label{6.26} \\
\text{Back side of }G\text{ is }\partial _{2}G &=&\left\{ z=0.32\right\}
\cap \overline{G},  \label{6.27} \\
\partial _{3}G &=&\partial G\diagdown \left( \partial _{1}G\cup \partial
_{2}G\right) .  \label{6.28}
\end{eqnarray}%
Therefore, we actually assume here that we measure the backreflected signal
at the distance of 4 cm off the belt. Although this is unrealistic, we can
justify this as follows. Suppose that we actually measure the backscattering
signal on a plane $P=\left\{ z=-Z,Z>0.08\right\} .$ We can approximately
assume that $w\left( x,y,-0.08,s\right) =w_{0}\left( x,y,-0.08,s\right) $
for $\left( x,y\right) \in \mathbb{R}^{2}\diagdown \Gamma ,s\in \left[ 
\underline{s},\overline{s}\right] ,$ see subsection 6.5. Recall that the
function $\varphi \left( x,y,z,s\right) $ is the Laplace transform of the
data $g\left( x,y,z,t\right) $ in (\ref{3.4}) (subsection 3.2). Using the
Green's function for the equation $\Delta w-s^{2}w=0$ in the half space $%
\left\{ z<-0.08\right\} ,$ we can obtain an integral equation of the first
kind with respect to the function $p\left( x,y,s\right) :=w\left(
x,y,-0.08,s\right) .$ The right hand side of this equation will be the
function $\varphi \left( x,y,-Z,s\right) ,~s$ will be a parameter and
integration will be carried out over the rectangle $\Gamma .$ This is a
convolution equation, which represents a linear ill-posed problem.
Algorithms of solving convolution equations using the Tikhonov
regularization are described in the book \cite{T}. Thus, solution of this
equation would provide us with an approximation of the function $w\left(
x,y,-0.08,s\right) $ for $\left( x,y\right) \in \Gamma ,s\in \left[ 
\underline{s},\overline{s}\right] .$ On the other hand, the latter is the
function which we consider as given data in our numerical experiments of
subsection 6.9. Thus, assuming below that we have the data at $\Gamma ,$ we
avoid the intermediate step of solving that integral equation.

\subsection{Data simulation in 2d}

\label{sec:6.3}

\begin{figure}[tbp]
\begin{center}
\begin{tabular}{cc}
{\includegraphics[scale=0.2,clip=]{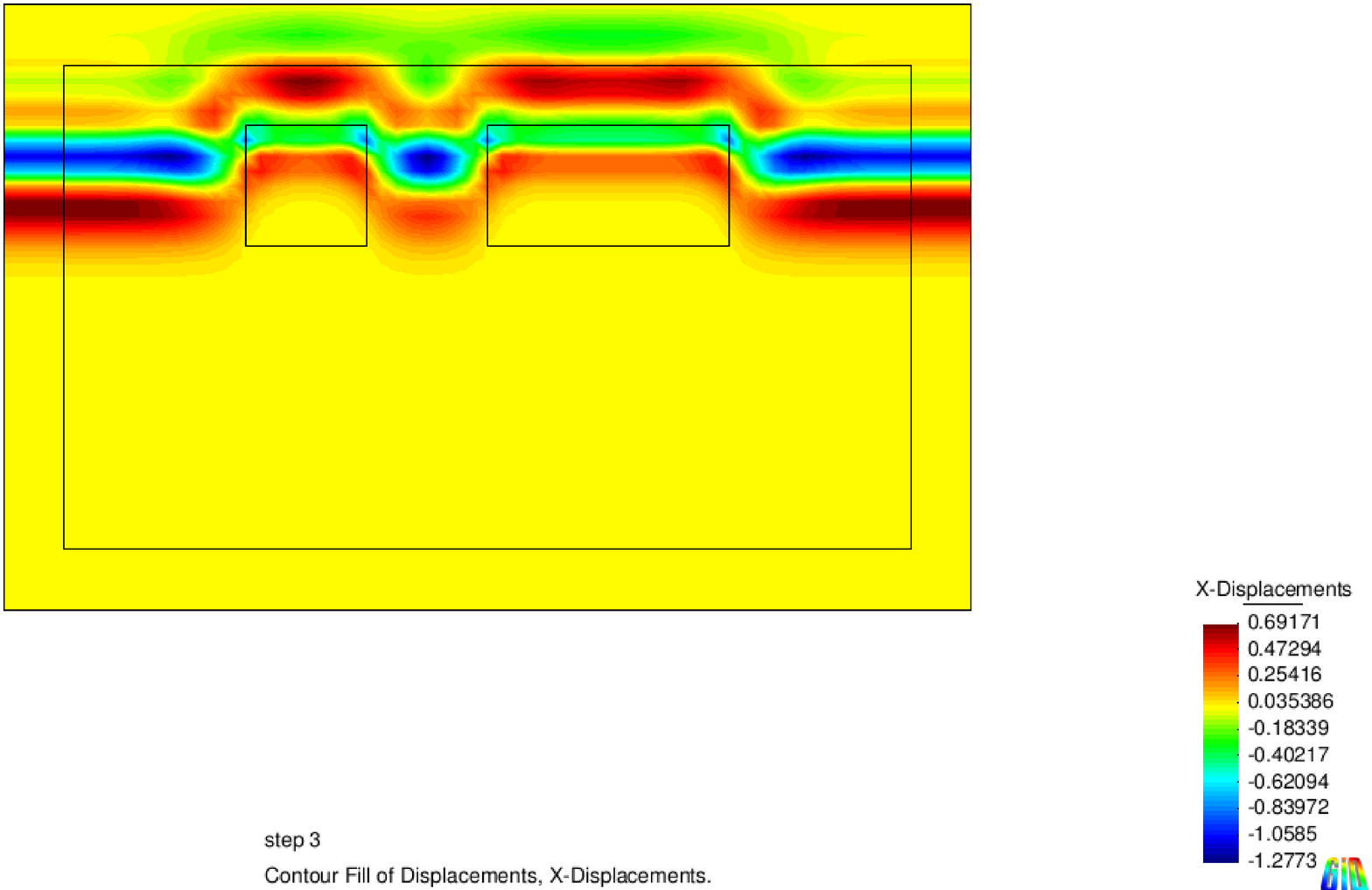}} & {%
\includegraphics[scale=0.2,clip=]{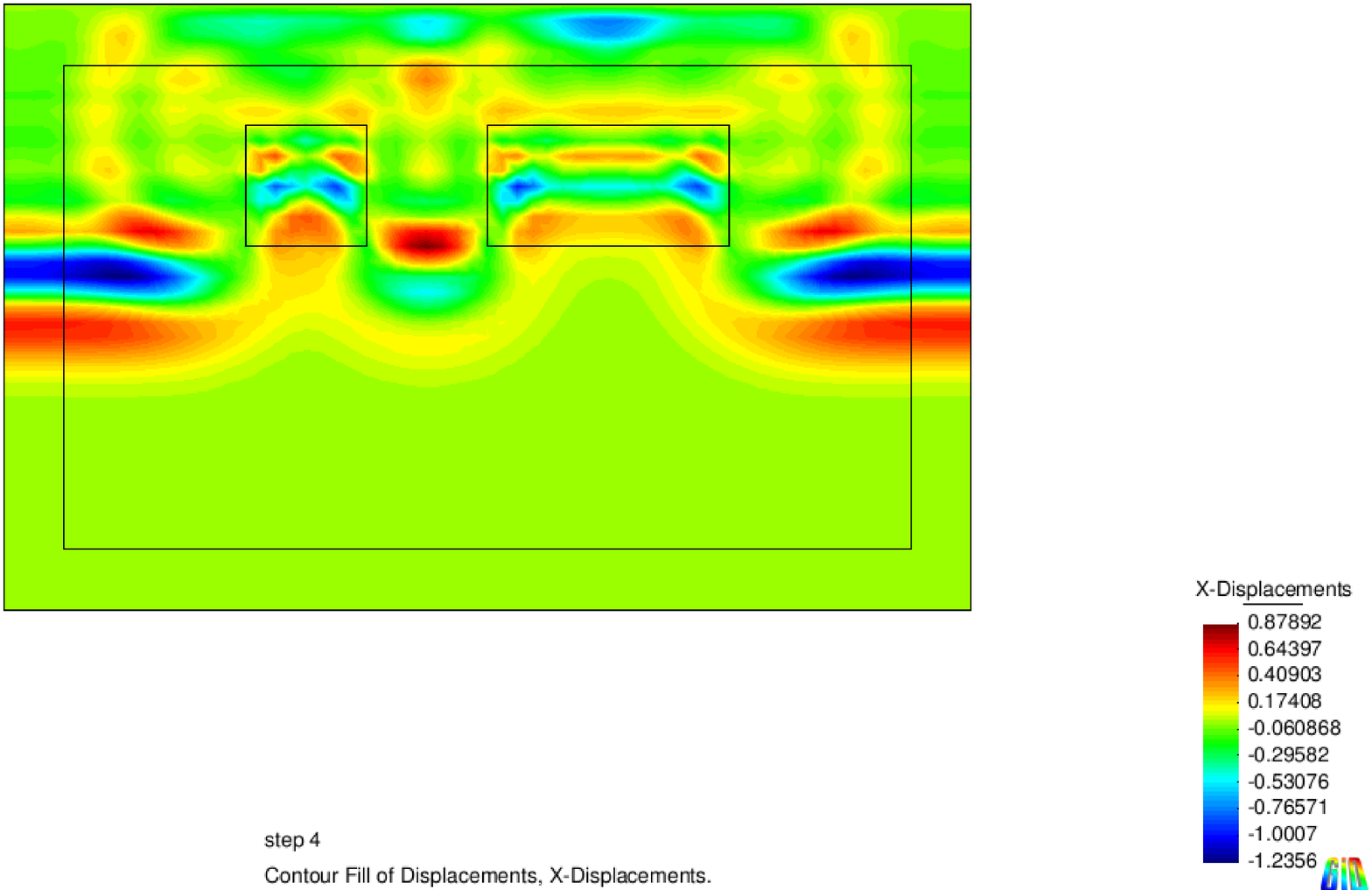}} \\ 
a) t= 3.0 & b) t= 4.0 \\ 
{\includegraphics[scale=0.2,clip=]{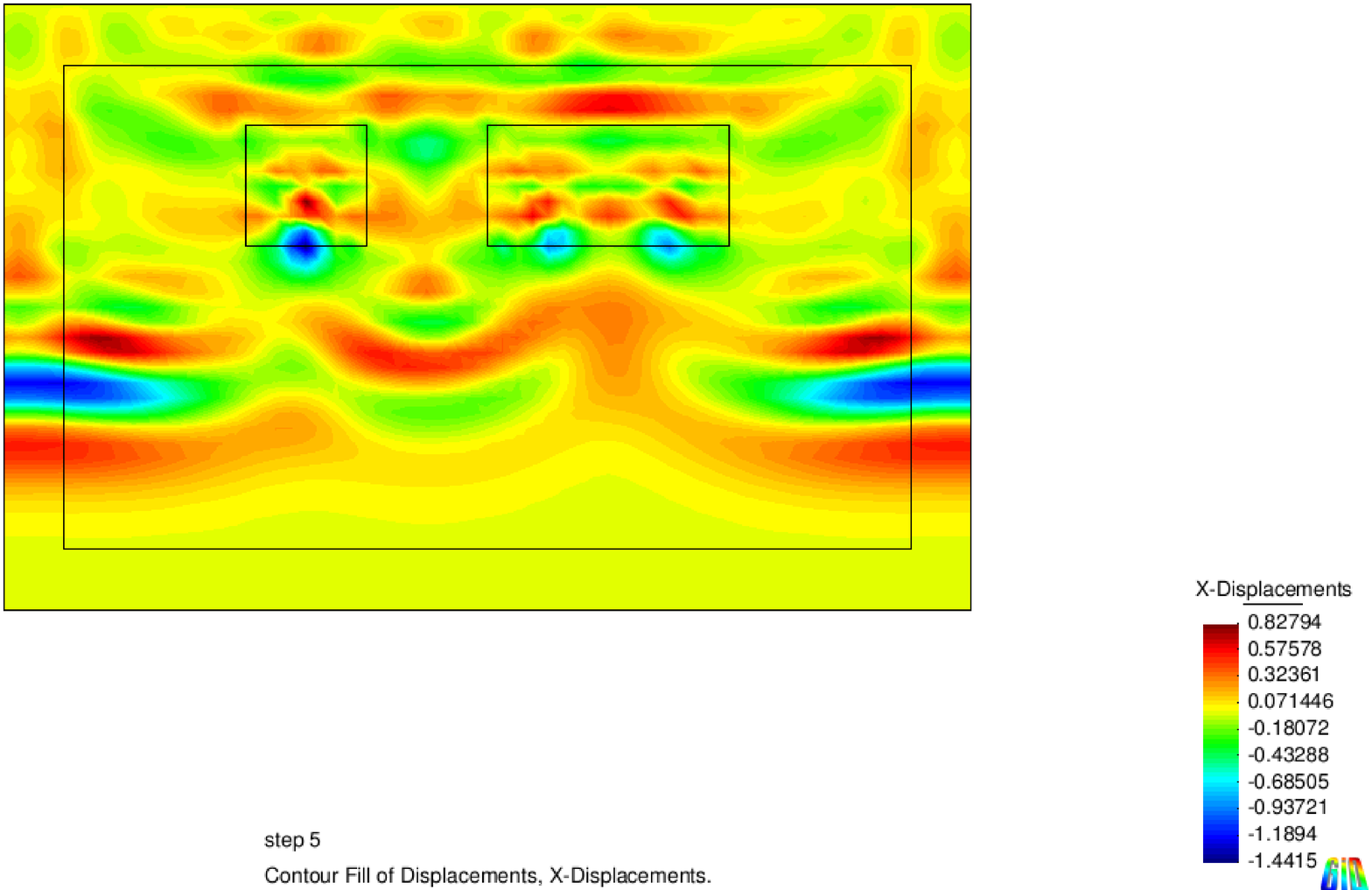}} & {%
\includegraphics[scale=0.2,clip=]{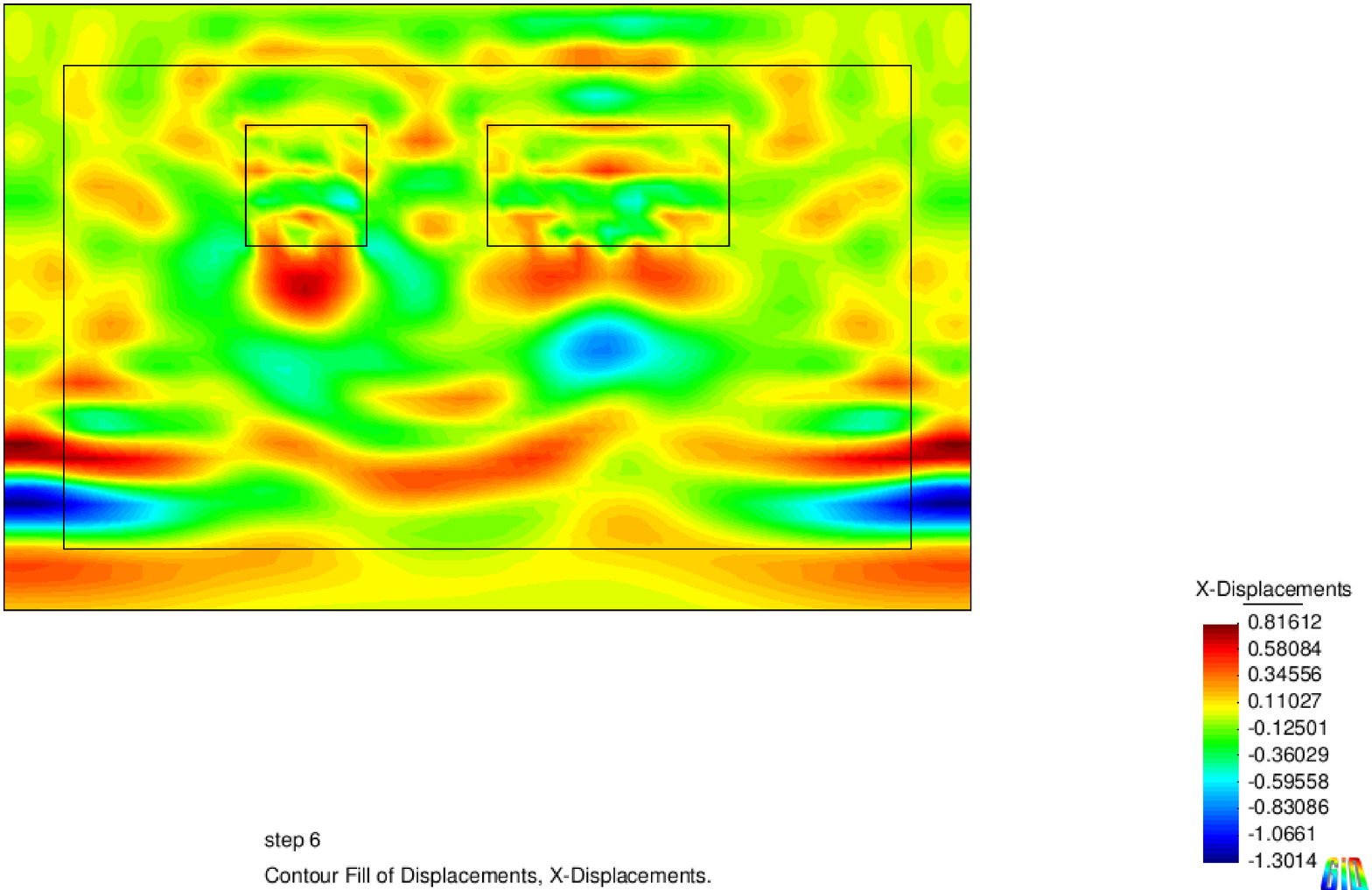}} \\ 
c) t= 5.0 & d) t= 6.0 \\ 
& 
\end{tabular}%
\end{center}
\caption{ \emph{Isosurfaces of the computed solution $u\left( x,t\right) $
of the forward problem (\protect\ref{6.4}), (\protect\ref{6.5}) for the case
of the mine-like targets of Figure \protect\ref{fig:F1} for different times.
One can observe that values of $u\left( x,t\right)$ at the backscattering
(top) side of the boundary are affected quite significantly by the presence
of these targets. Also, values at the bottom side are significantly
affected. However, since this side is located far away from the
backscattering side, then the secondary reflected wave does not provide a
significant impact on the backscattering side, see d). Values at lateral
sides are almost the same as ones for the uniform background with $c\left(
x\right) \equiv 1.$ These observations are the basis for our decision about
the change of boundary conditions, see (\protect\ref{6.6}). }}
\label{fig:F2}
\end{figure}

\begin{figure}[tbp]
\begin{center}
\begin{tabular}{cc}
{\includegraphics[scale=0.4,clip=]{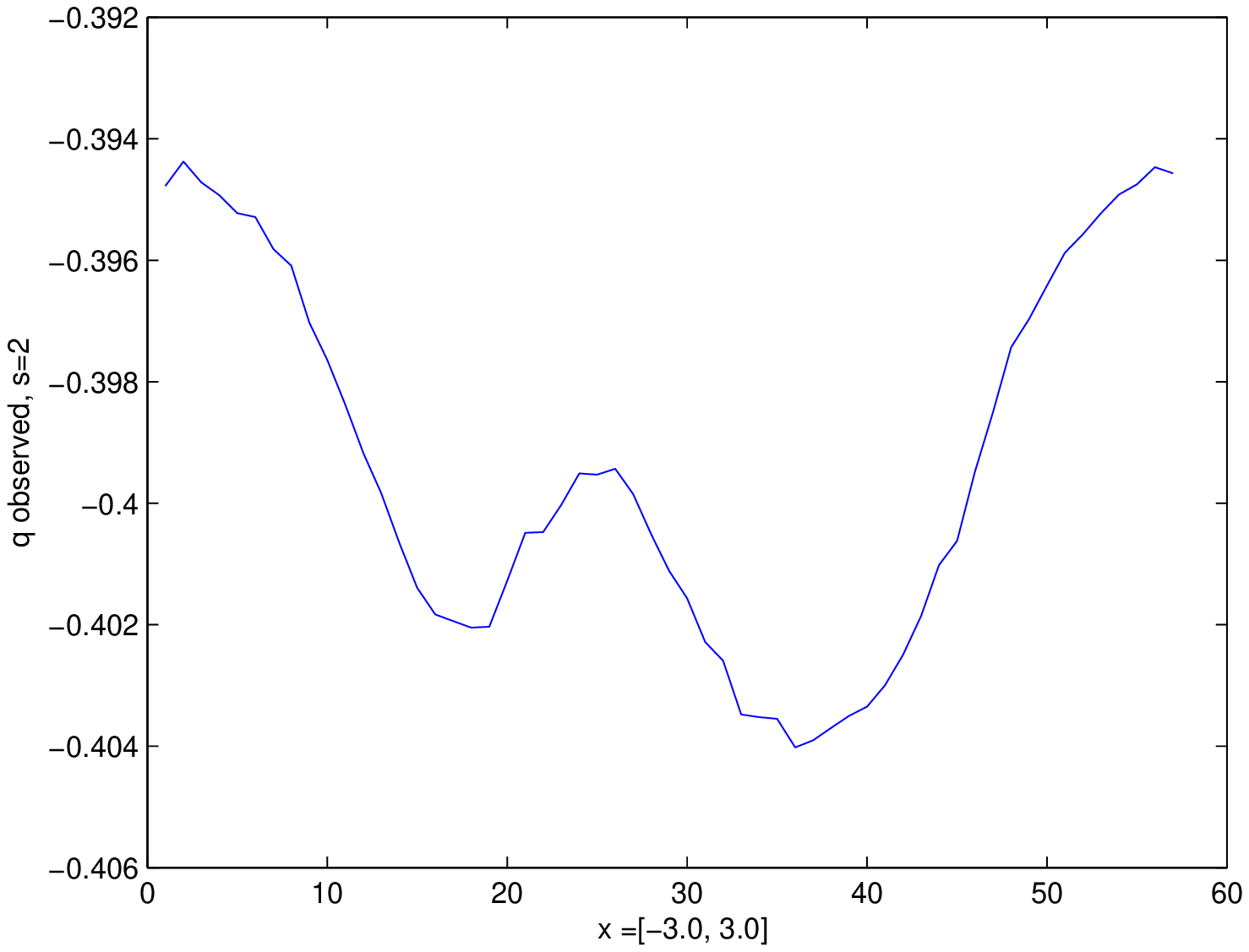}} & {\ %
\includegraphics[scale=0.4,clip=]{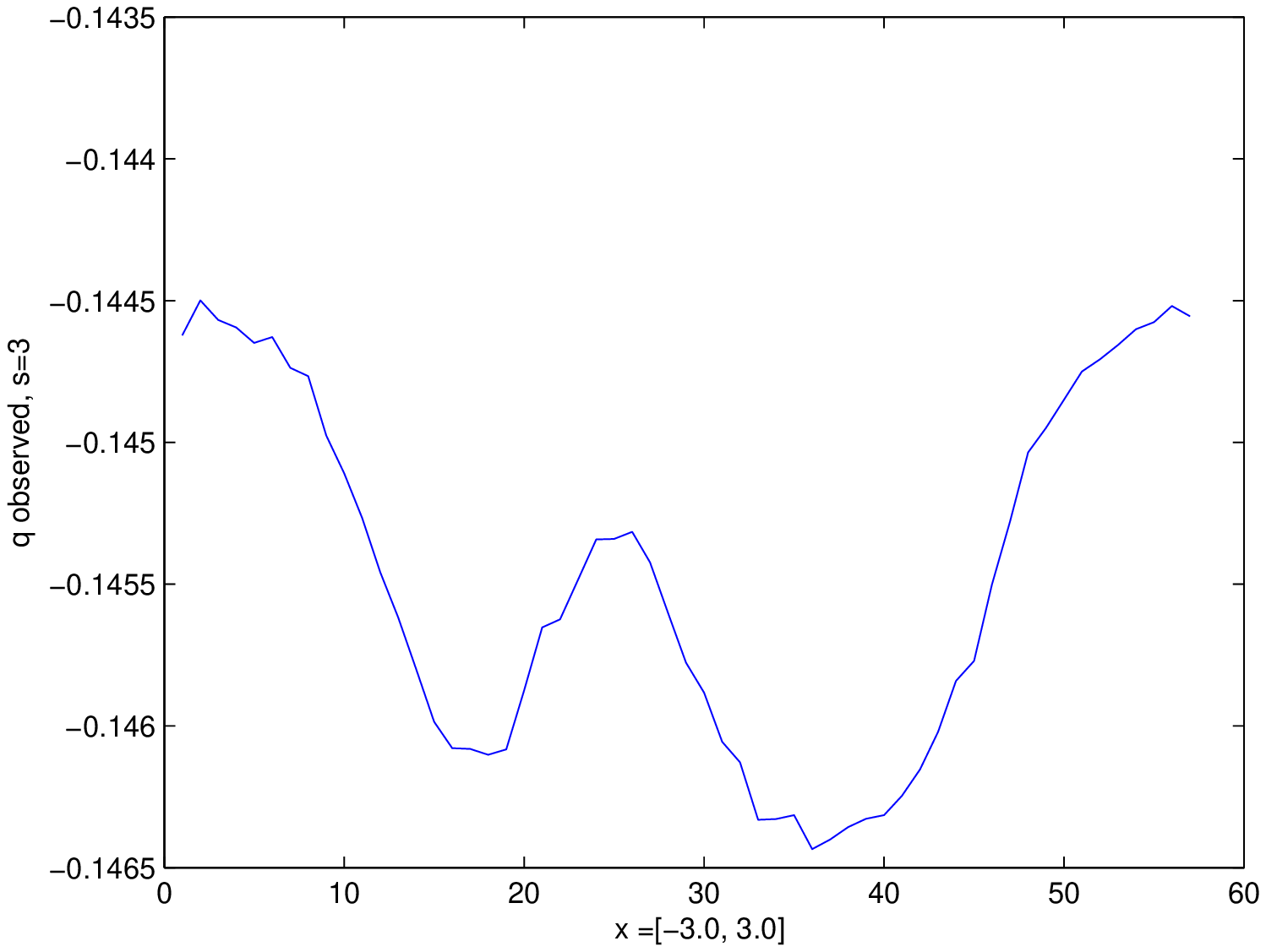} } \\ 
a) $s=2$ & b) $s=3$ \\ 
{\includegraphics[scale=0.4,clip=]{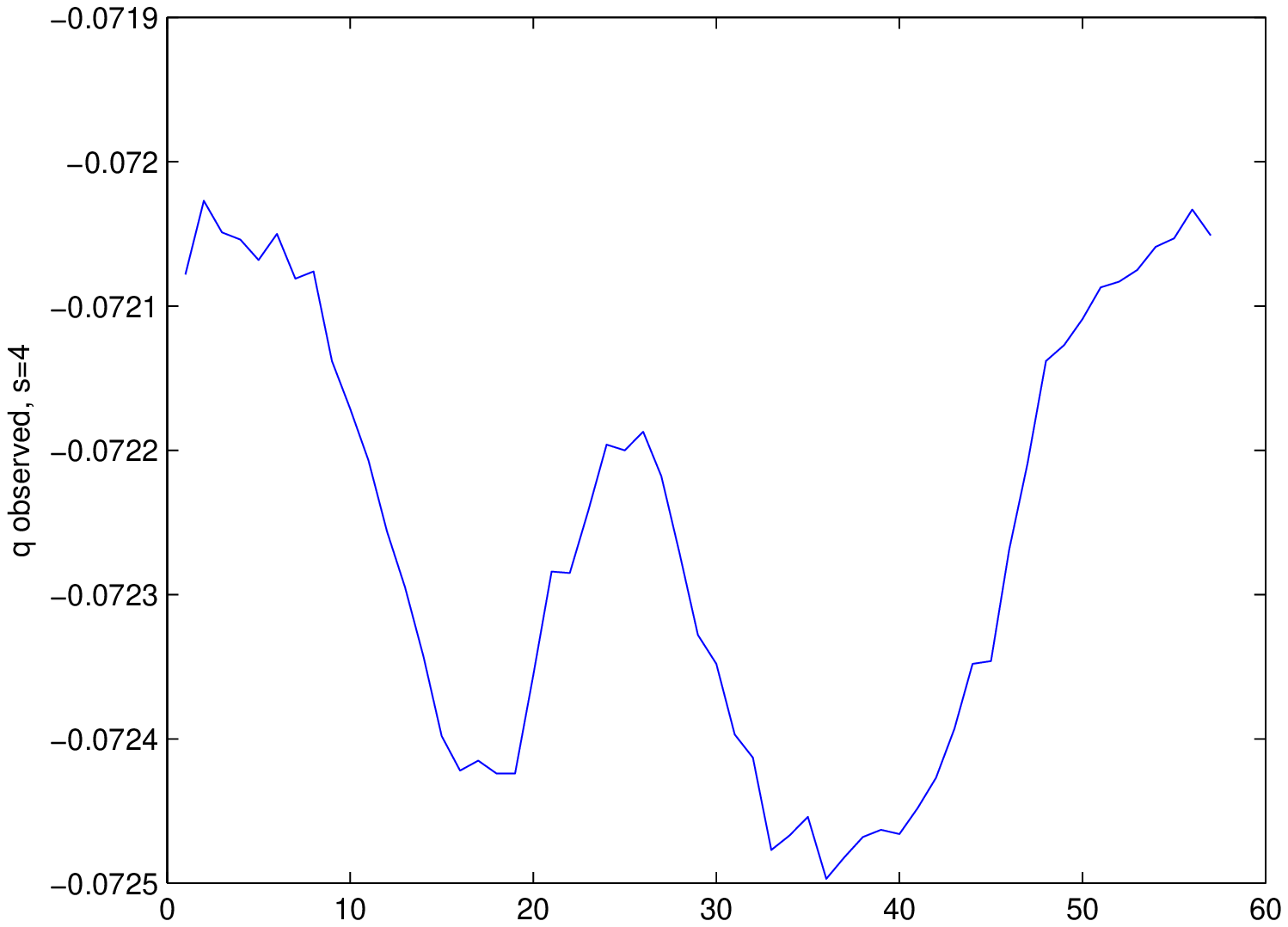}} & {\ %
\includegraphics[scale=0.4,clip=]{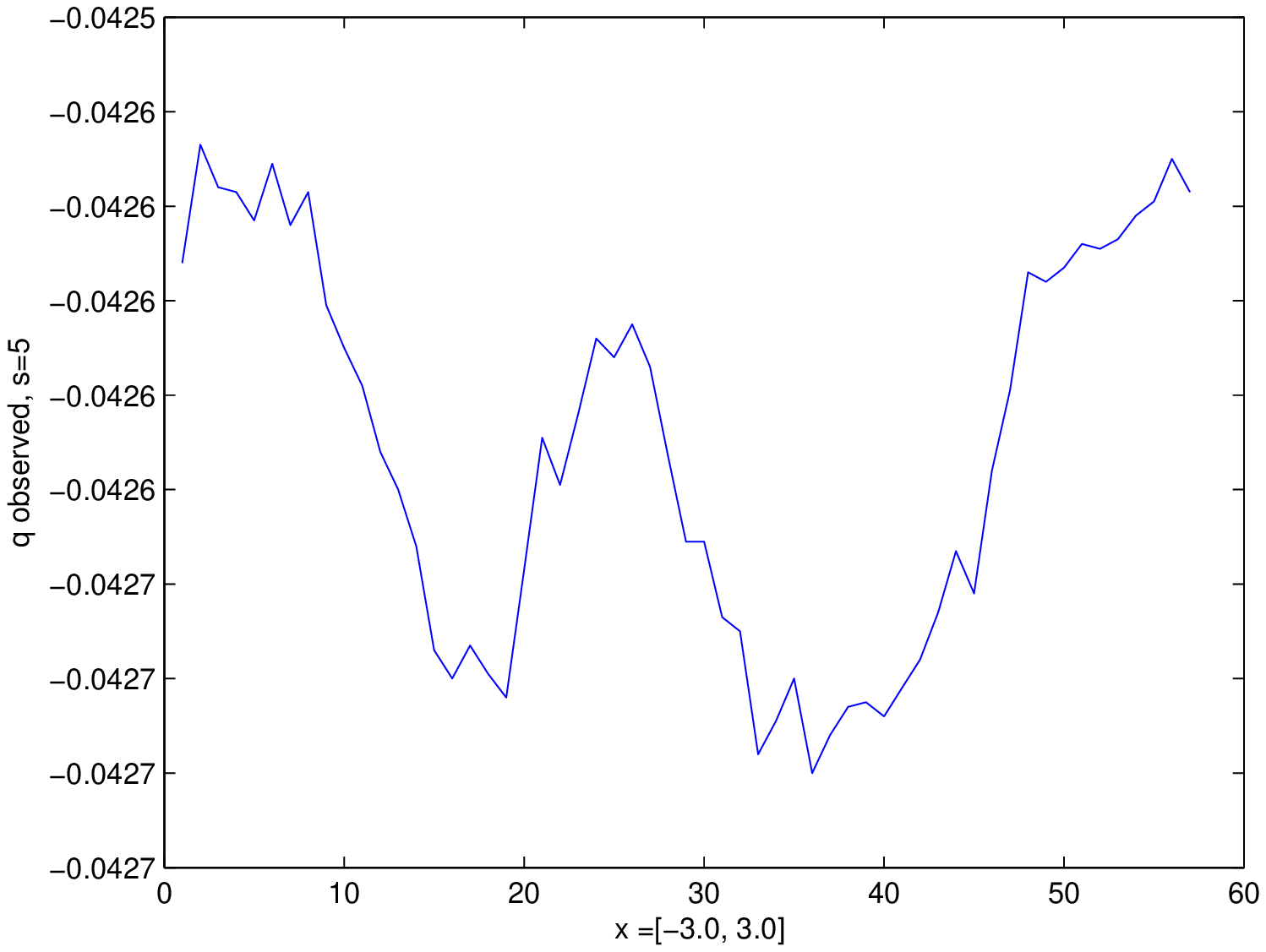} } \\ 
c) $s=4$ & d) $s=5$ \\ 
{\includegraphics[scale=0.4,clip=]{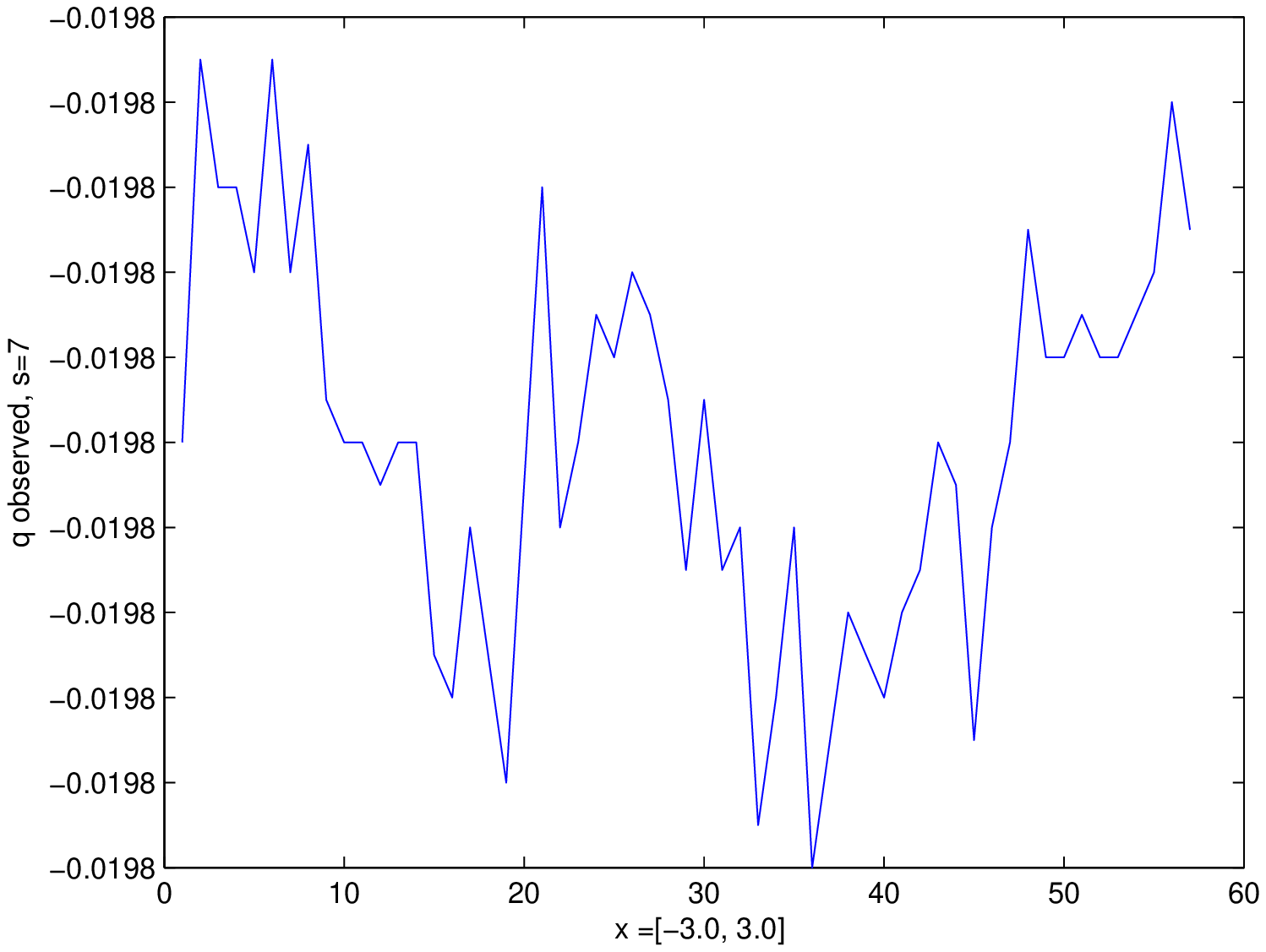}} & {\ %
\includegraphics[scale=0.4,clip=]{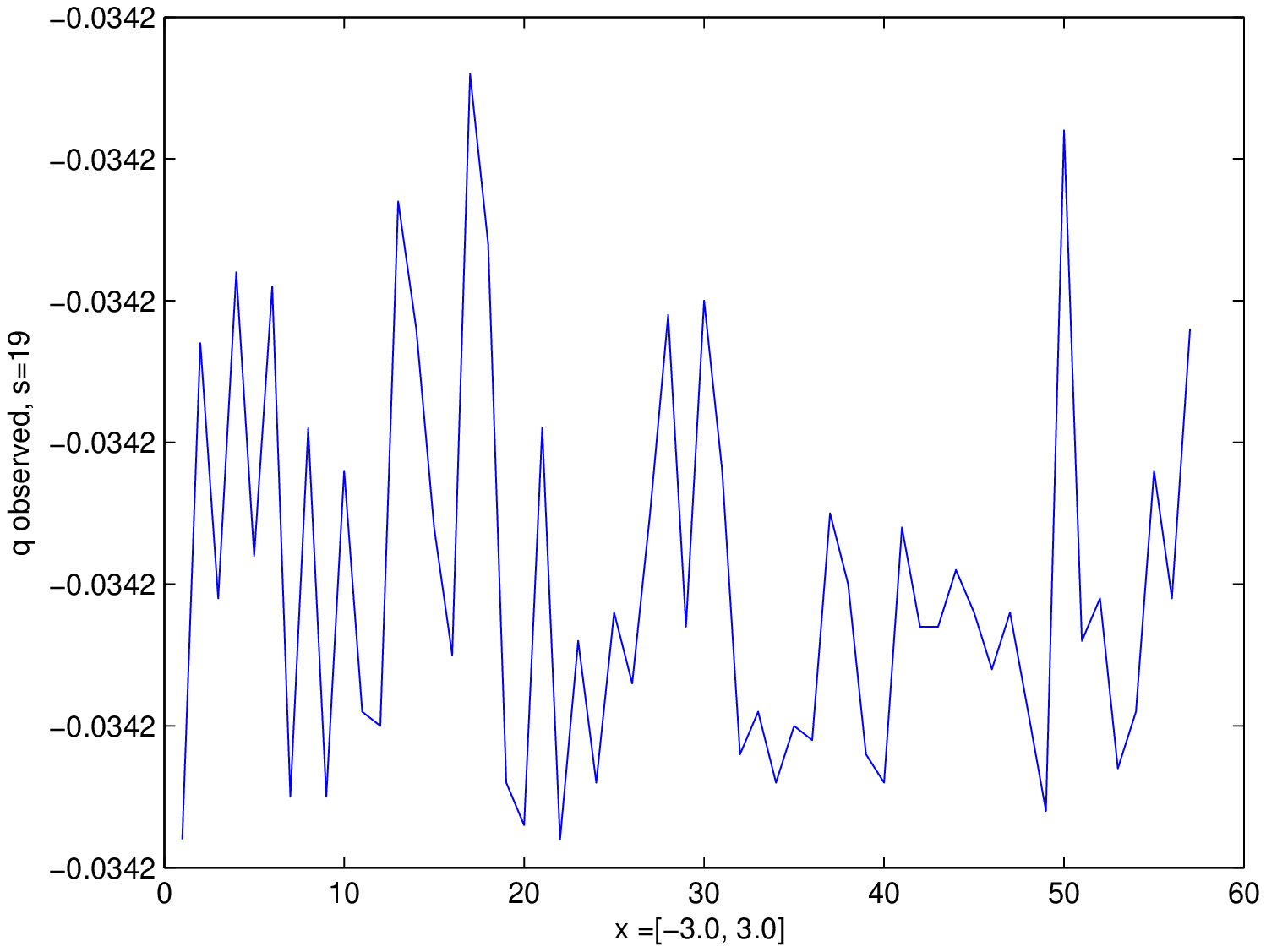} } \\ 
a) $s=7$ & b) $s=19$%
\end{tabular}%
\end{center}
\caption{ \emph{Backscattered data at the top boundary $\Gamma$ of the
function $q(x,s)$ at different values pseudo-frequencies $s$.}}
\label{fig:F3}
\end{figure}

To simulate the data for our CIP in 2d, we solve the forward problem for
equation (\ref{6.1}) for the case of the incident plane wave propagating
along the negative direction of the $x_{2}$ axis. This plane wave is
initialized on the top boundary of the rectangle $G$ of Figure ~\ref{fig:F1}%
. We simulate the data for the inverse problem using the software package
WavES \cite{waves}. To do that we solve the forward problem via the hybrid
FEM/FDM method described in \cite{BSA}. In this method the computational
domain $G$ is split in two subdomains, $G=G_{FDM}\cup G_{FEM},$ where 
\begin{equation*}
G=[-4,4]\times \lbrack -1,4],G_{FEM}:=\Omega =\left( -3.5,3.5\right) \times
\left( -0.5,3.5\right) ,G_{FDM}=G\diagdown G_{FEM},
\end{equation*}%
see Figure ~\ref{fig:F1}. Thus the subdomain $G_{FEM}:=\Omega $ is the same
as in (\ref{6.2}). We use structured mesh and FDM in $G_{FDM}$ and
non-structured mesh and FEM in $G_{FEM}=\Omega .$ The space mesh in $\Omega $
consists of triangles and it consists of squares in $G_{FDM}$, with the mesh
size $\tilde{h}=0.125$ in the overlapping regions. At the top and bottom
boundaries of $G$ we use first-order absorbing boundary conditions. These
conditions are exact in our case since we initialize a plane wave in a
normal direction to the top boundary of $G$. At the lateral boundaries, the
zero Neumann boundary condition is used. Since the incident plane wave
propagates downwards, then the zero Neumann boundary condition allows us to
model an infinite space domain in the lateral direction.

Small square and small rectangle of Figure \ref{fig:F1} are mine-like
targets with $c(x)=4$ inside of them, see (\ref{6.3}). Thus, 
\begin{equation}
c(x)=\left\{ 
\begin{array}{c}
4\text{ in mine-like targets of Figure \ref{fig:F1},} \\ 
1\text{ otherwise.}%
\end{array}%
\right.  \label{6.3_1}
\end{equation}%
When solving the inverse problem, we assume that the coefficient $c(x)$ is
unknown in the rectangle $\Omega \subset G$ and has a known constant value $%
c(x)=1$ in $G\diagdown \Omega ,$ see Figure \ref{fig:F1}. The boundary of
the rectangle $G$ is $\partial G=\partial G_{1}\cup \partial G_{2}\cup
\partial G_{3}.$ Here, $\partial G_{1}$ and $\partial G_{2}$ are
respectively top and bottom sides of the largest rectangle of Figure \ref%
{fig:F1}, and $\partial G_{3}$ is the union of left and right sides of this
rectangle. Let $T$ be the final time for data generation, see the paragraph
after (\ref{3.4}) in subsection 3.1. We generate the data via solution of
the following forward problem 
\begin{equation}
\begin{split}
c\left( x\right) u_{tt}-\Delta u& =0,~~~\mbox{in}~G\times (0,T), \\
u(x,0)& =0,~u_{t}(x,0)=0,~\mbox{in}~G, \\
\partial _{n}u\big \vert_{\partial \Omega _{1}}& =f\left( t\right) ,~%
\mbox{on}~\partial G_{1}\times (0,t_{1}], \\
\partial _{n}u\big \vert_{\partial \Omega _{1}}& =\partial _{t}u,~\mbox{on}%
~\partial G_{1}\times (t_{1},T), \\
\partial _{n}u\big \vert_{\partial G_{2}}& =\partial _{t}u,~\mbox{on}%
~\partial G_{2}\times (0,T), \\
\partial _{n}u\big \vert_{\partial \Omega _{3}}& =0,~\mbox{on}~\partial
G_{3}\times (0,T),
\end{split}
\label{6.4}
\end{equation}

The plane wave with the wave form $f\left( t\right) $ is initialized at the
top boundary $\partial G_{1}$ of the computational domain $G$ during the
time period $t\in (0,2\pi /\omega ]=\left( 0,t_{1}\right] $, propagates
downwards into $G$ and is absorbed at the bottom boundary $\partial G_{2}$
for all times $t\in (0,T).$ In addition, it is also absorbed at the top
boundary $\partial G_{1}$ for times $t\in (2\pi /\omega ,T)$. Here 
\begin{equation}
f\left( t\right) =\left\{ 
\begin{array}{c}
\frac{1}{10}(\sin {(}\omega t{-\pi /2)}+1)\text{ for }t\in (0,\frac{2\pi }{%
\omega }], \\ 
0\text{ for }t\in \left( \frac{2\pi }{\omega },T\right).%
\end{array}%
\right.  \label{6.5}
\end{equation}%
We took $\omega =7$ and $T=6$ in (\ref{6.5}) for 2d tests. To update tails,
we have solved on each iterative step the forward problem (\ref{6.4}). Next,
we have calculated the Laplace transform (\ref{3.5}) to obtain the function $%
w_{n,i}\left( x,\overline{s}\right) ,$ see (\ref{3.12}) and (\ref{3.30}).

The trace $g\left( x,t\right) $ of the solution $u\left( x,t\right) $ of the
forward problem (\ref{6.4}), (\ref{6.5}) is recorded at the top boundary $%
\Gamma $ of the domain $\Omega $ where we solve the inverse problem$,$ see (%
\ref{6.2_1}). This trace generates the Dirichlet boundary data $\psi \left(
x,s\right) ,x\in \Gamma $ in (\ref{3.15}) (after the Laplace transform).
Next, the coefficient $c(x)$ is \textquotedblleft
forgotten\textquotedblright , and our goal is to reconstruct this
coefficient for $x\in \Omega $ from the data $\psi \left( x,s\right) .$

\subsection{Data simulation in 3d}

\label{sec:6.4}

\begin{figure}[tbp]
\begin{center}
\begin{tabular}{ccc}
{\includegraphics[scale=0.4,clip=]{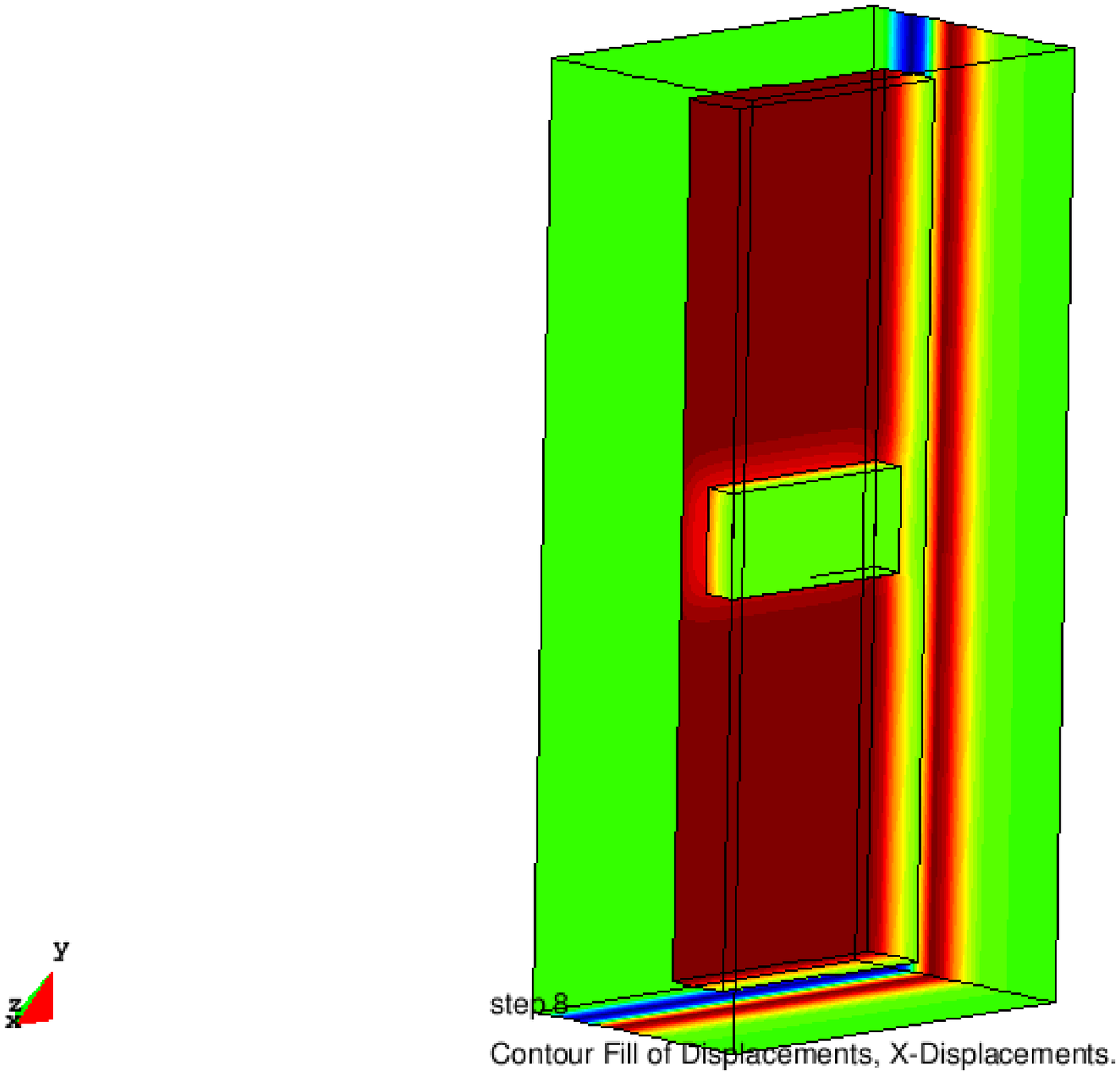}} & {%
\includegraphics[scale=0.4,clip=]{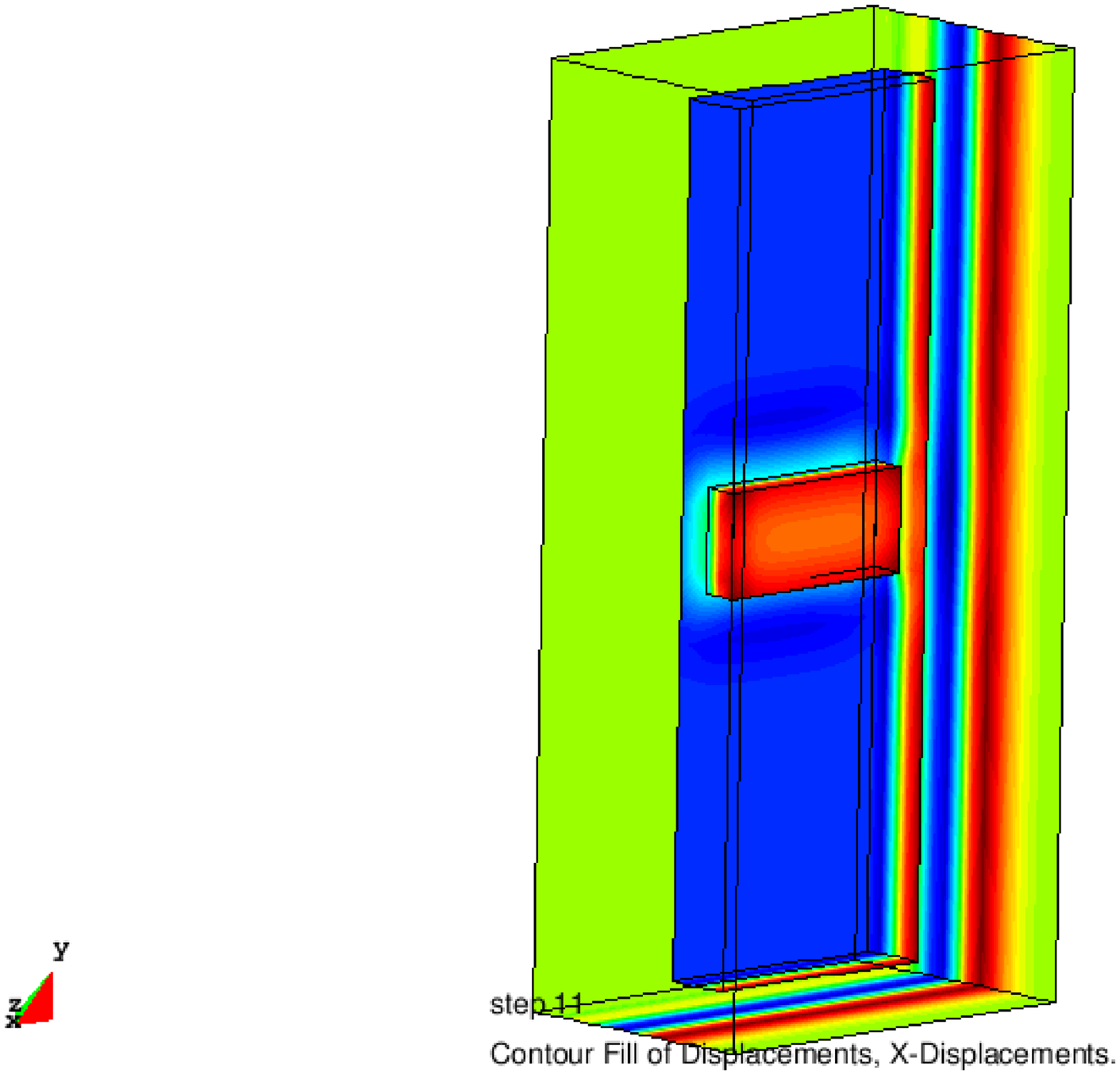}} & {%
\includegraphics[scale=0.4,clip=]{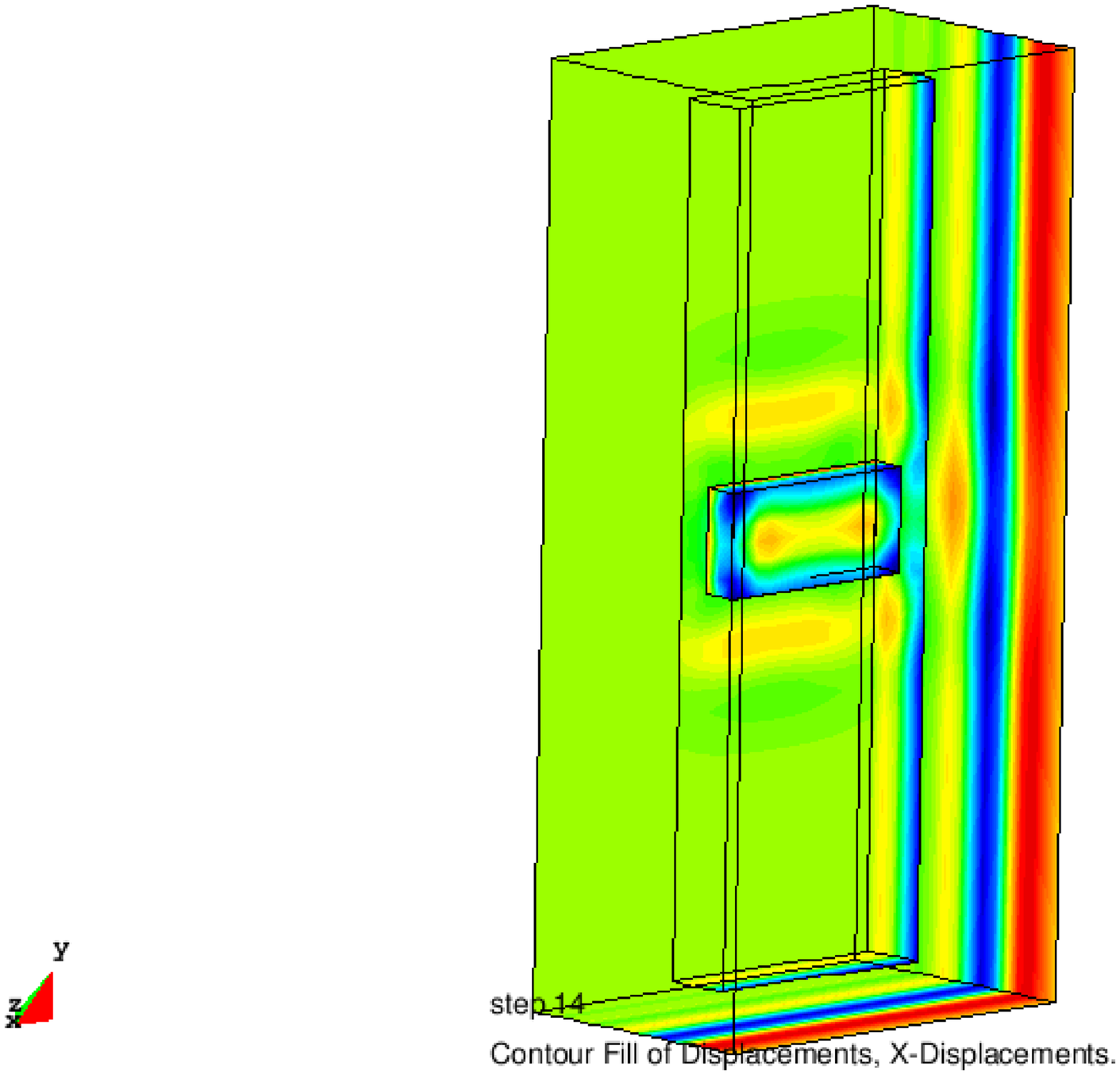}} \\ 
a) $t=0.4$ & b) $t=0.55$ & c) $t=0.7$ \\ 
{\includegraphics[scale=0.4,clip=]{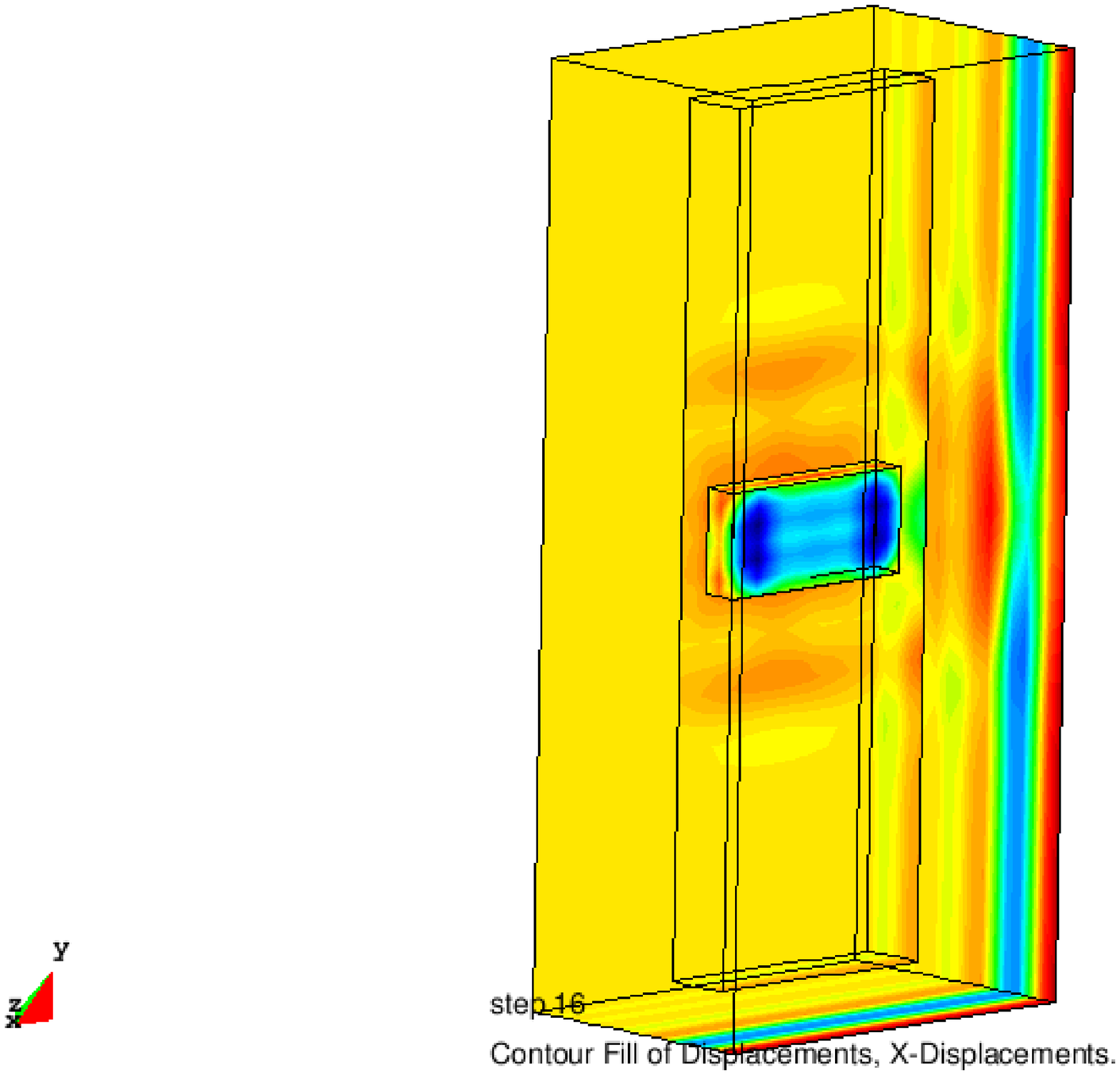}} & {%
\includegraphics[scale=0.4,clip=]{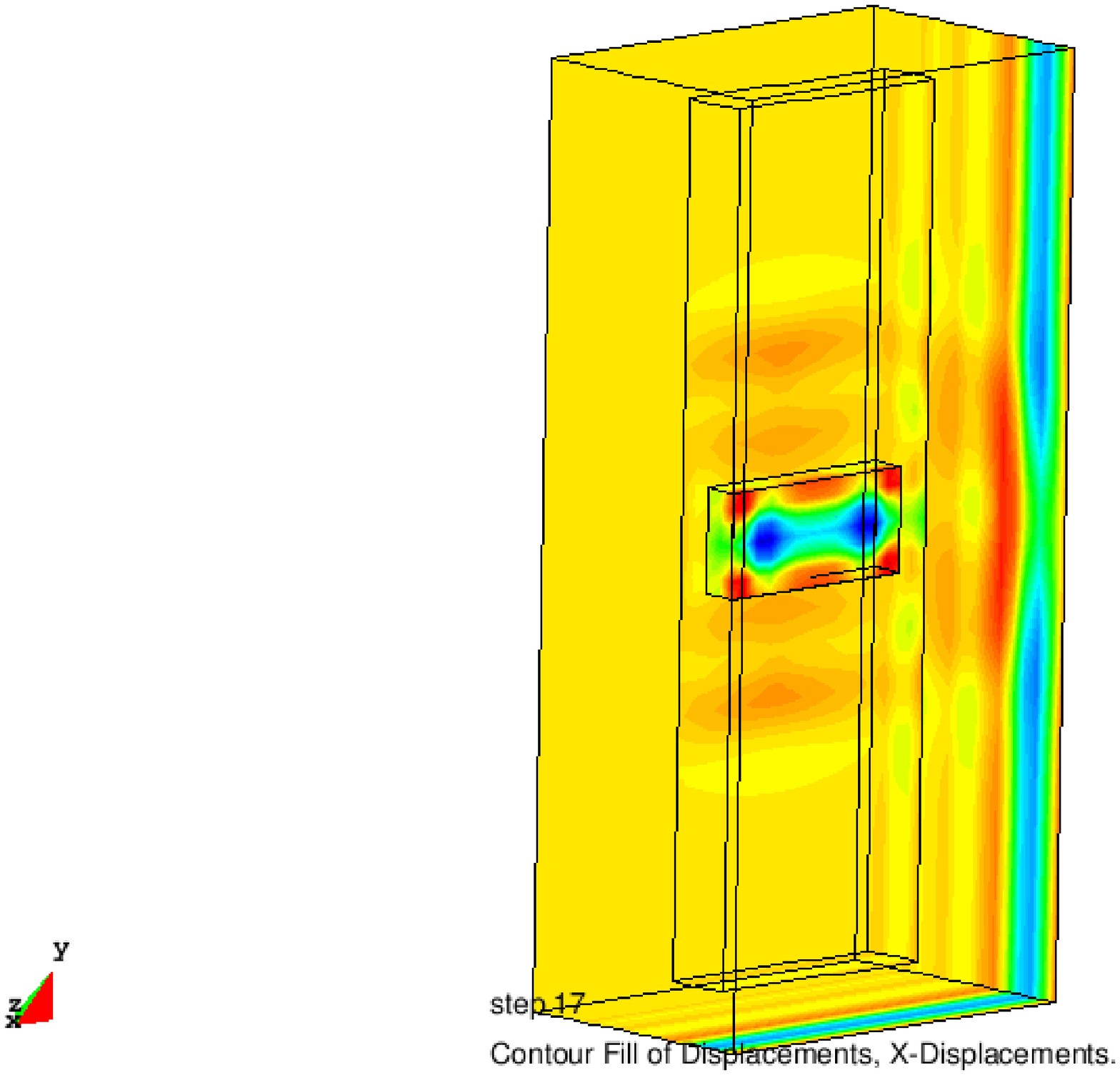}} & {%
\includegraphics[scale=0.4,clip=]{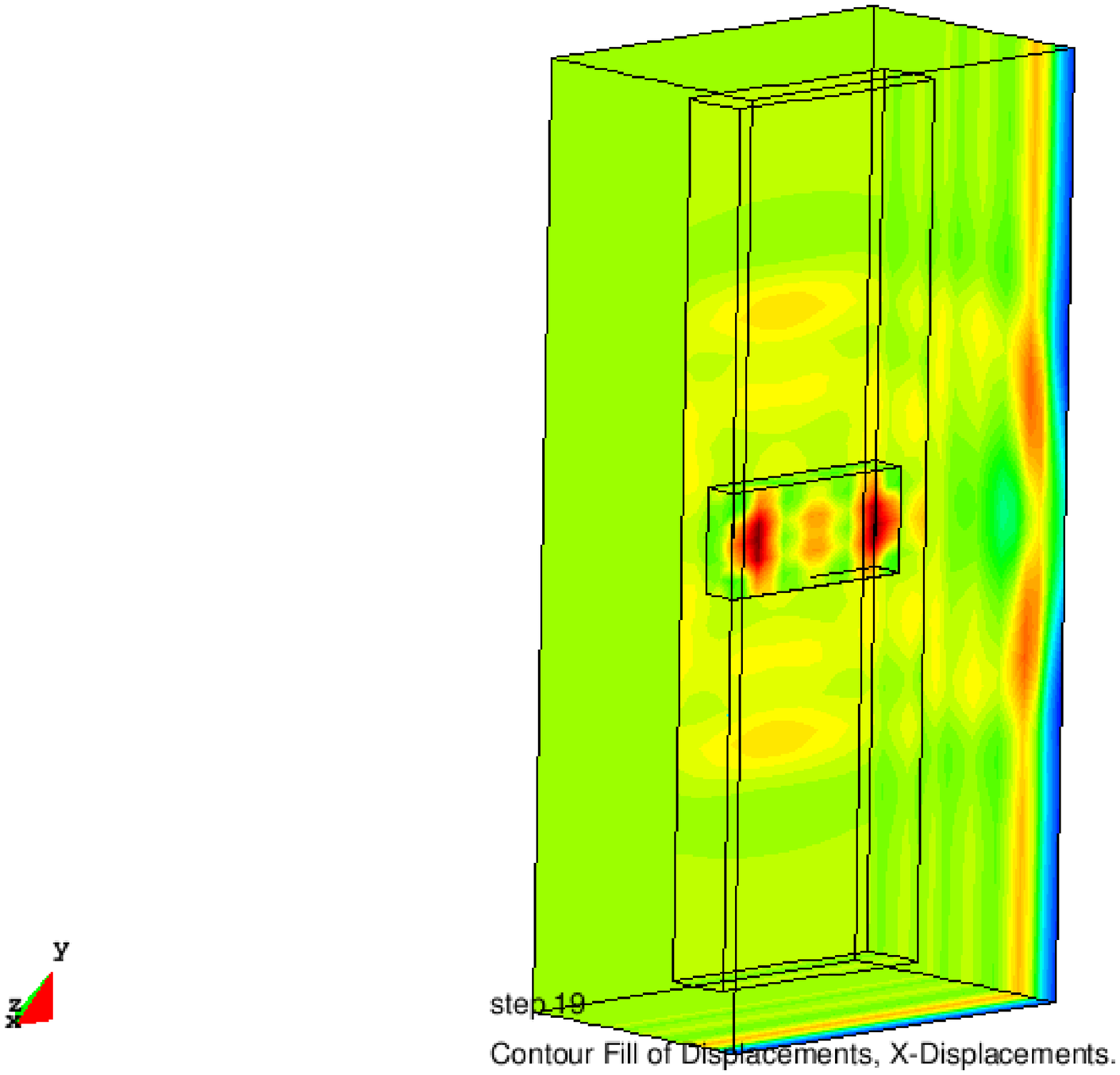}} \\ 
a) $t=0.8$ & b) $t=0.85$ & c) $t=0.95$ \\ 
&  & 
\end{tabular}%
\end{center}
\caption{{\protect\small \emph{Isosurfaces of the computed solution $u(x,t)$
of the forward problem (\protect\ref{6.4}), (\protect\ref{6.5}) at different
times $t$ with the plane wave initialized at the front boundary of $G$ on
the mesh with the mesh size $h=0.04$. Test was performed in time $t=[0,1]$
with time step $\protect\tau = 0.001$. }}}
\label{fig:F3D_2}
\end{figure}

In this case domains $G$ and $\Omega $ are those of (\ref{6.23}) and (\ref%
{6.20}) respectively. Since the human body consists mostly of water, and the
dielectric constant of water is about 80 \cite{Tables}, we set in Test 5
below 
\begin{equation}
c\left( x\right) =\left\{ 
\begin{array}{c}
80,x\in \Omega \diagdown \Omega _{belt}, \\ 
1,x\in G\diagdown \Omega , \\ 
3.2,x\in \Omega _{belt}.%
\end{array}%
\right.  \label{6.29}
\end{equation}%
Hence, (\ref{6.29}) is a quite heterogeneous and, therefore, a very
complicated case. Because of this, we start from a simpler problem in our
Tests 3,4 via choosing%
\begin{equation}
c\left( x\right) =\left\{ 
\begin{array}{c}
3.2,x\in \Omega _{belt}, \\ 
1,x\in G\diagdown \Omega _{belt}.%
\end{array}%
\right.  \label{6.30}
\end{equation}
As to the subdomain $\Omega _{belt}\subset \Omega ,$ we assume that it is
filled with an improvised explosive device (IED). Analyzing dielectric
constants of some materials which might form IEDs \cite{Dan}, we came to the
conclusion that we can take $c\left( x\right) =3.2$ to model an IED. This
value of the dielectric constant is close to RDX
Hexahydro-1,3,5-trinitro-1,3,5-triazine \cite{Dan}. Given notations (\ref%
{6.20})-(\ref{6.28}), (\ref{6.29}), (\ref{6.30}), we have simulated the data
via solving the forward problem (\ref{6.4}), (\ref{6.5}). We have used the
mesh step size $\widetilde{h}=0.04$ in $G$.

In our 3d tests we took $\omega =21$ and $T=1$ in (\ref{6.5}). To generate
backscattered data we solve the forward problem (\ref{6.4}), (\ref{6.5}) in
time $t=[0,1]$ with the time step $\tau=0.001$ using the software package
WavES \cite{waves}. Figure \ref{fig:F3D_2} shows isosurfaces of the computed
solution $u(x,t)$ of the forward problem (\ref{6.4}), (\ref{6.5}) for
different times $t\in \left( 0,1\right) $ for the case when the belt with
explosives was as the one on Figure \ref{fig:F3D_1}. The trace $g\left(
x,t\right) $ of the solution $u\left( x,t\right) $ of the forward problem (%
\ref{6.4}), (\ref{6.5}) is recorded at the front boundary $\Gamma $ of the
domain $\Omega ,$ which is the backscattering side of $\Omega ,$ see (\ref%
{6.25_1}) for $\Gamma $. Again, this trace generates the Dirichlet boundary
data $\psi \left( x,s\right) ,x\in \Gamma $ in (\ref{3.15}) (after the
Laplace transform). Next, the coefficient $c(x)$ is \textquotedblleft
forgotten\textquotedblright , and our goal is to reconstruct this
coefficient for $x\in \Omega$ from the data $\psi \left( x,s\right).$

\subsection{ Boundary conditions on $\partial \Omega \diagdown \Gamma $ and
the choice of the $s-$interval in 2d}

\label{sec:6.5}

Although the above theory requires the knowledge of the function $u\left(
x,t\right) :=g\left( x,t\right) $ at the entire boundary $\partial \Omega ,$
the backscattering data are given only on the top part $\Gamma $ of the
rectangle $\Omega .$ To see how we can complement these data, we analyze the
time dependent behavior of the function $u\left( x,t\right) ,$ which is
calculated as the solution of the problem (\ref{6.4}), (\ref{6.5}). Figure %
\ref{fig:F2} displays this function for different times $t\in \left(
0,6\right) $ for all $x\in \Omega $ for the case of two mine-like targets of
Figure \ref{fig:F1} with $c\left( x\right) =4$ inside of them and $c\left(
x\right) =1$ everywhere else, see (\ref{6.3}). We see that values of $%
u\left( x,t\right) $ for $x\in G$ are substantially affected by the presence
of these inclusions. On the other hand, values at lateral sides of the
rectangle $\Omega $ are affected insignificantly. Values at the lower part
of the boundary $\partial \Omega $ are also significantly affected by the
presence of those targets. On the other hand, that lower part of $\partial
\Omega $ is located rather far away from the top part of $\partial \Omega .$
This means that waves reflected from the lower part reach the to part of $%
\partial \Omega $ at larger times $t>8.$ On the other hand, the Laplace
transform (\ref{3.5}) actually discounts values of the function $u\left(
x,t\right) $ for large $t$: because of the rapid decay of the kernel $%
e^{-st}.$

These observations provide a numerical justification for assigning the
following boundary condition at $\partial \Omega $ 
\begin{equation}
w\left( x,s\right) \mid _{\partial \Omega }=\left\{ 
\begin{array}{c}
w_{calc}\left( x,s\right) ,x\in \Gamma , \\ 
w_{unif}\left( x,s\right) ,x\in \partial \Omega \diagdown \Gamma .%
\end{array}%
\right.  \label{6.6}
\end{equation}%
Here $w_{calc}\left( x,s\right) $ is the function $w\left( x,s\right) $
which is calculated as the Laplace transform (\ref{3.5}) of the solution of
the forward problem (\ref{6.4}), (\ref{6.5}). On the other hand $%
w_{unif}\left( x,s\right) $ is the the Laplace transform of the solution of
this problem for the uniform medium with $c\left( x\right) \equiv 1$.

Consider now the function $,x\in \Omega .$ Figure \ref{fig:F3} displays
graphs of the function $q\left( x,s\right) $ along the top boundary of the
rectangle $\Omega $ for different values of the pseudo frequency $s$. One
can observe that for $s=2,3,4,5,7$ each graph has dents. The locations of
these dents exactly correspond to projections of two targets of Figure \ref%
{fig:F1} on the top boundary of $\Omega $. Therefore, values of the function 
$q\left( x,s\right) ,x\in \Gamma $ carry an information about the horizontal
coordinate of this inclusion. However, figuring out vertical coordinates of
targets is a more difficult task. To do this, one needs to apply the above
algorithm. We also observe that values of $\left\vert q\left( x,s\right)
\right\vert $ for $s\geq 7$ are much lower than those for $s\leq 5.$
Therefore, to solve the inverse problem in 2d, we have chosen the $s-$%
interval as 
\begin{equation}
s\in \left[ 2,3\right] ,h=0.05.  \label{6.7}
\end{equation}

\subsection{3d case: boundary conditions on $\partial \Omega \diagdown
\Gamma $ and the choice of the $s-$interval}

\label{sec:6.6}

\begin{figure}[tbp]
\begin{center}
\begin{tabular}{cc}
{\includegraphics[scale=0.4,clip=]{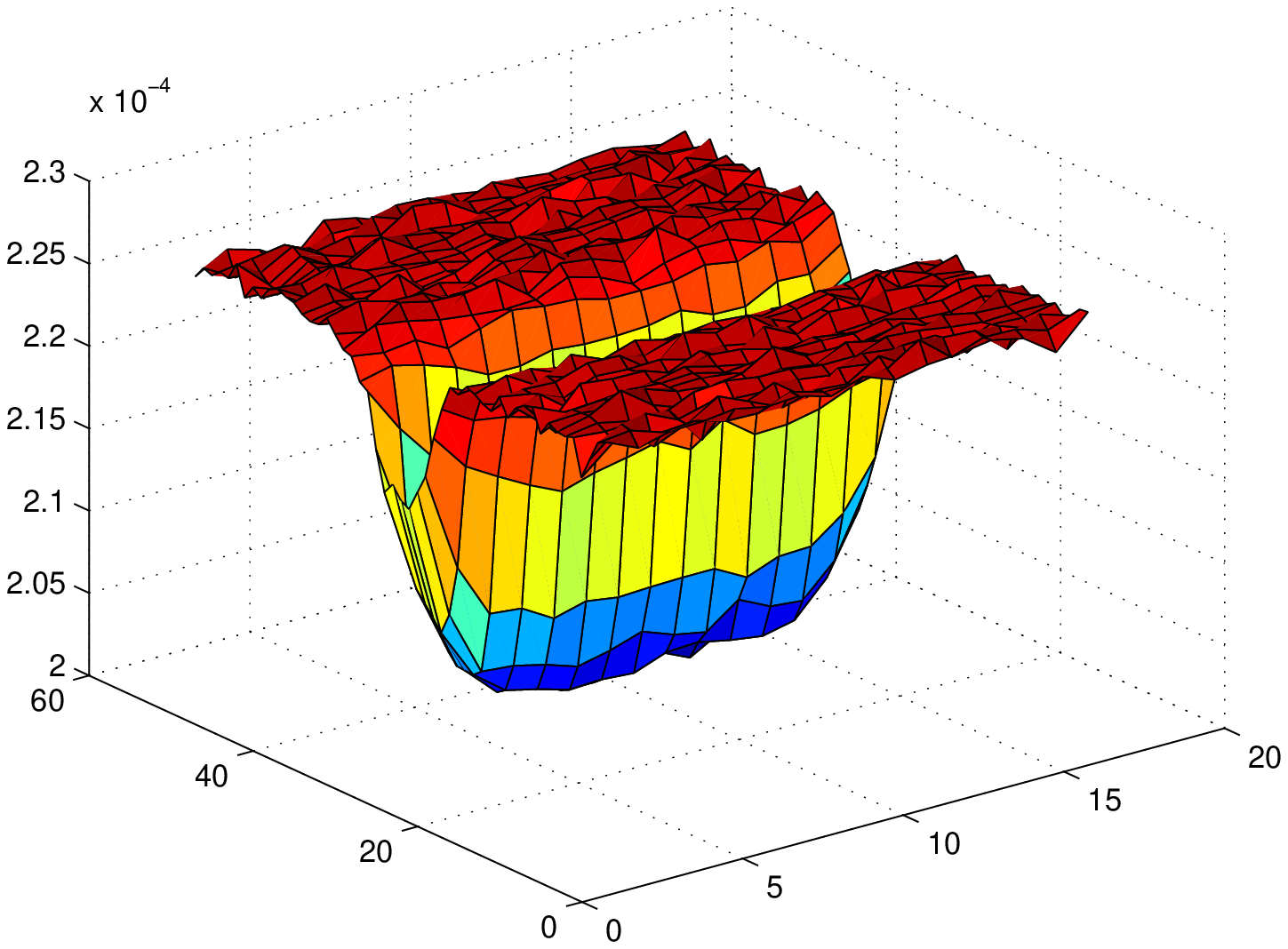}} & {\ %
\includegraphics[scale=0.4,clip=]{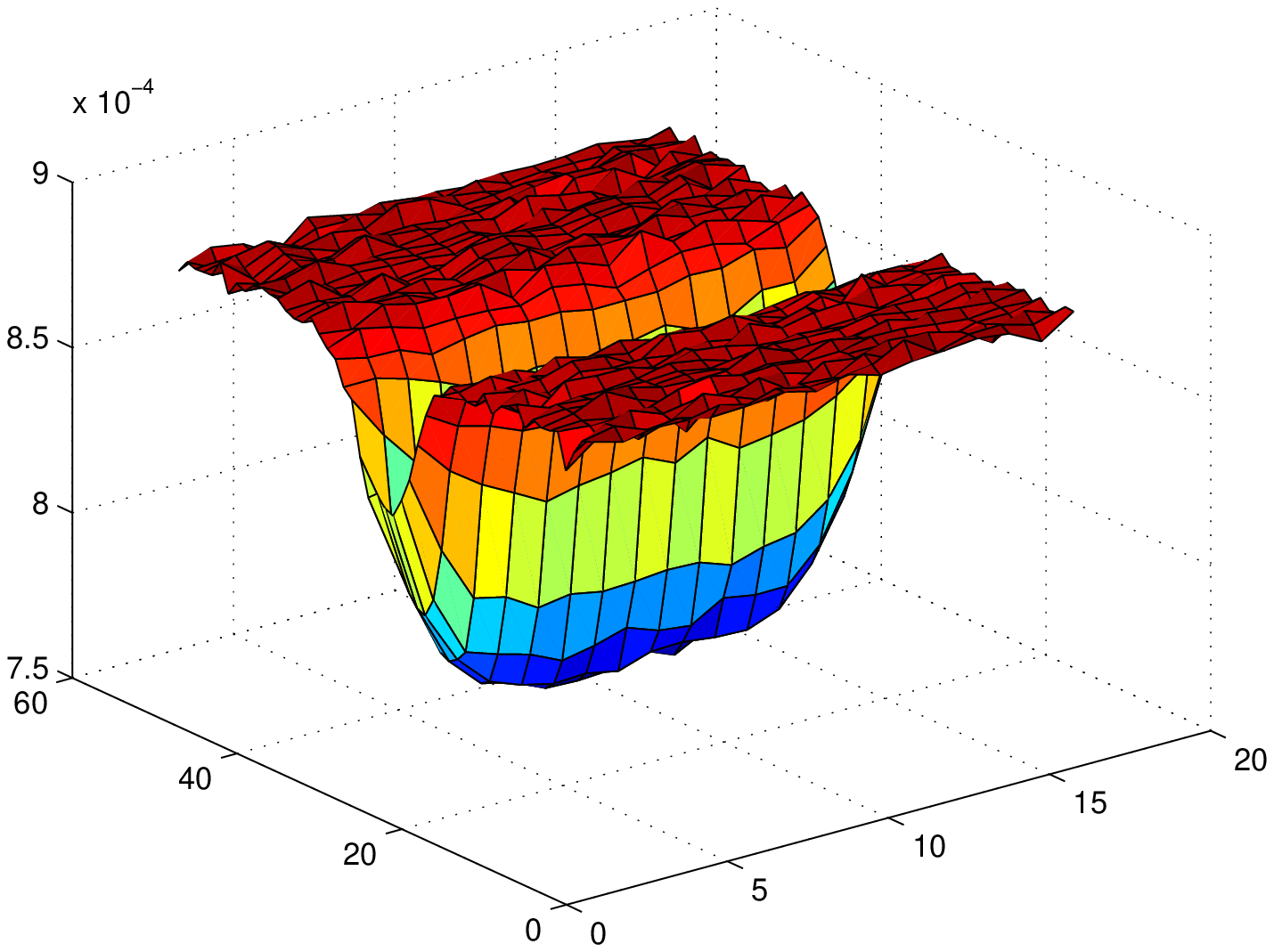} } \\ 
a) $s=20$ & b) $s=15$ \\ 
{\includegraphics[scale=0.4,clip=]{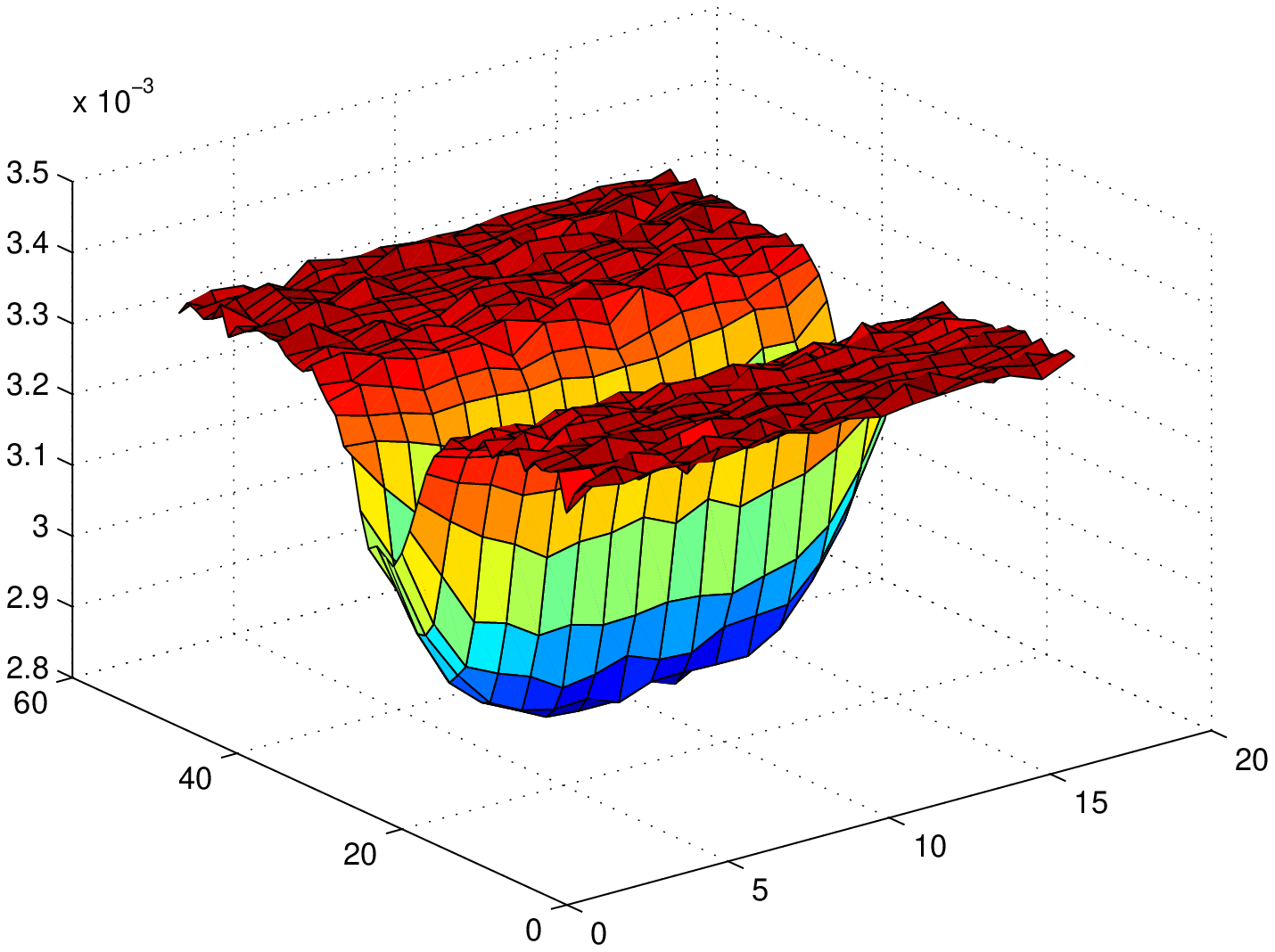}} & {\ %
\includegraphics[scale=0.4,clip=]{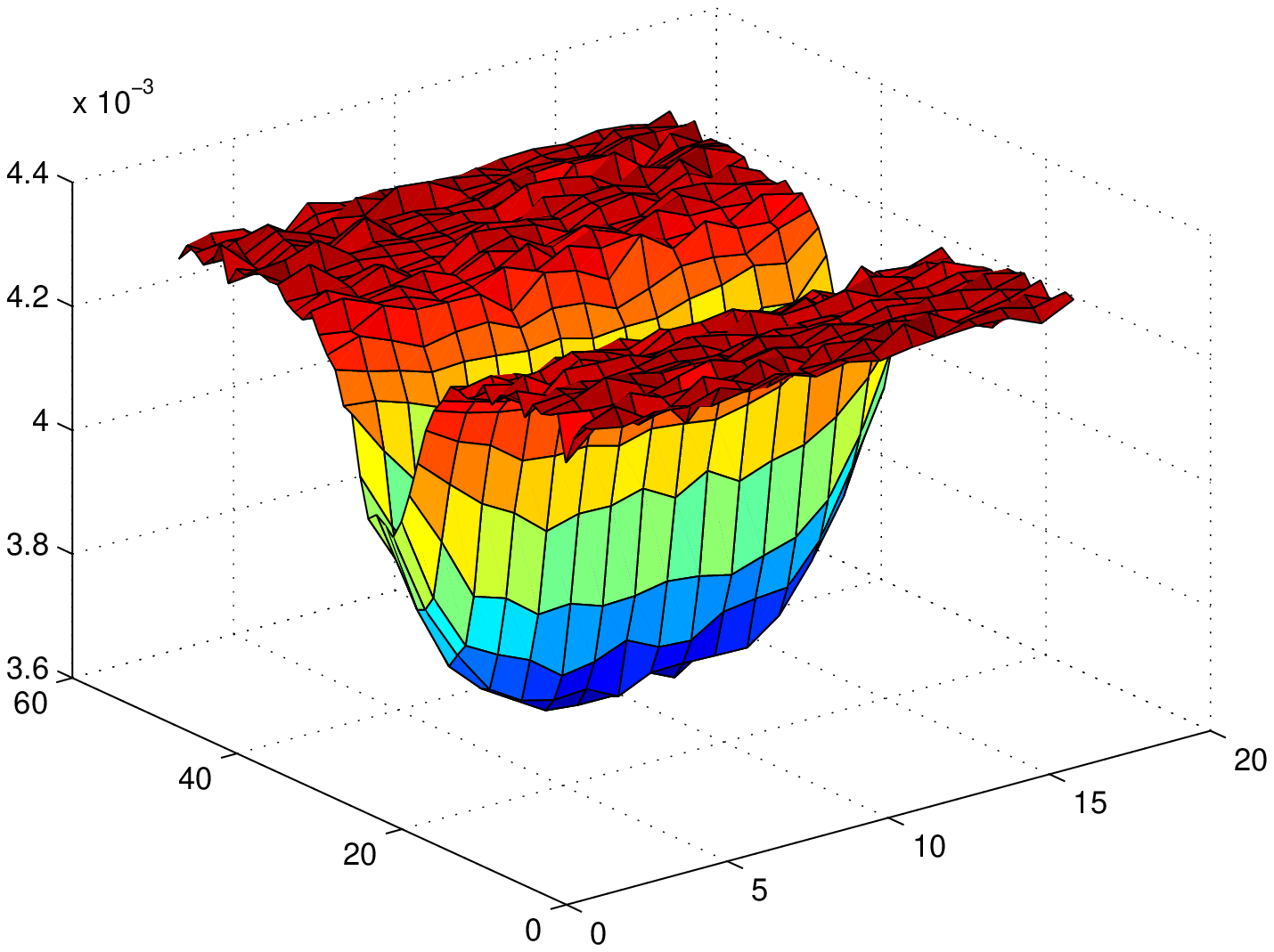} } \\ 
c) $s=10$ & d) $s=9$ \\ 
{\includegraphics[scale=0.4,clip=]{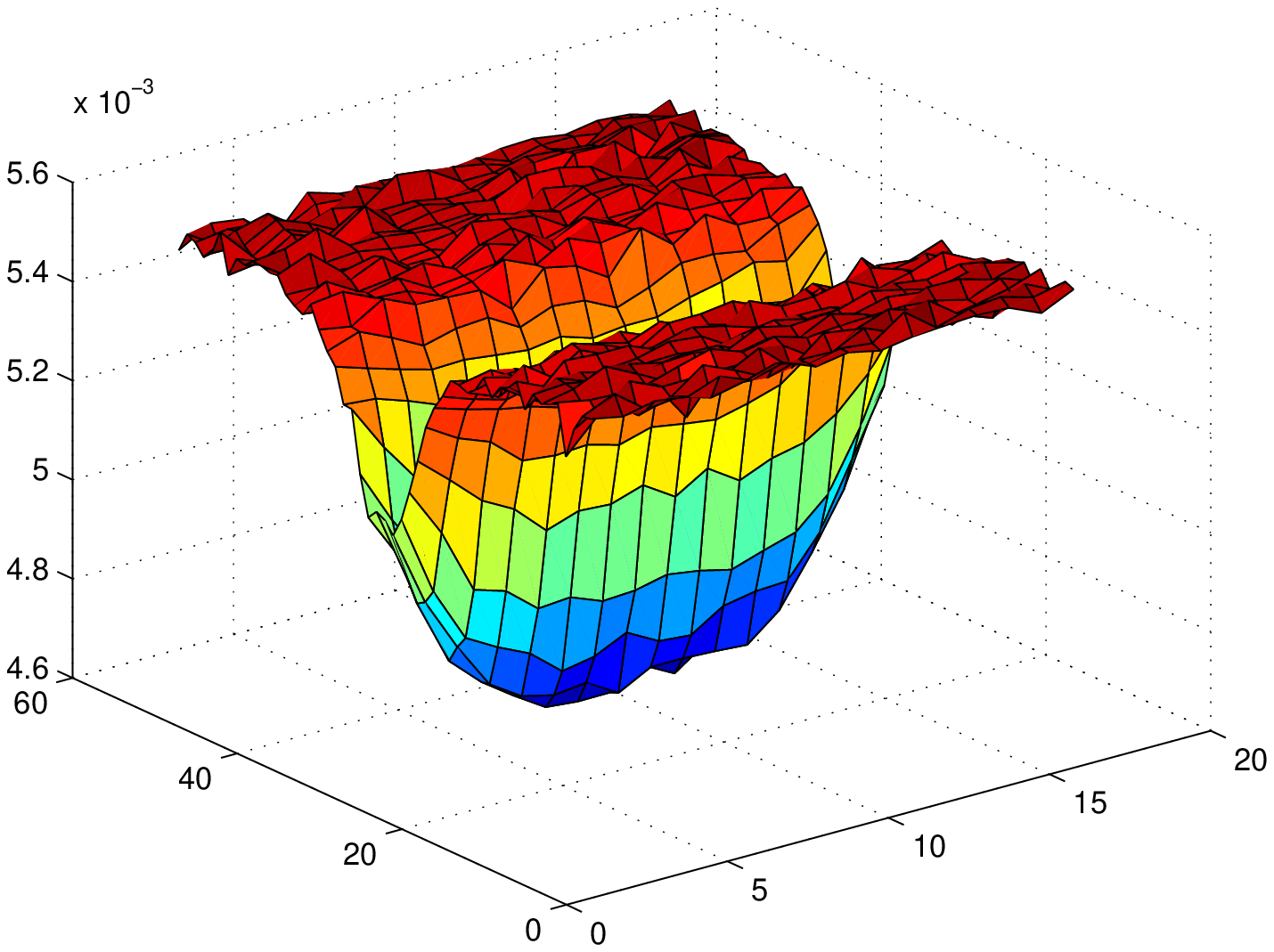}} & {\ %
\includegraphics[scale=0.4,clip=]{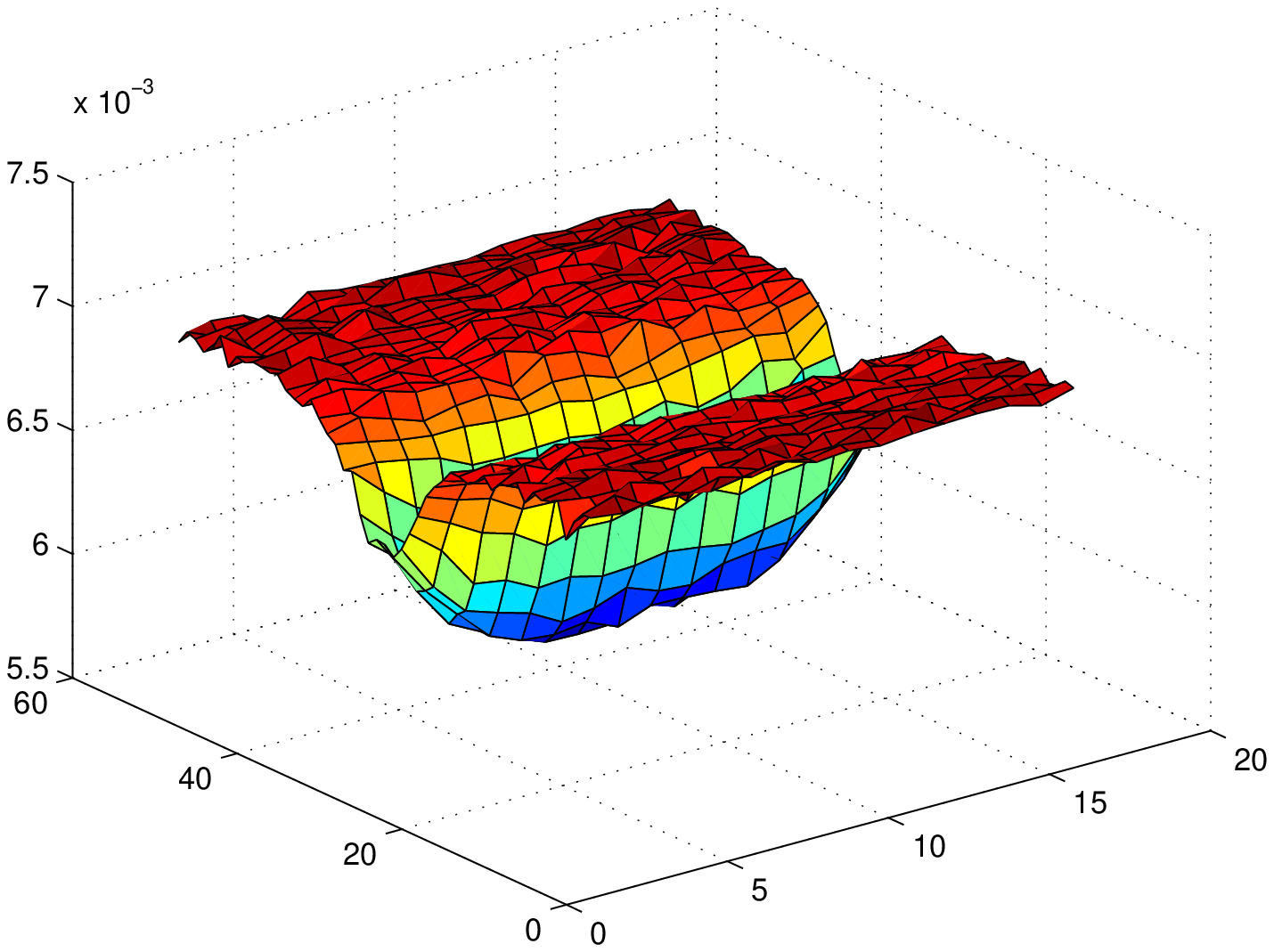} } \\ 
a) $s=8$ & b) $s=7$%
\end{tabular}%
\end{center}
\caption{{\protect\small \emph{Backscattered data $\protect\psi(x,s), x \in
\Gamma $ at different pseudo-frequencies $s \in [7; 20]$.}}}
\label{fig:F3D_3}
\end{figure}

\begin{figure}[tbp]
\begin{center}
\begin{tabular}{cc}
{\includegraphics[scale=0.4,clip=]{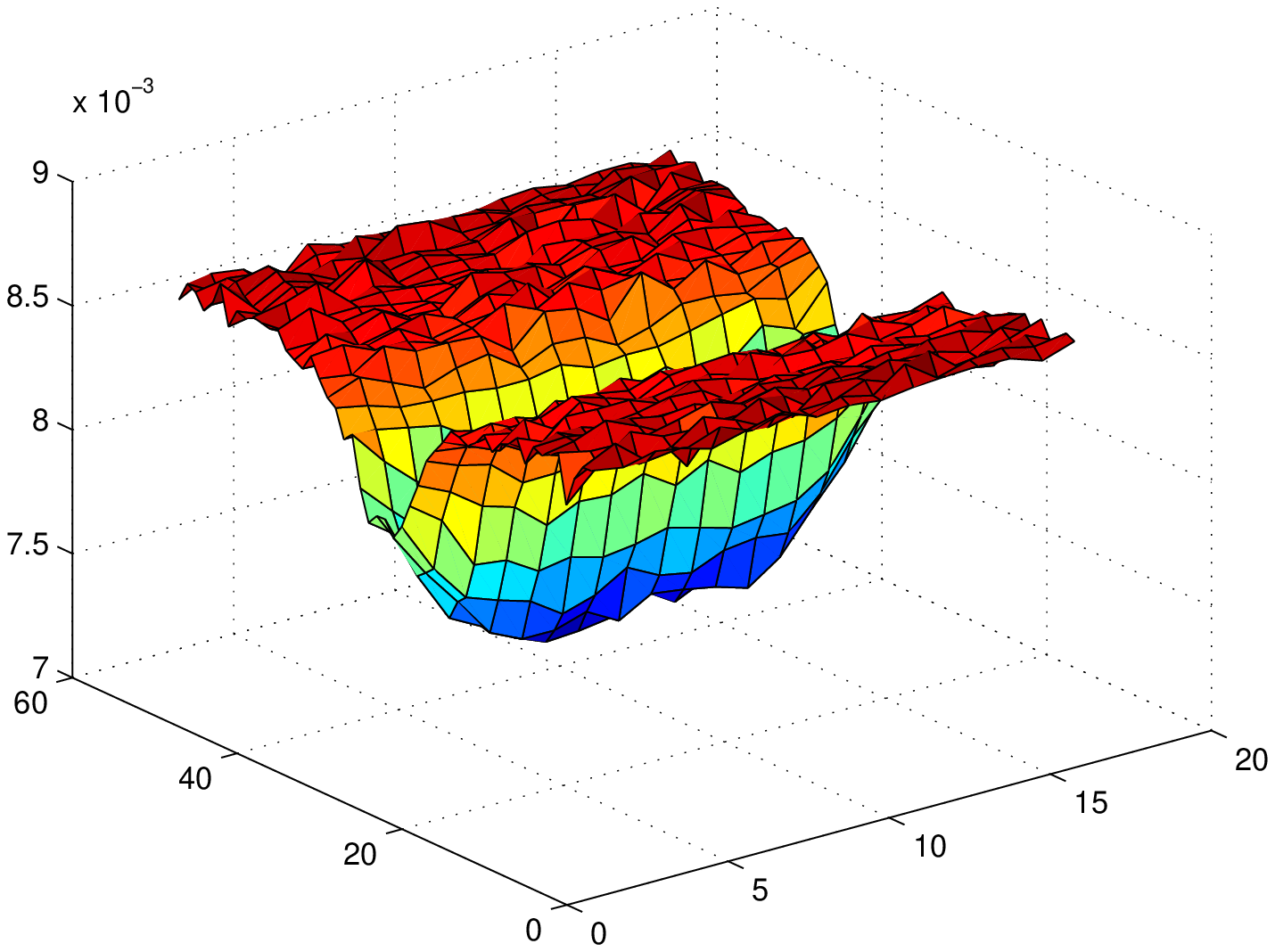}} & {\ %
\includegraphics[scale=0.4,clip=]{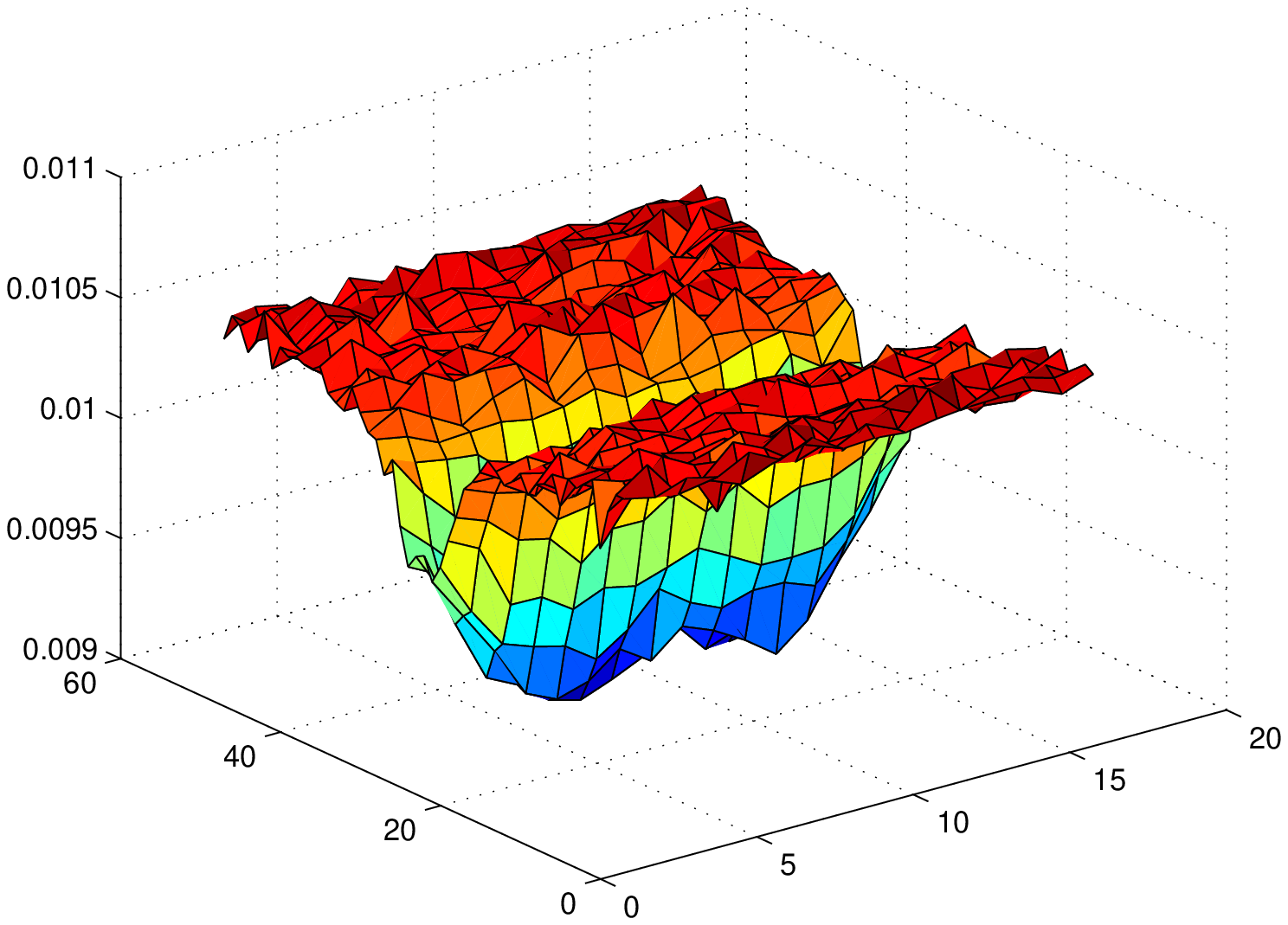} } \\ 
a) $s=6$ & b) $s=5$ \\ 
{\includegraphics[scale=0.4,clip=]{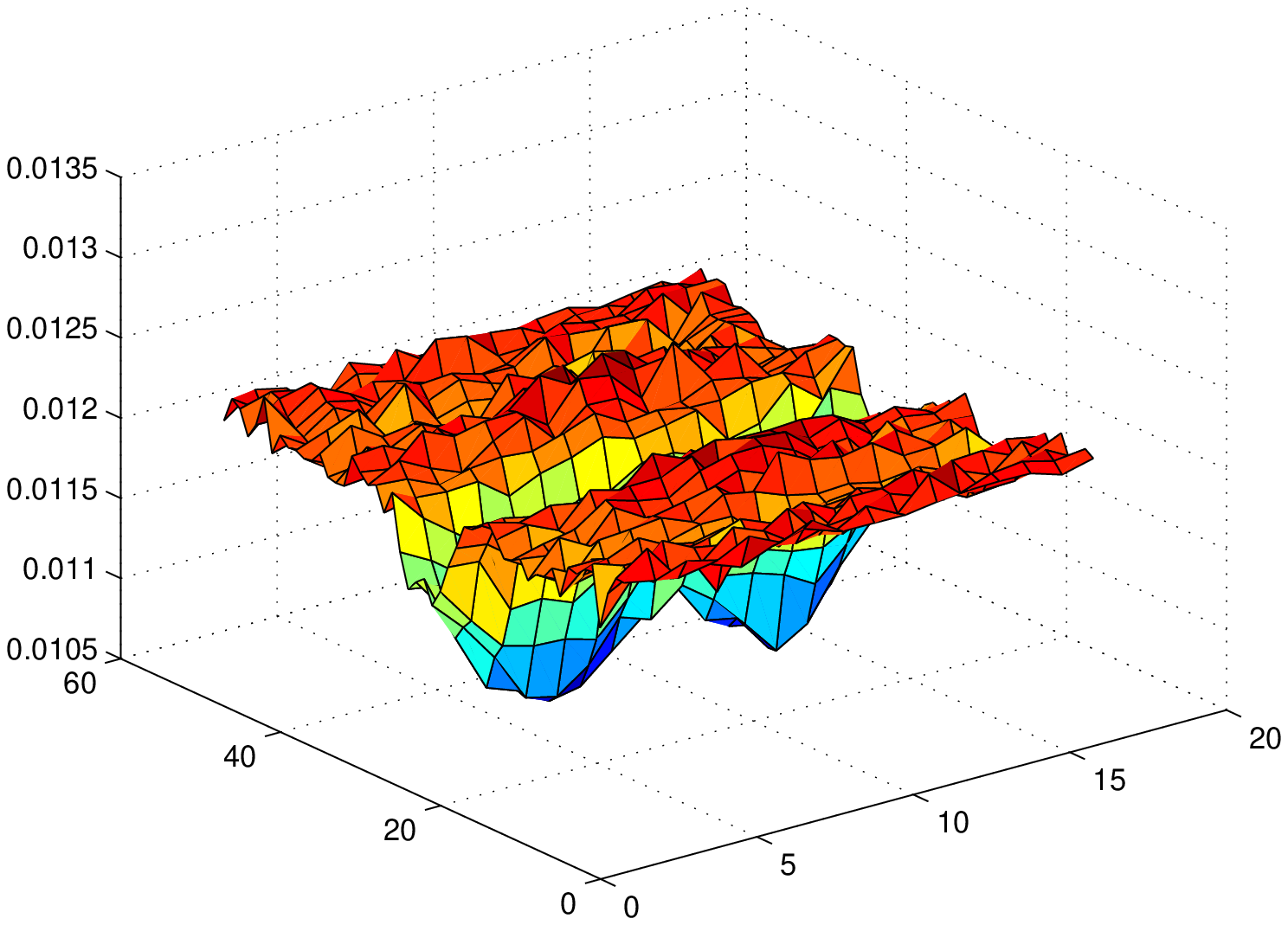}} & {\ %
\includegraphics[scale=0.4,clip=]{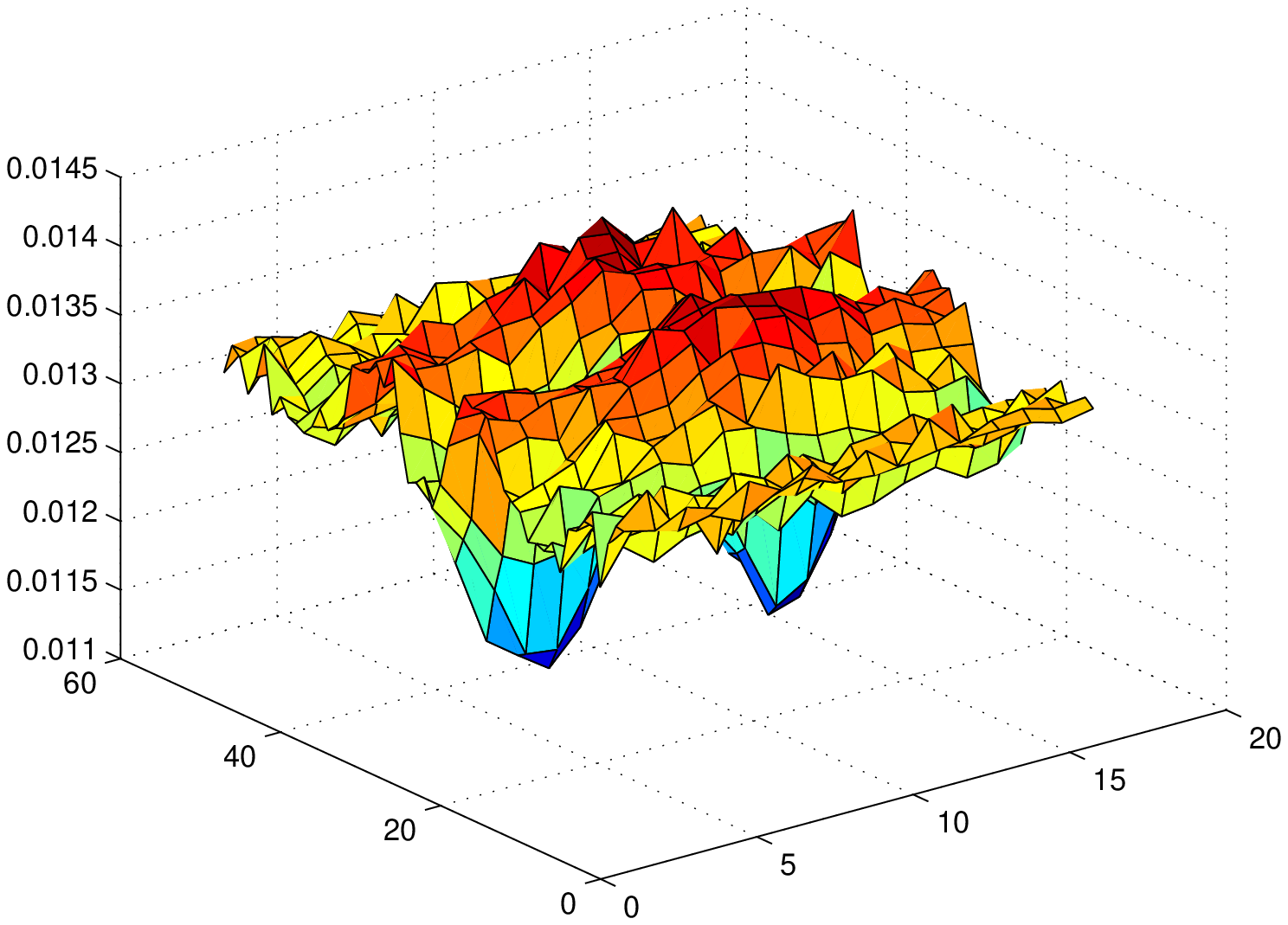} } \\ 
c) $s=4$ & d) $s=3$ \\ 
{\includegraphics[scale=0.4,clip=]{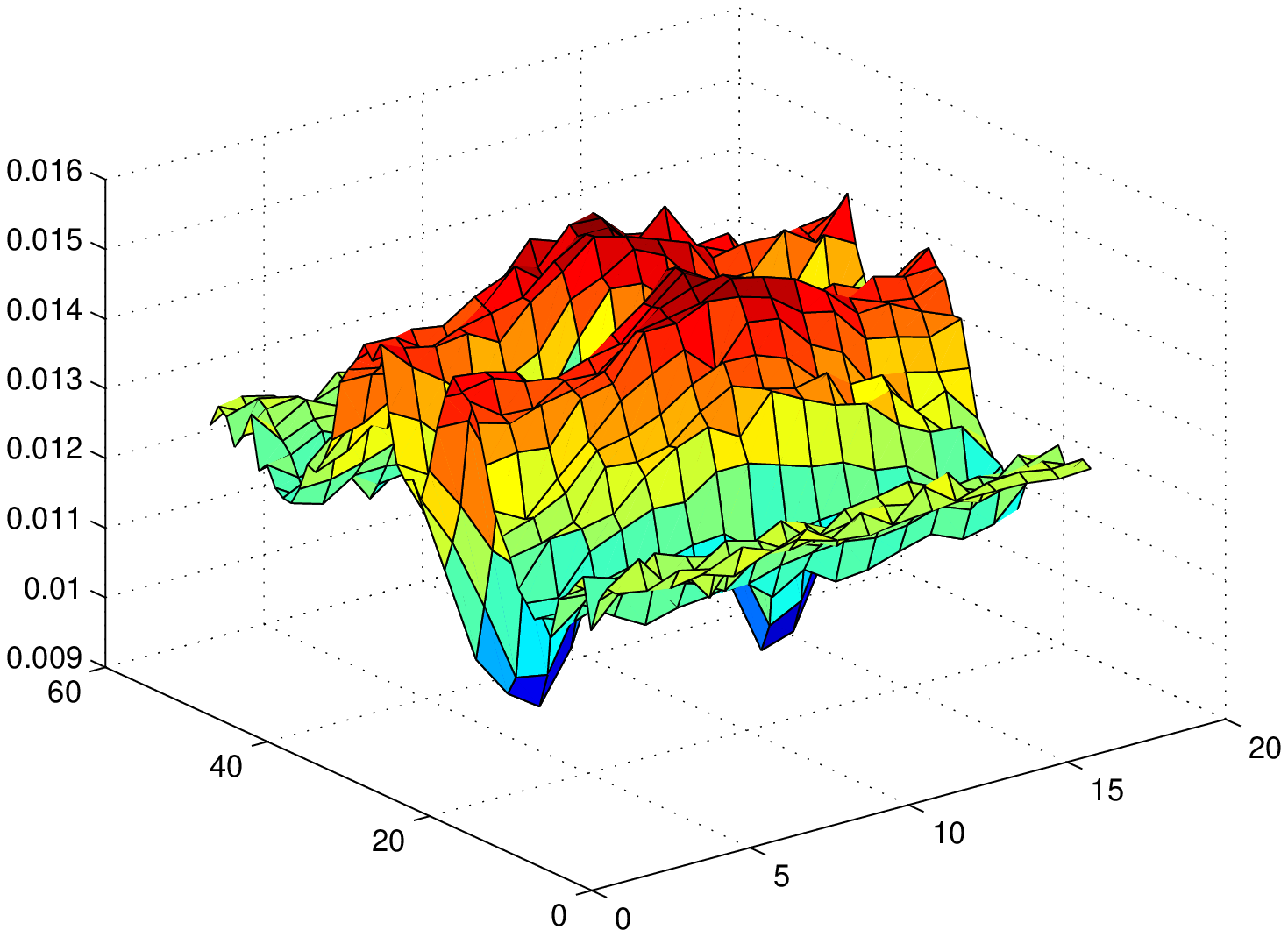}} & {\ %
\includegraphics[scale=0.4,clip=]{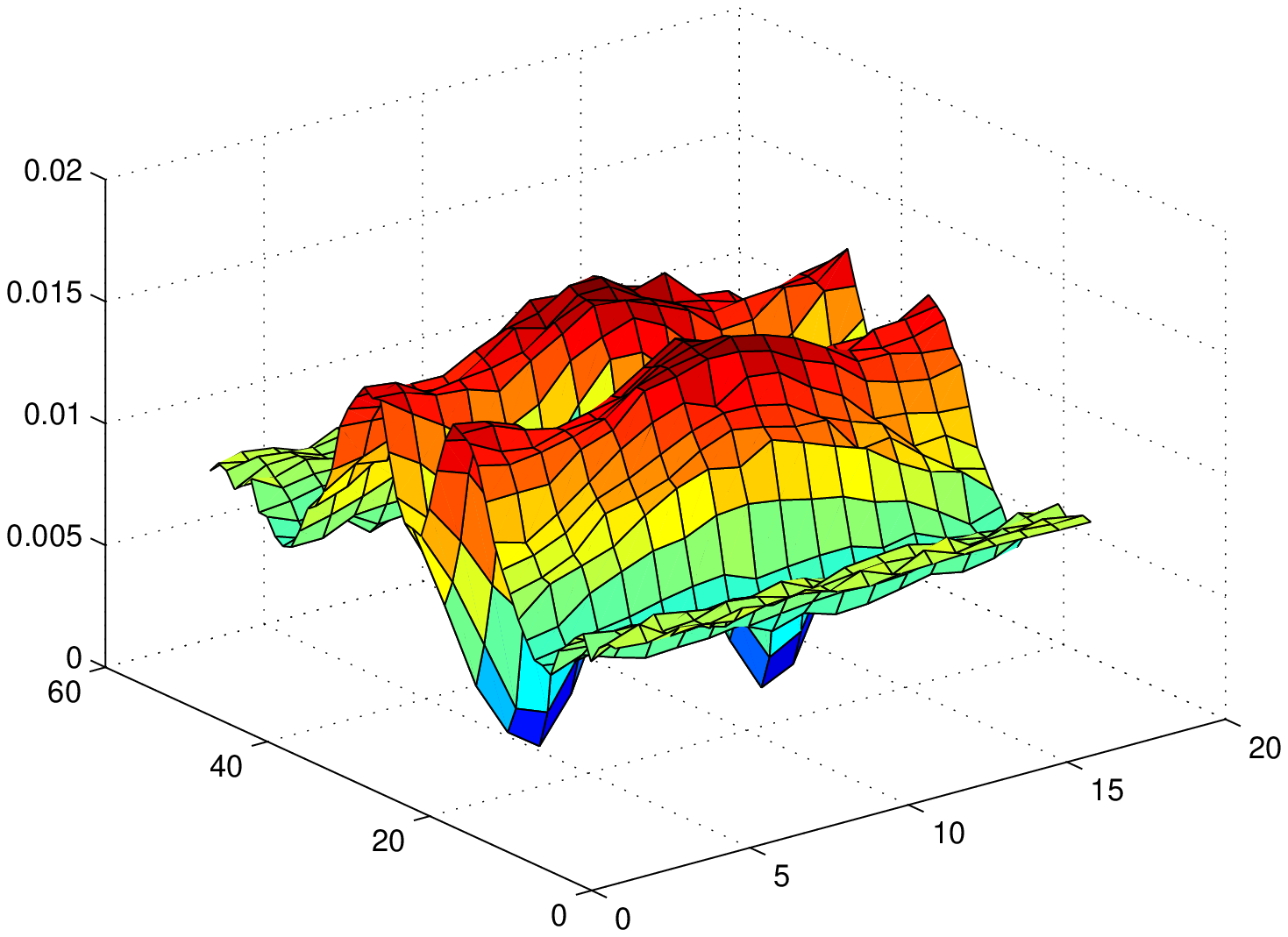} } \\ 
a) $s=2$ & b) $s=1$%
\end{tabular}%
\end{center}
\caption{{\protect\small \emph{Backscattered data $\protect\psi(x,s), x \in
\Gamma $ at different pseudo-frequencies $s \in [1, 6]$. One can see from
Figures \protect\ref{fig:F3D_3} and \protect\ref{fig:F3D_4} that one should
take $s \geq 4 $ in the reconstruction algorithm in 3d tests.}}}
\label{fig:F3D_4}
\end{figure}

\begin{figure}[tbp]
\begin{center}
\begin{tabular}{c}
{\includegraphics[scale=0.4,clip=]{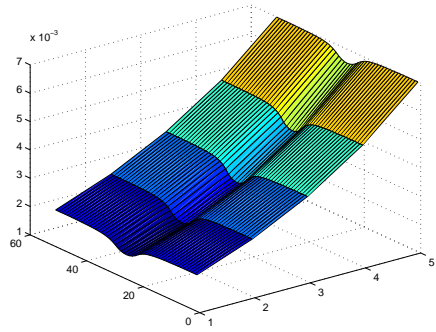}} \\ 
a) \\ 
{\includegraphics[scale=0.4,clip=]{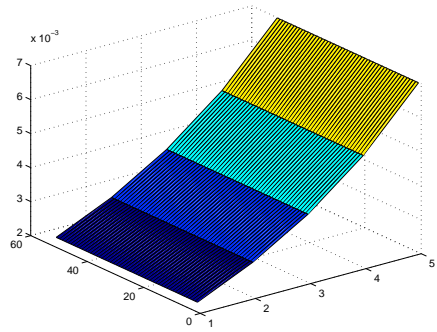} } \\ 
b) \\ 
{\includegraphics[scale=0.4,clip=]{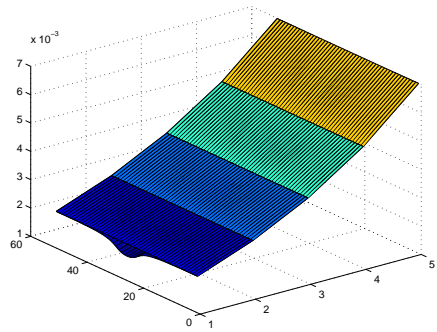} } \\ 
c)%
\end{tabular}%
\end{center}
\caption{{\protect\small \emph{Analysis of the scattered data at the left $%
\Gamma_l$ and right $\Gamma_r$ boundaries of the $G_{FEM}$ domain: a)
scattered data $\protect\psi(x,s), x \in \Gamma_l $ superimposed with the
scattered data $\protect\psi(x,s), x \in \Gamma_r$: one can see that the
dent is very small. The data are scattered from the belt modeling and
explosive, see Figure \protect\ref{fig:F3D_1}. Here, $c=3.2$ inside the belt
and $c=1$ at all other points of $G_{FEM}$; b) homogeneous data $\protect\psi%
(x,s)$ with $c=1$ in $G_{FEM}$; c) Superimposed data of a) and b). }}}
\label{fig:F3D_5}
\end{figure}

\begin{figure}[tbp]
\begin{center}
\begin{tabular}{c}
{\includegraphics[scale=0.4,clip=]{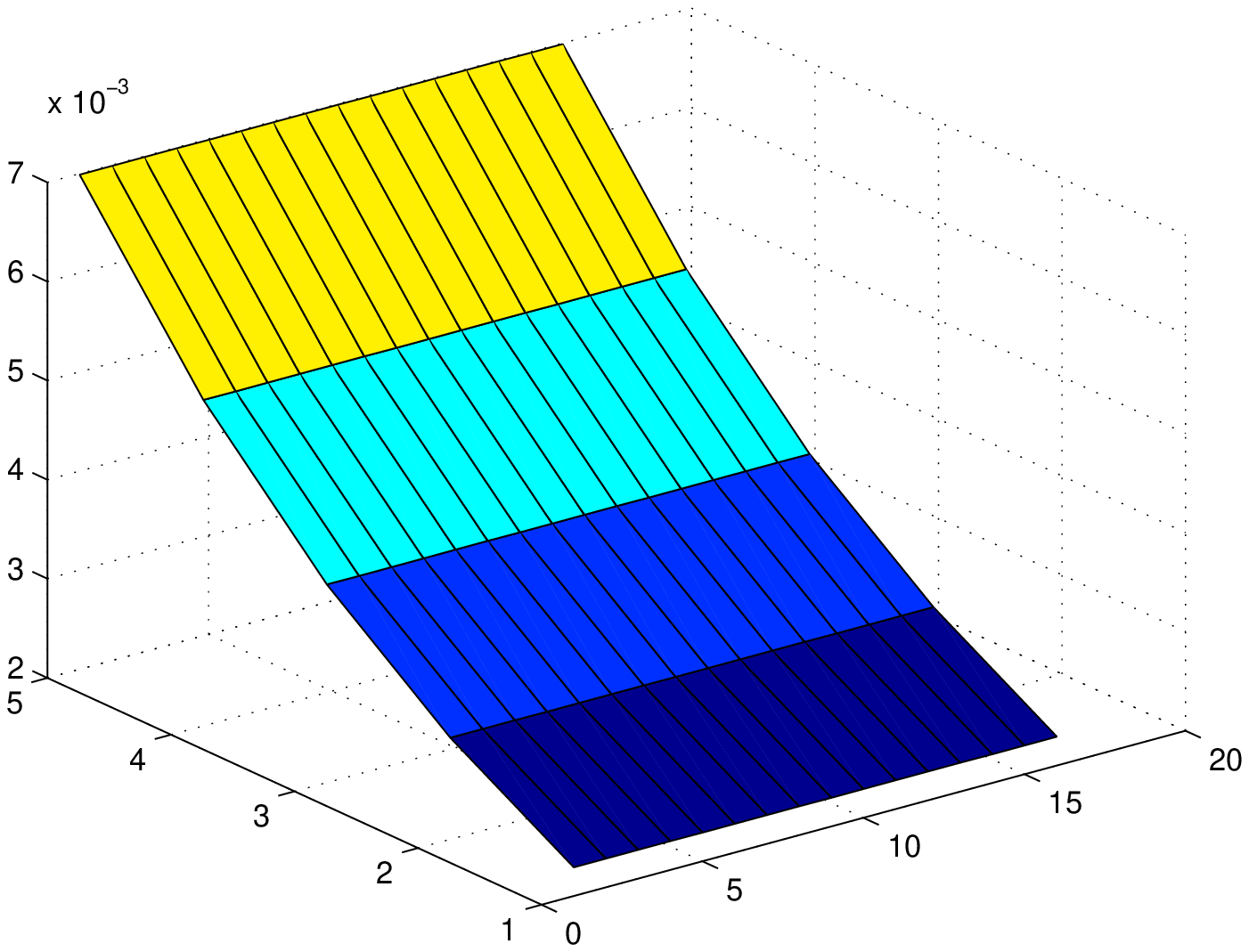}} \\ 
a) \\ 
{\includegraphics[scale=0.4,clip=]{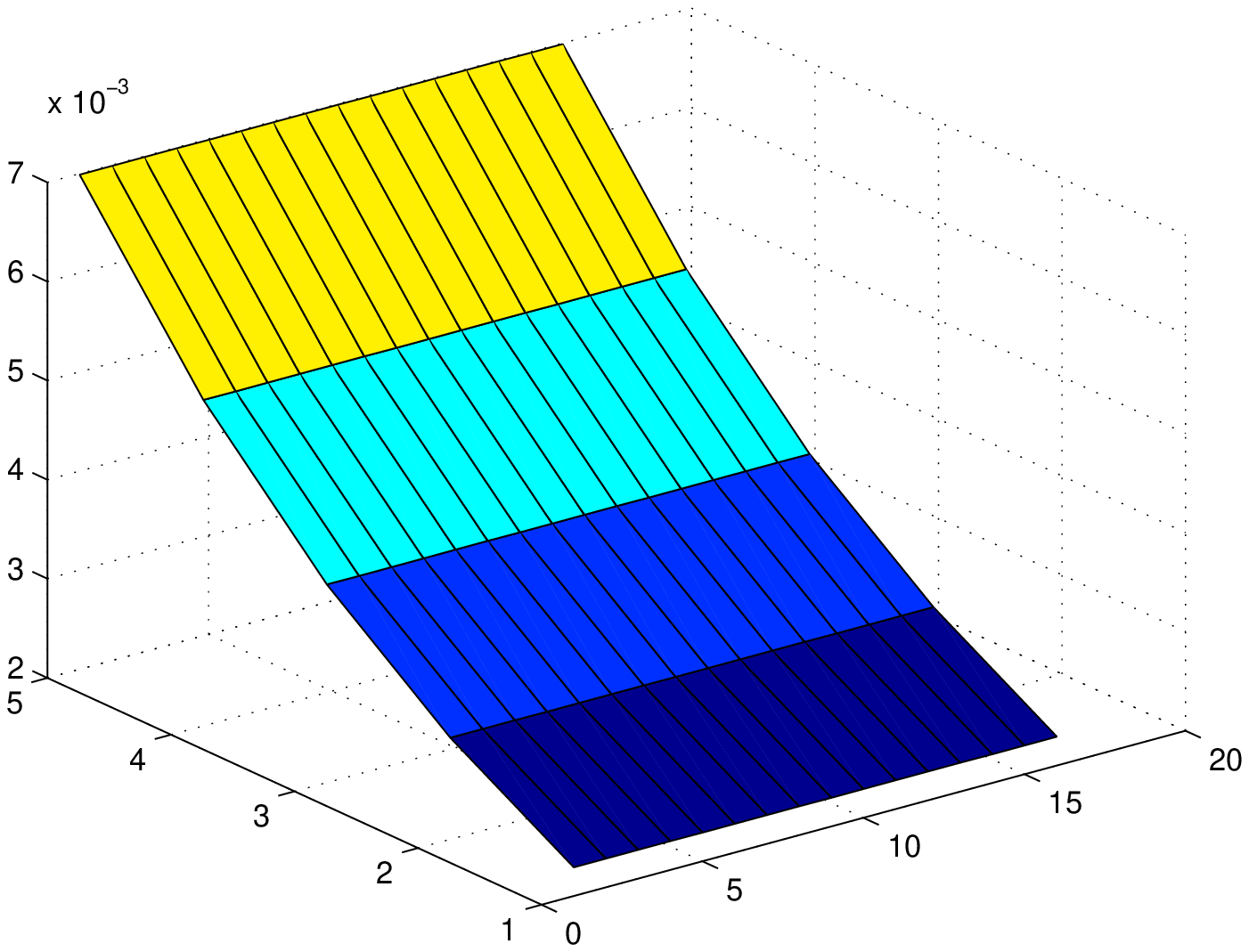} } \\ 
b) \\ 
{\includegraphics[scale=0.4,clip=]{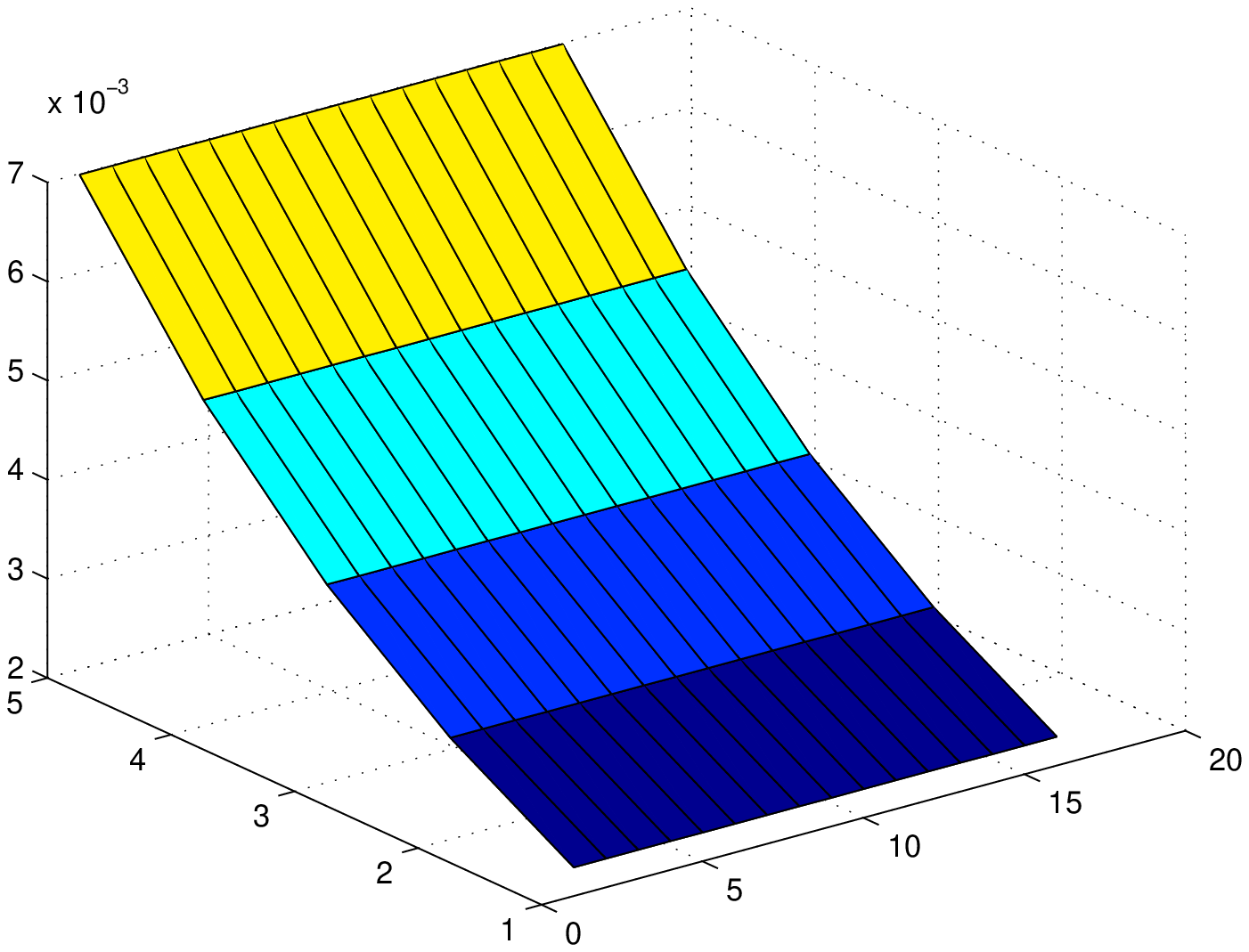} } \\ 
c)%
\end{tabular}%
\end{center}
\caption{{\protect\small \emph{Analysis of the scattered data at the top $%
\Gamma_t$ and bottom $\Gamma_{bot}$ boundaries of the $G_{FEM}$ domain: a)
scattered data $\protect\psi(x,s), x \in \Gamma_t$ superimposed with
scattered data $\protect\psi(x,s), x \in \Gamma_{bot}$. The data are
scattered from the belt with explosive of Figure \protect\ref{fig:F3D_1}
with $c=3.2$ inside the belt and $c=1$ at all other points of $G_{FEM}$; b)
homogeneous data $\protect\psi(x,s)$ with $c=1$ in $G_{FEM}$; c)
Superimposed data of a) and b).}}}
\label{fig:F3D_6}
\end{figure}

\begin{figure}[tbp]
\begin{center}
\begin{tabular}{c}
{\includegraphics[scale=0.5,clip=]{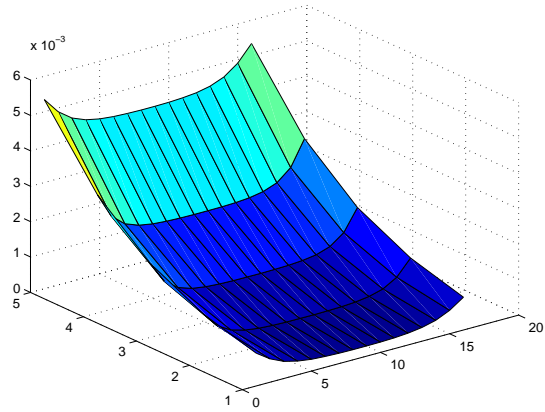}} \\ 
a) \\ 
{%
\includegraphics[scale=0.5,clip=]{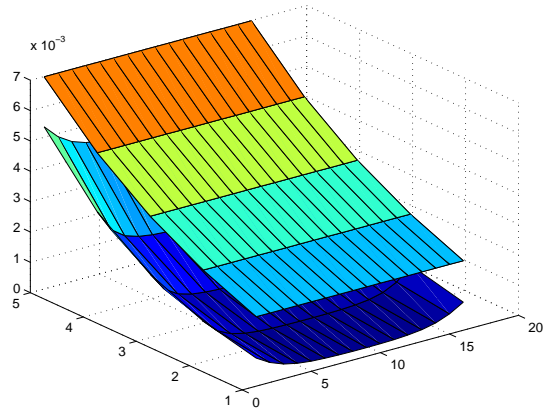}
} \\ 
b)%
\end{tabular}%
\end{center}
\caption{{\protect\small \emph{\ Test 5. Analysis of scattered data at the
top $\Gamma_t$ and bottom $\Gamma_{bot}$ boundaries of the $G_{FEM}$: a)
scattered data $\protect\psi(x,s), x \in \Gamma_t$ superimposed with
scattered data $\protect\psi(x,s), x \in \Gamma_{bot}$. b) Superimposed data
of a) (low figure) and homogeneous data $\protect\psi(x,s)$ (top figure)
computed with $c=1$ in $G_{FEM}$.}}}
\label{fig:Test5_1}
\end{figure}

\begin{figure}[tbp]
\begin{center}
\begin{tabular}{cc}
{\includegraphics[scale=0.25,clip=]{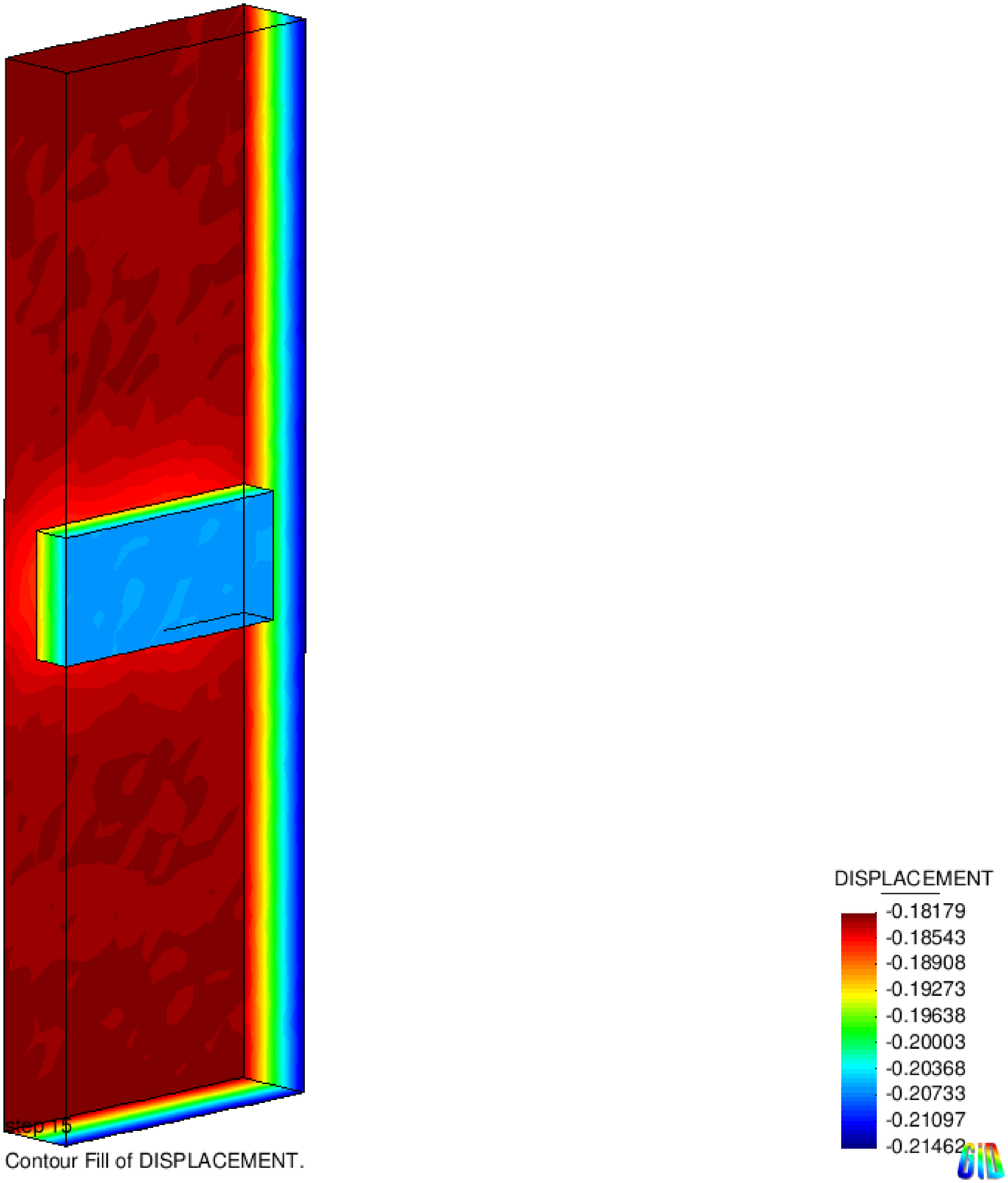}} & {%
\includegraphics[scale=0.25,clip=]{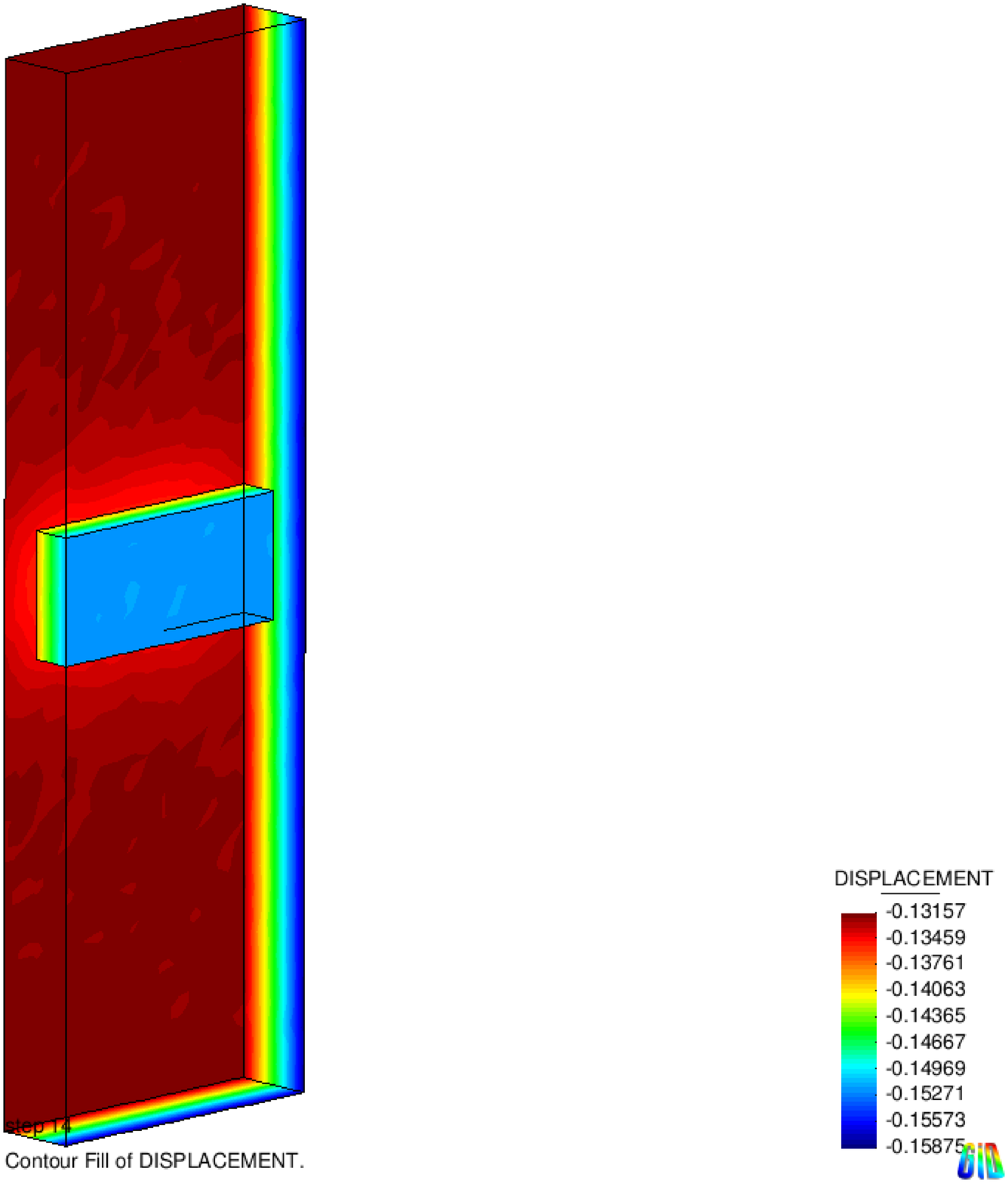}} \\ 
a) $s=6 $ & b) $s=7$ \\ 
{\includegraphics[scale=0.25,clip=]{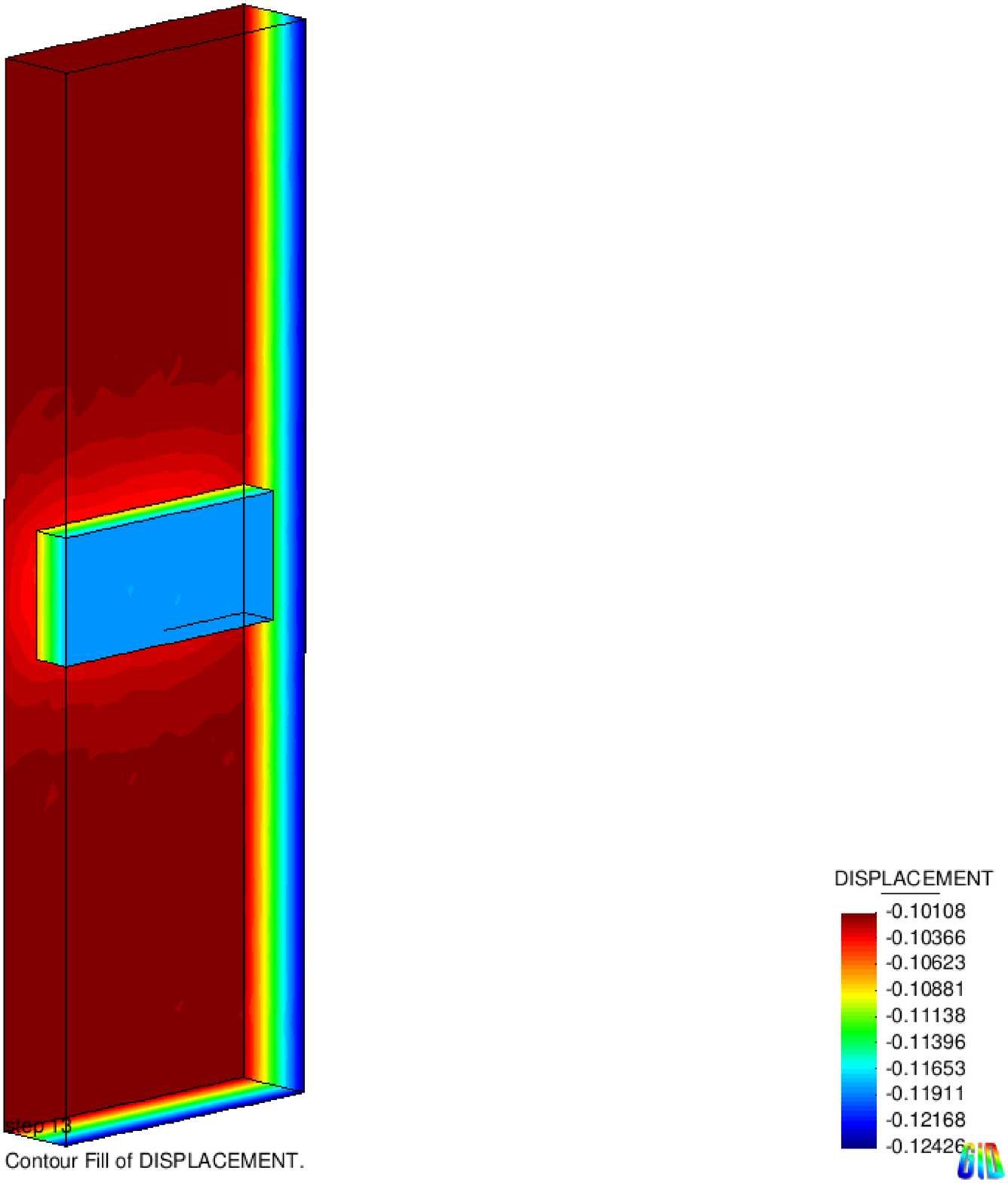}} & {%
\includegraphics[scale=0.25,clip=]{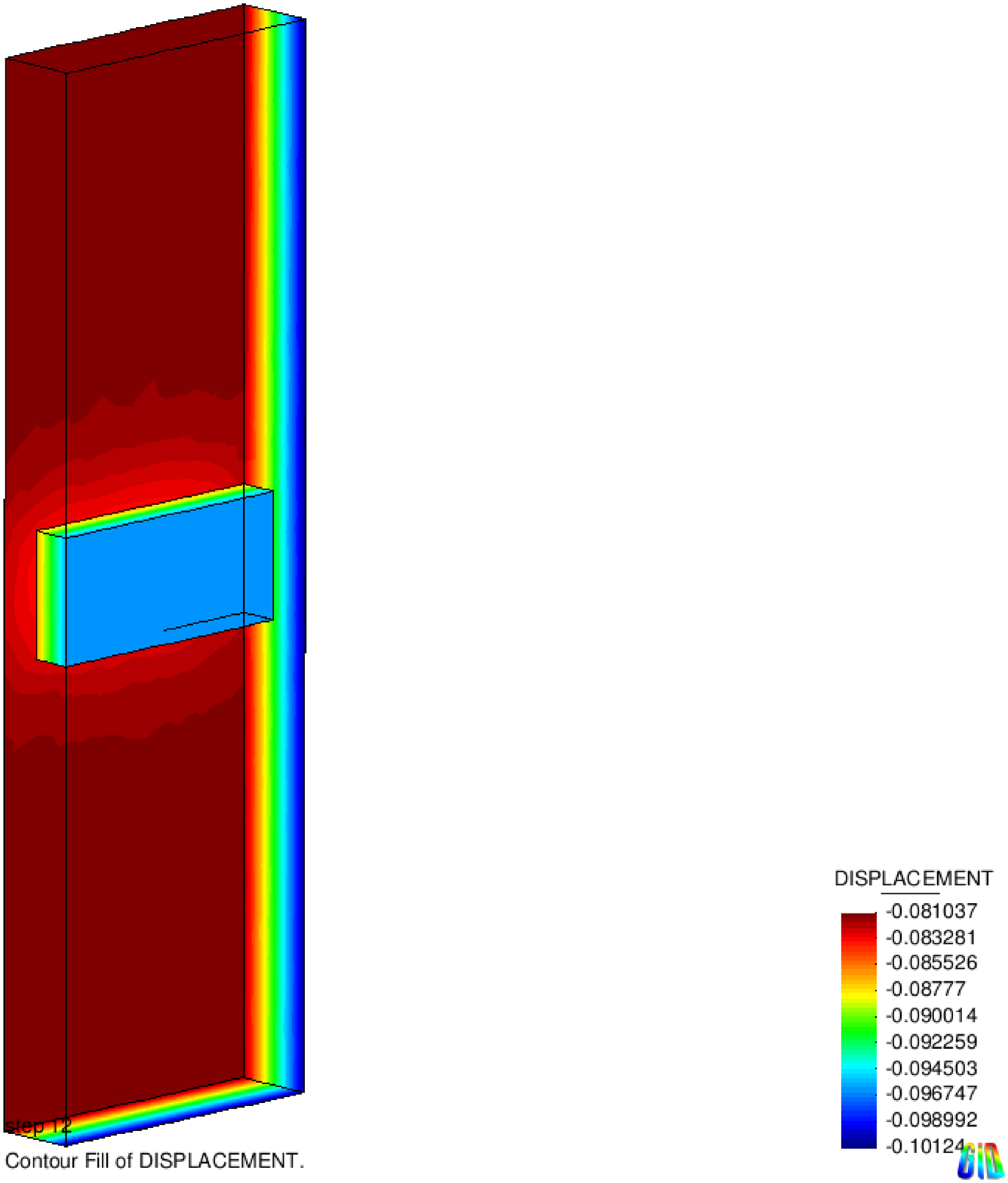}} \\ 
a) $s=8 $ & b) $s=9$ \\ 
& 
\end{tabular}%
\end{center}
\caption{{\protect\small \emph{\ Test 5. Backscattered data $\protect\psi %
(x,s), x \in \Gamma $ immersed into data $\protect\psi(x,s), x \in \partial
\Omega \diagdown \Gamma $ computed with $c=1$ in $\Omega = G_{FEM}$. Data
are presented at different pseudo-frequencies $s=6,7,8,9$.}}}
\label{fig:Test5_2}
\end{figure}

\begin{figure}[tbp]
\begin{center}
\begin{tabular}{cc}
{\includegraphics[scale=0.2,clip=]{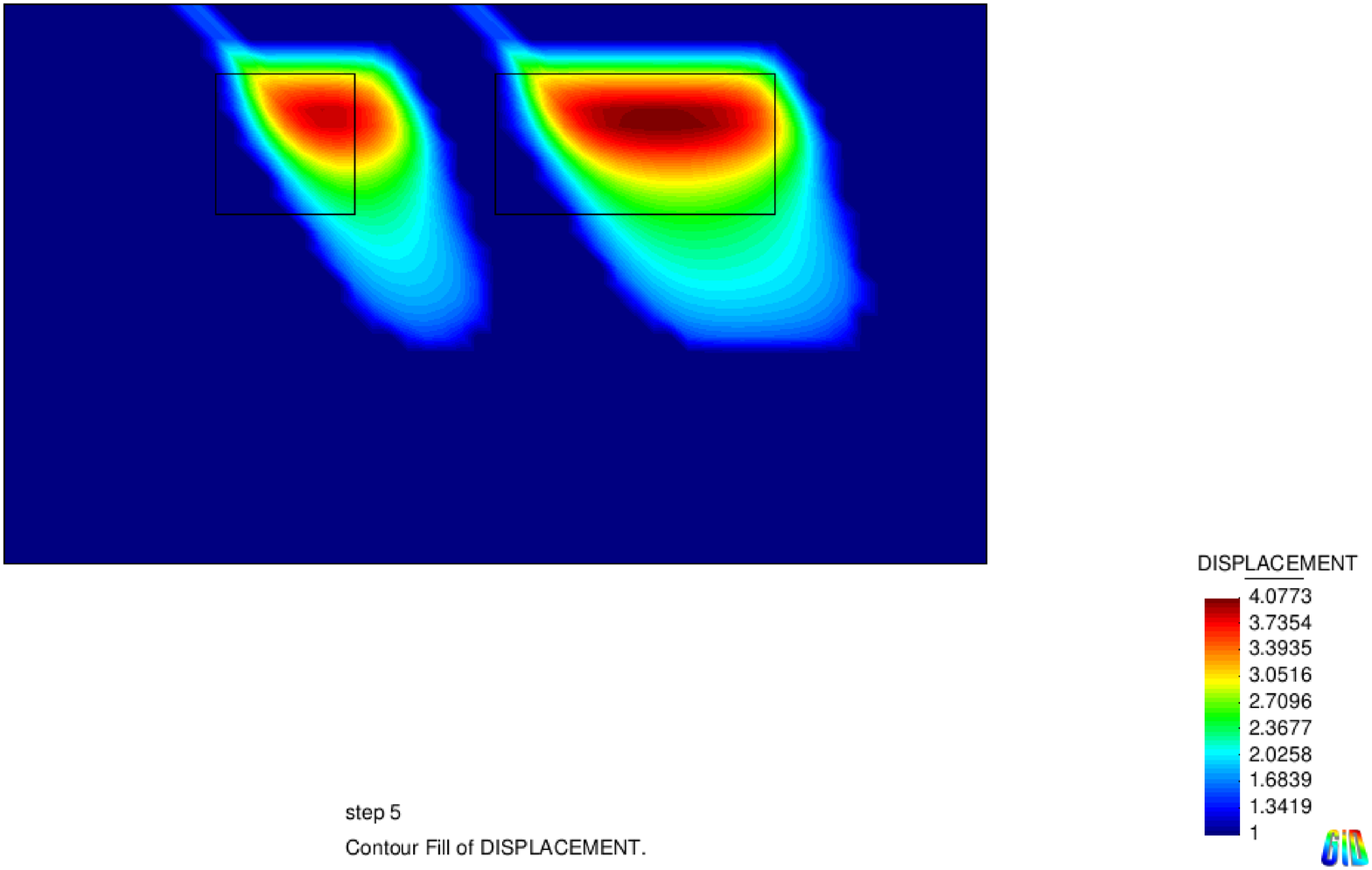}} & {%
\includegraphics[scale=0.2,clip=]{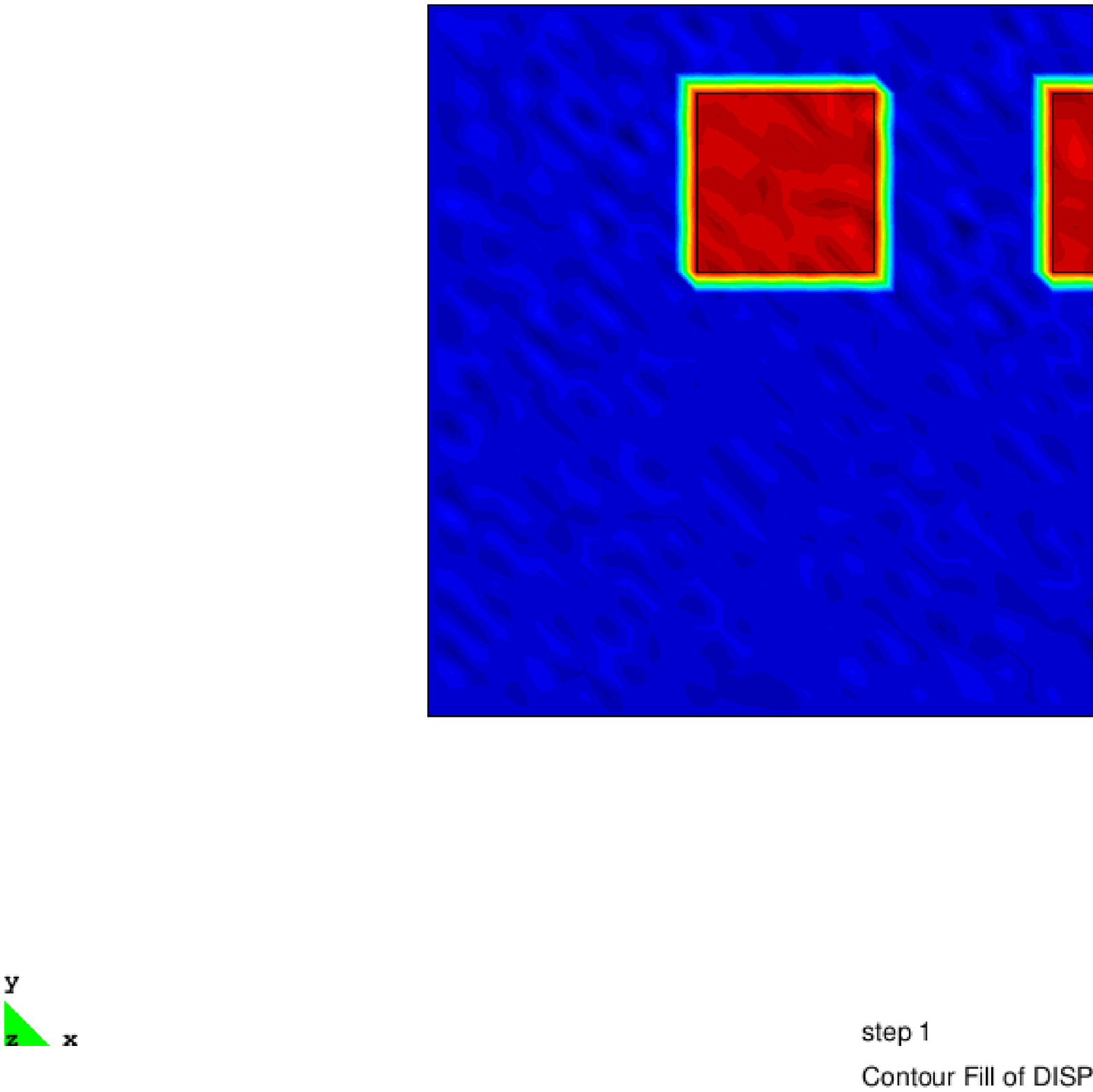}} \\ 
a) & b) \\ 
& 
\end{tabular}%
\end{center}
\caption{\emph{\ Test 1: a) The computed function $c_{3,5}\left( x\right).$
Maximal values of this function are 4.07 in both imaged mine-like targets
and $c_{3,5}\left( x\right) =1$ outside of imaged targets. The image is
accurate: compare with Figure \protect\ref{fig:F1}-c) and with (\protect\ref%
{6.3_1}). b) The reconstructed function $c_{1,1}\left( x\right) $ for the
the case when the exact tail function $V^{\ast }\left( x\right) $ is known.
However, this is an unrealistic case, which is presented here only to
demonstrate the accuracy of our method in the ideal case. }}
\label{fig:F4}
\end{figure}

\begin{figure}[tbp]
\begin{center}
\begin{tabular}{cc}
{\includegraphics[scale=0.2,clip=]{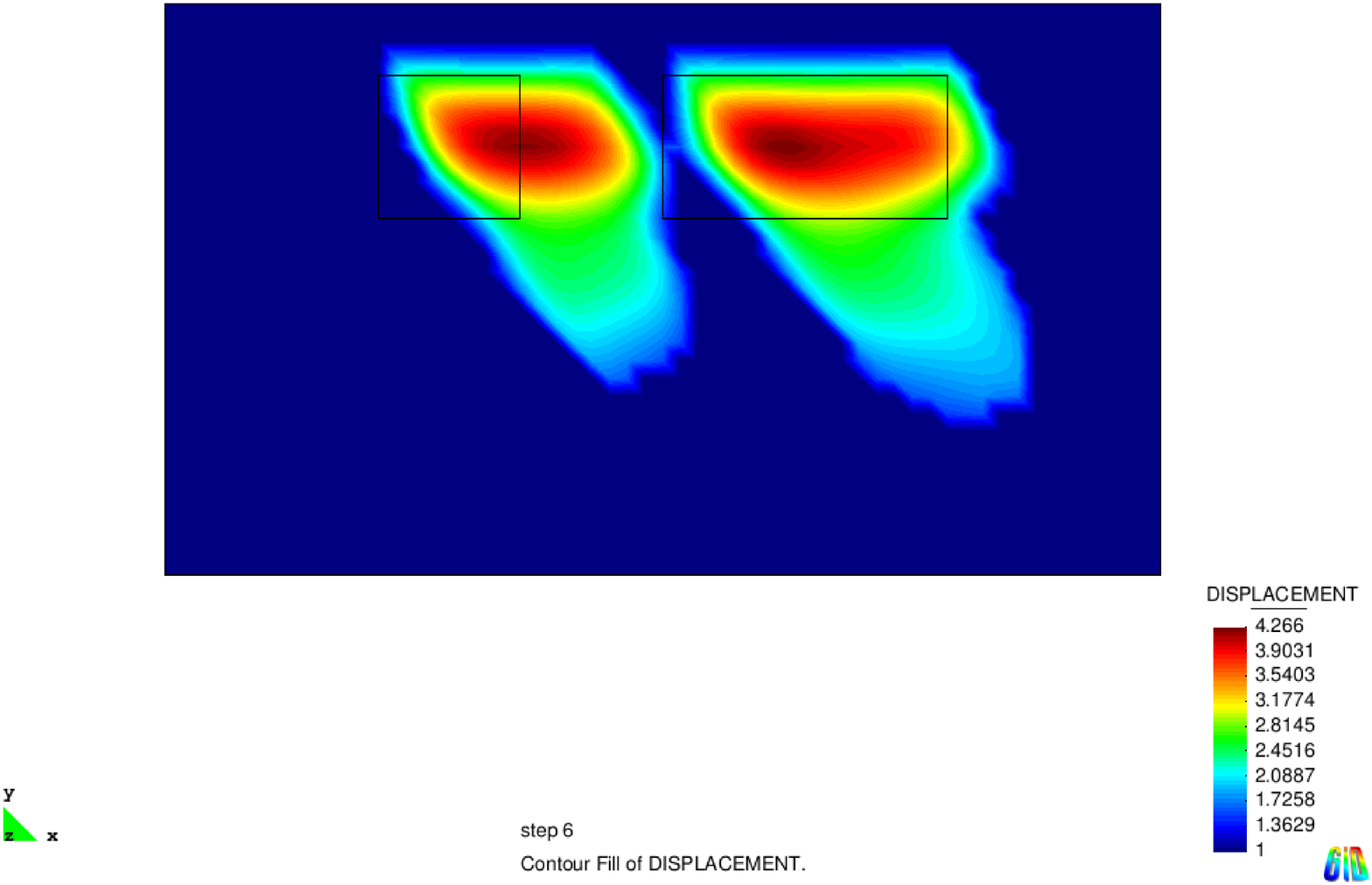}} & {%
\includegraphics[scale=0.2,clip=]{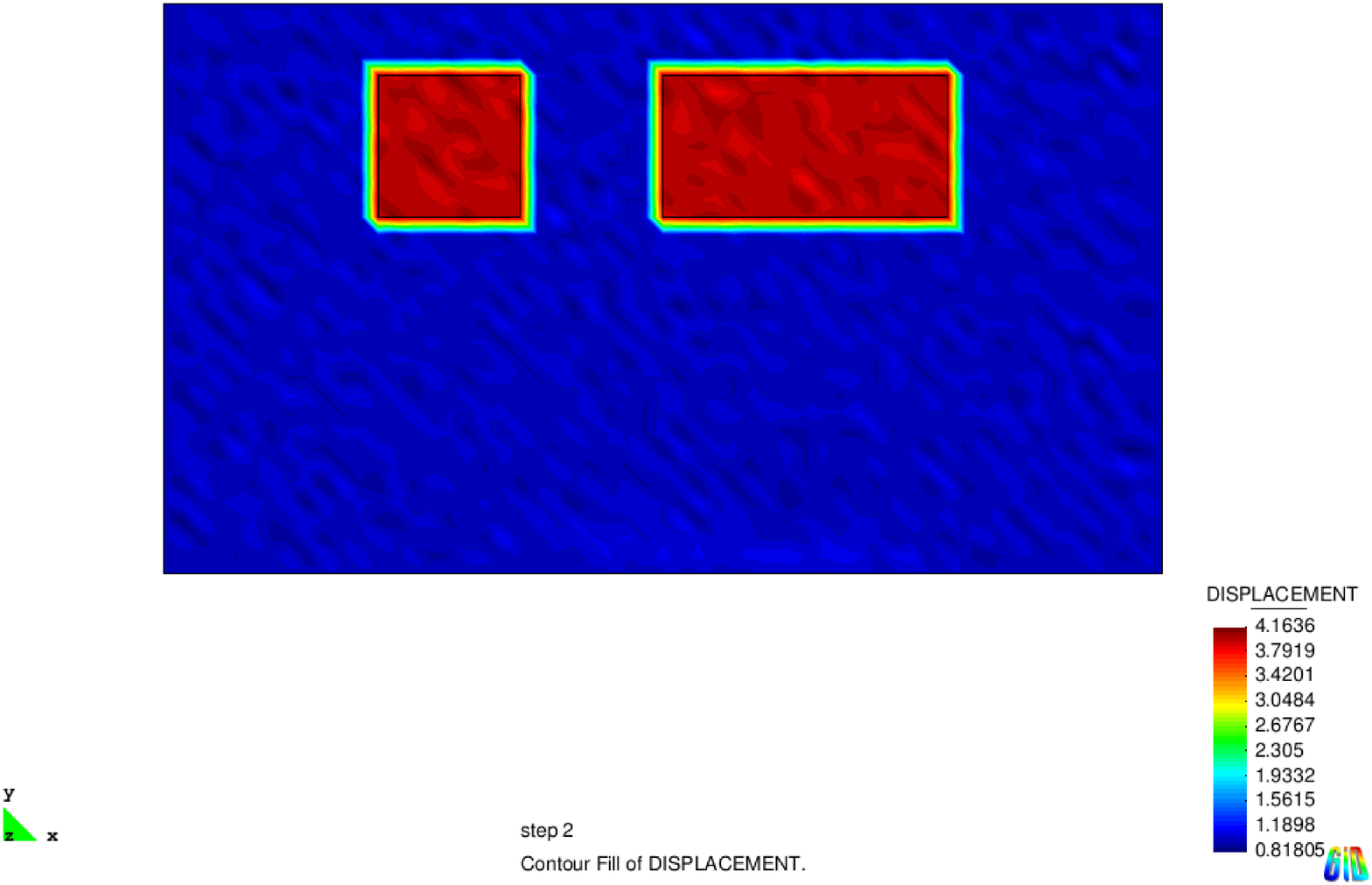}} \\ 
a) & b) \\ 
& 
\end{tabular}%
\end{center}
\caption{\emph{\ Test 2: a) The computed function $c_{8,6}\left( x\right).$
Maximal values of this function are 4.27 in both imaged mine-like targets
and $c_{8,6}\left( x\right) =1$ outside of imaged targets. The image is
accurate: compare with Figure \protect\ref{fig:F1}-c) and with (\protect\ref%
{6.3_1}). b) The reconstructed function $c_{1,1}\left( x\right) $ for the
the case when the exact tail function $V^{\ast }\left( x\right) $ is known.}}
\label{fig:F5}
\end{figure}

\begin{figure}[tbp]
\begin{center}
\begin{tabular}{c}
{\includegraphics[scale=0.5, angle=90, clip=]{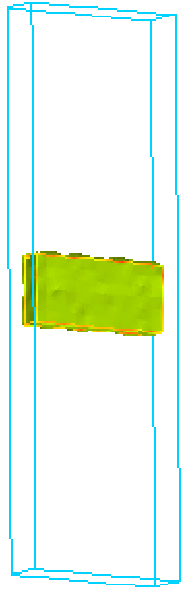}} \\ 
a) $\max {c}_{1,1} \approx 3.2 $ \\ 
{\includegraphics[scale=0.4, angle=90, clip=]{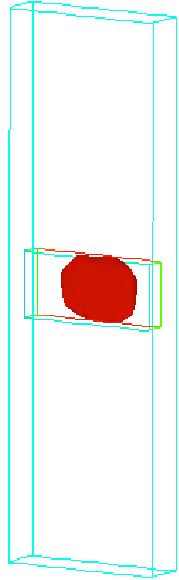}} \\ 
b) $\max {c}_{3,2} \approx 3.54$%
\end{tabular}%
\end{center}
\caption{{\protect\small \emph{\ Test 3. Reconstruction of a belt with
explosives for the case (\protect\ref{6.30}). a) The computed function $%
c_{1,1}\left( x\right) $ for the case when the exact tail function $V^{\ast
}\left( x\right) $ is known. The reconstruction is perfect. However, this is
an unrealistic scenario. We display it here only to show that our method is
accurate in an ideal case. b) The reconstruction for the case when the
initial tail function $V_{1,1}\left( x\right) $ is taken the same as the one
for the uniform medium with $c\left( x\right) \equiv 1.$ The computed
function $c_{3,2}\left( x\right) $ is shown. Observe that $\max
c_{3,2}\left( x\right) =3.54$ inside of the imaged inclusion (belt with
explosives) and $c_{3,2}\left( x\right) =1$ everywhere else. The image is
accurate.}}}
\label{fig:F3D_7}
\end{figure}

\begin{figure}[tbp]
\begin{center}
\begin{tabular}{c}
{\includegraphics[scale=0.4, angle=90, clip=]{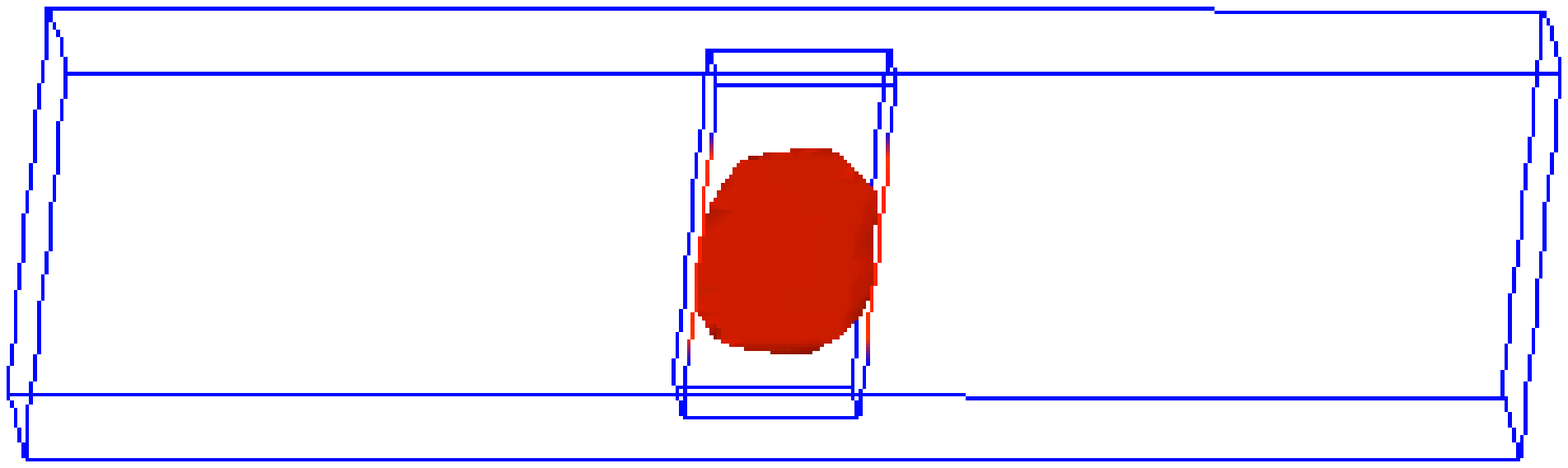}} \\ 
a) $\max c_{3,2} \approx 3.54$ \\ 
{\includegraphics[scale=0.4, angle=90, clip=]{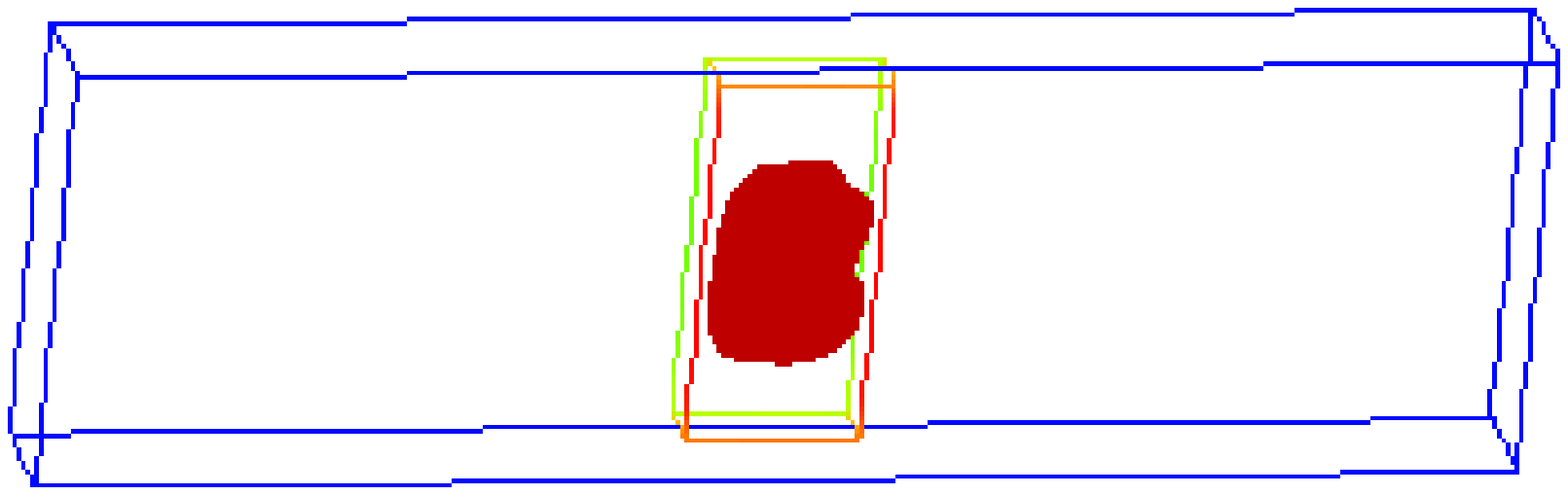}} \\ 
b) $\max {c}_{2,3} \approx 3.09$ \\ 
\end{tabular}%
\end{center}
\caption{{\protect\small \emph{Reconstruction with back-scattered data. a)
Test 4. The computed image of the function $c_{3,2}$. Observe that $\max
c_{3,2}\left( x\right) =3.54$ inside of the imaged inclusion (belt with
explosives) and $c_{3,2}\left( x\right) =1$ everywhere else. b) Test 5. This
is the most challenging test since the medium is a quite heterogeneous one,
see (\protect\ref{6.29}). The computed image of the function $c_{2,3}\left(
x\right).$ One can see that $\max c_{2,3}\left( x\right) =3.09$ inside of
the imaged inclusion (``belt" with explosives) and $c_{2,3}(x) =1$
everywhere else. Both images are quite accurate ones. }}}
\label{fig:F3D_8}
\end{figure}

Just as in the 2d case, we have chosen boundary conditions as in (\ref{6.6}%
). To justify this, we present Figures \ref{fig:F3D_3}-\ref{fig:F3D_6}. One
can see from these figures that values of the function $w\left( x,s\right) $
at all parts of $\partial \Omega \diagdown \Gamma $, except of the back side 
$\Gamma _{b}$ of the prism $\Omega $, are about the same as ones for the
case $c\left( x\right) \equiv 1,$ which is the value of this coefficient
outside of our domain of interest $\Omega .$ As to the surface $\Gamma _{b},$
it corresponds to the transmitted signal and values of $w\left( x,s\right) $
here are far from those of the uniform background outside of $\Omega $.
However, just as in the 2d case, the transmitted side $\Gamma _{b}$ is
located far from the backscattering side $\Gamma .$ Therefore, the Laplace
transform (\ref{3.5}) diminishes the influence of waves reflected from $%
\Gamma _{b},$ at least for large values of the parameter $s$, see Figures %
\ref{fig:F3D_3}, \ref{fig:F3D_4}. Hence, the amplitude of reflected waves
from this side is small when they reach $\Gamma,$ compared with reflections
at $\Gamma$ from the target $\Omega_{belt}.$ This provides a numerical
justification of (\ref{6.6}) in the 3d case. We call the resulting boundary
function $\psi \left( x,s\right)$ ``immersed boundary data", see Figure \ref%
{fig:Test5_2}.

\subsection{Numerical tests for the 2d case}

\label{sec:6.8}

Let $u_{calc}\left( x,t\right) ,x\in \Gamma $ be the calculated solution of
the forward problem (\ref{6.4}), (\ref{6.5}) at the backscattering side $%
\Gamma $ of the boundary $\partial \Omega .$ We have introduced a random
noise in the function $u_{calc}\left( x,t\right) ,x\in \Gamma $ as%
\begin{equation}
u_{\sigma }\left( x^{\left( i\right) },t^{\left( j\right) }\right)
=u_{calc}\left( x^{\left( i\right) },t^{\left( j\right) }\right) \left[
1+\alpha _{j}\sigma \left( u_{\max }-u_{\min }\right) \right] ,  \label{6.10}
\end{equation}
where $\left( x^{\left( i\right) },t^{\left( j\right) }\right) \in \Gamma
\times \left( 0,T\right) $ are mesh points, $u_{\max }$ and $u_{\min }$ are
maximal and minimal values of $u_{calc}\left( x,t\right) $ for $x\in \Gamma
, $ numbers $\alpha _{j}\in \left( -1,1\right) $ are randomly distributed
and $\sigma =0.05$. Thus, the noise level was 5\%. %To find the boundary
%condition $\psi \left( x,s\right) ,x\in \Gamma $ in (\ref{3.15}) for the
%function $q\left( x,s\right) ,$ we need to differentiate the function $%
%w\left( x,s\right) ,x\in \Gamma $ with respect to $s$. To calculate the
%derivative $\partial _{s}w\left( x,s\right) ,x\in \Gamma ,$ we have used the
%finite difference formula, see \cite{BK1} and page 176 of \cite{BK}. This
%worked well, regardless on the noise in (\ref{6.10}). The latter can be
%explained by the fact that Laplace transform smooths out the noise.
In both Test 1 and Test 2 the correct coefficient $c\left( x\right) $ is the
same as in (\ref{6.3_1}).

\textbf{Test 1}. In this test the initial guess for the tail function $%
V_{1,1}$ the function was computed via (\ref{3.12}) for the case of the
homogeneous domain $G$ with $c(x)\equiv 1$. Next, the algorithm of
subsection 3.4 was applied to reconstruct the true function $c(x)$ in (\ref%
{6.3_1}).\ The computed image is presented on Figure \ref{fig:F4}-a). We
observe that both the location and the contrast of both mine-like targets
are reconstructed accurately. The number of inner iterations with respect to
tails was $m=5.$ The stopping criterion (\ref{crit1})-(\ref{crit2}) was
achieved at $c_{3,5}\left( x\right) ,$ i.e. we have stopped at $n=3.$ By (%
\ref{6.7}), (\ref{3.17}) and (\ref{3.18}) this corresponds to $s\in \left[ 
2.85,2.90\right].$ The reconstructed dielectric constant in this test is $%
c_{3,5}\left( x\right) =4.07$ inside of both imaged mine-like targets and $%
c_{3,5}\left( x\right) =1 $ at all other points of $\Omega $. To see what
happens in an ideal case when the exact tail function $V^{\ast }\left(
x\right) $ is known, we refer to Figure \ref{fig:F4}-b), which corresponds
to the function $c_{1,1}\left( x\right).$ Figure \ref{fig:F4}-b) confirms
that the reconstruction is perfect in this case.

\textbf{Test 2. }In this test we choose the initial guess for the tail
function $V_{1,1}$as an initial guess for the tail function $V_{1,1}$ we
take the function computed via (\ref{3.43}) and use the algorithm of
subsection 3.4 to reconstruct the dielectric constant of Figure \ref{fig:F5}%
-a). In this test the reconstructed dielectric constant is $c_{8,6}\left(
x\right) =4.27$ inside mine-like targets and $c_{8,6}\left( x\right) =1$ at
all other points of $\Omega $. This reconstruction was obtained on the
pseudo-frequency interval $s\in \left[ 2.6,2.65\right] $ and after 6
iterations with respect to the tail function. In other words, we took the
number of inner iterations with respect to tails $m=6$ and the stopping
criterion (\ref{crit1})-(\ref{crit2}) was achieved at $n=8$.

Thus, in both Tests 1,2 reconstructions were accurate ones.

\subsection{Numerical test for the 3d case}

\label{sec:6.9}

In Test 3 and Test 4 we present results for the case (\ref{6.30}), and in
Test 5 - for the case (\ref{6.29}). The same random noise of 5\% was
introduced as the one in (\ref{6.10}).

\textbf{Test 3}. In this test we took the first guess for the tail function $%
V_{1,1}\left( x\right) $ the same as the one for the uniform background when 
$c\left( x\right) \equiv 1$ for $x\in G$. Using Figures \ref{fig:F3D_3}, \ref%
{fig:F3D_4} and analyzing the backscattered data $\psi (x,s)$ for $x\in
\Gamma $ we have decided to choose the interval of pseudo frequencies as 
\begin{equation*}
s\in \left[ 4,11\right] ,h=1.
\end{equation*}%
Then we have used the algorithm of subsection 3.4 to reconstruct the
dielectric constant in the belt of Figure \ref{fig:F3D_1}-c). Figure \ref%
{fig:F3D_7}-a) presents reconstruction of the dielectric constant $c(x)$ for
the unrealistic case when we know the exact tail function. In this case we
observe that the reconstruction is perfect.

Figure \ref{fig:F3D_7}-b) shows the maximal values of the reconstructed
function $c(x)$ when the initial tail $V_{1,1}\left( x\right) $ was computed
from the homogeneous domain with $c\left( x\right) \equiv 1$ for $x\in G$.
We observe that the location and the contrast of the explosive-like target
are reconstructed accurately. The reconstructed dielectric constant in this
test is $c_{3,2}\left( x\right) =3.54$ inside the \textquotedblleft belt",
and $c\left( x\right) =1$ at all other points of $\Omega $. We took the
number of inner iterations with respect to tails $m=2,$ and the stopping \
criterion (\ref{crit1})-(\ref{crit2}) was achieved at $n=3$, which
corresponds to $s\in \left[ 8,9\right] =\left[ s_{4},s_{3}\right] $ in (\ref%
{3.17}), (\ref{3.18}). We conclude that this reconstruction is accurate.

\textbf{Test 4}. In this test we took the tail $V_{1,1}\left( x\right) $ the
same as in our above theory, see (\ref{3.40}), (\ref{3.41}), (\ref{3.43}).
Analyzing results of Test 3 we have also decided to refine the
pseudo-frequency interval in this test. Indeed, we got our final image of
Test 3 for $s\in \left[ 8,9\right].$ Hence, we decided to take the interval
of pseudo frequencies 
\begin{equation}
s\in \left[ 8,8.85\right] ,h=0.05.  \label{6.11}
\end{equation}%
Next, we have used the algorithm of subsection 3.4. We took $m=2.$ The
stopping criterion (\ref{crit1})-(\ref{crit2}) was achieved at $n=3,$ which
corresponds to $s\in \left[ 8.7,8.75\right] =\left[ s_{4},s_{3}\right] $ in (%
\ref{3.17}), (\ref{3.18}). The reconstructed function $c_{3,2}\left(
x\right) $ is depicted on Figure \ref{fig:F3D_8}-a). In this test the
reconstructed dielectric constant is $c_{3,2}\left( x\right) =3.54$ inside
the belt and $c_{3,2}\left( x\right) =1$ at all other points of $\Omega $.
Thus, the reconstruction was again a quite accurate one.

\textbf{Test 5}. \emph{This is the most challenging test, because the medium
is a quite heterogeneous one:} \emph{there are substantial contrasts between
three values of the target function }$c(x).$ Indeed, we have used the case (%
% \ref{6.29}) for data simulation. In
this test we took the first guess for the tail $V_{1,1}\left( x\right) $ as
in our above theory see (\ref{3.40}), (% \ref{3.41}),
(\ref{3.43}), i.e. the same as the one in Test 4. Used results of Test 4 we
took now the interval of pseudo-frequencies 
\begin{equation*}
s\in \left[ 8.0,8.8\right] ,h=0.05.
\end{equation*}
Next, we have used the algorithm of subsection 3.4. However, since we know
that the dielectric constant of the human body is large, $c=80,$ then we
have truncated to 1 those values of computed functions $c_{n,i}\left(
x\right) ,$ which exceeded 10. In other words (\ref{6.100}), was replaced
with%
\begin{equation*}
\overline{c}_{n,i}\left( x\right) =\left\{ 
\begin{array}{c}
c_{n,i}\left( x\right) \text{ if }c_{n,i}\left( x\right) \in \left[ 1,10%
\right] \text{ and }x\in \overline{\Omega }, \\ 
1\text{ if either }c_{n,i}\left( x\right) <1,\text{ or }c_{n,i}\left(
x\right) >10,\text{ or }x\notin \overline{\Omega }.%
\end{array}%
\right.
\end{equation*}
We took $m=3$ and the stopping criterion (\ref{crit1})-(\ref{crit2}) was
achieved at $n=2$. The latter corresponds to $s\in \left[  8.70,8.75\right] =%
\left[ s_{3},s_{2}\right] $ in (\ref{3.17}), (\ref{3.18}). The reconstructed
function $c_{2,3}\left( x\right) $ is depicted on Figure \ref{fig:F3D_8}-b).
The reconstructed dielectric constant is $c=3.09$ inside the belt, and $c=1$
at all other points of $\Omega $. Therefore, the reconstruction is again a
quite accurate one even in this most difficult case.

\section{Summary}

\label{sec:7}

We have presented a new approximate mathematical model. This model amounts
to the truncation of the asymptotic series with respect to $1/\overline{s},$
where $\overline{s}>>1$ is the upper limit of the positive parameter of the
Laplace transform of the solution of the Cauchy problem (\ref{3.1}), (\ref%
{3.2}). However, this truncation is done only on the first iteration of our
method to ensure estimate (\ref{5.322}) for the accuracy of the first tail
function $\left\vert \nabla \widetilde{V}_{1,1}\right\vert _{1+\alpha }$. No
other \textquotedblleft special" simplifying assumptions are made. On the
basis of this new model, we have developed a new convergence analysis, which
is more realistic than the one of our first publications \cite{BK1,BK2}
about this topic. This time we estimate tail functions. Tails were not
estimated in our previous publications, and this is a significantly new
element of the convergence analysis here.

We have modified our approximately globally convergent algorithm for the
case of backscattering data. To do so, we have used a computational
observation that one can replace the unknown Dirichlet boundary condition on
the non-backscattering part of the boundary with the data obtained for the
case of the uniform background, which is assumed to be known outside of the
domain of interest (but not inside of it), see (\ref{6.6}). Therefore, our
previously developed technique for the case when the Dirichlet data are
given at the entire boundary, works. Our numerical tests confirm this.

Our numerical tests 1-4 demonstrate that the case when the first tail is
taken the same as the one for the uniform medium with $c\left( x\right)
\equiv 1$ provides almost the same results as ones for the new tail
function. Numerical studies demonstrate the accuracy of our technique. It is
worthy to note that we have obtained an accurate image even in the most
difficult case of Test 5 when the medium was quite a heterogeneous one, see (%
\ref{6.29}).

We believe that results of Tests 1-5 combined with results for blind
experimental data of \cite{KBKSNF,IEEE} and section 6.9 of \cite{BK} confirm
the validity of our approximate mathematical model, as indicated in Steps
4-6 of section 2.

\begin{center}
\textbf{Acknowledgments}
\end{center}

This research was supported by US Army Research Laboratory and US Army
Research Office grant W911NF-11-1-0399, the Swedish Research Council, the
Swedish Foundation for Strategic Research (SSF) in Gothenburg Mathematical
Modelling Centre (GMMC) and by the Swedish Institute, Visby Program.

\bigskip

\bigskip

\bigskip

\bigskip


\begin{thebibliography}{99}
\bibitem{Dan} D. J. Daniels, UWB radar for the detection of buried ordnance,
NATO/OTAN, unclassified/unlimited, can be found online at
ftp.rta.nato.int/public/PubFullText/RTO/.../\$MP-SET-120-KN.doc
\end{thebibliography}
\end{document}